# Black Holes, Singularity Theorems
# &
# The Global Structure of Spacetime
# in General Relativity

Rohan Kulkarni
Matriculation number : 3753615

Group : Elementary Particle Theory,
Institute of Theoretical Physics, University of Leipzig

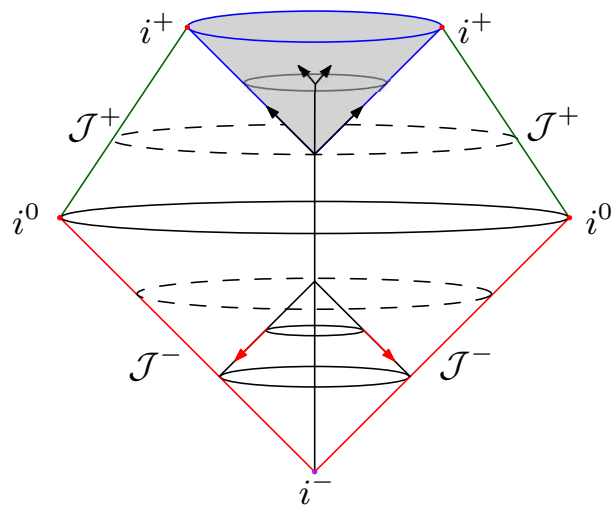

# Author's Declaration

I wrote this thesis on my own and it is primarily an extended literature review with my own comments all over the document. In order to achieve 10 ECTS for the course 12-PHY-BIPPP : Project Oriented Course, the sections 1.1, 1.2, 1.3, 1.4 and chapter 3 in a draft format were successfully submitted along with a presentation at this very same institution under Prof. Dr. Stefan Hollands. The goal of this course was to prepare for my Bachelor's thesis in which I self-studied the first seven chapters from [1] and submitted a document with the sections and chapters mentioned above.

Other than that I hereby declare that I have not used any sources other than those listed in the bibliography and identified as references. I further declare that I have not submitted this thesis at any other institution in order to obtain a degree.

Thesis submitted at the University of Leipzig in partial fulfillment of the requirements for the degree of
Bachelor of Physics IPSP
(International Physics Studies Program).

**Supervisor : Prof. Dr. Stefan Hollands**

Month and Year of Submission
April 2019

# ACKNOWLEDGMENTS

First and foremost, I thank Prof. Dr. Stefan Hollands for accepting to be my supervisor and for suggesting me this topic for my Bachelor's thesis. I also thank him for his guidance and helpful advice during this work. Apart from being the supervisor for my thesis, Prof. Hollands was also my lecturer for theoretical physics lectures in Quantum mechanics and Statistical mechanics at the University of Leipzig. I learned a great deal of Physics and Mathematics under him in these courses, project, and my thesis. I would also like to thank Dr. Zahn who agreed to be the second person to asses this thesis.

Finally, I would like to thank my family for all their support during my studies. They have always encouraged me to reach my dreams. They kept this task up even after the fact that most of them live on a different continents and I couldn't be more grateful about it.

# List of figures

The picture on the front page represents a Black hole. It was redrawn but inspired from a diagram in these lecture notes https://web.math.princeton.edu/~aretakis/columbiaGR.pdf.

# TABLE OF CONTENTS

# INTRODUCTION

In just a few years after Einstein published his famous field equations (2) in 1915, several solutions to these complicated equations were discovered. These solutions described some interesting physical scenarios and also solved big physics problems of that time such as the precession of the perihelion of Mercury. This was a big boost to the *General theory of relativity* as an experimental evidence had been found.

After this, the number of people starting to accept *General Relativity* (GR) as the model for gravity increased exponentially. Among many of the solutions, *Karl Schwarzschild's* solution was a particularly interesting one, as it described a physical object called the *Black hole*. A black hole in simple words is a super massive point-like object whose *gravity* is so strong that even light cannot escape. In the first two chapters named *Schwarzschild black holes* and *Reissner Nordström Black holes* respectively, we discuss such an object and it's properties with respect to Einstein's field equations. In the first chapter we assume a stationary, non-rotating, uncharged Black hole which is known as the *Schwarzschild Black hole*. In the second chapter we discuss a similar object but instead assume a *charged black hole* which is also known as the *Reissner-Nordström black hole* named after *Hans Reissner and Gunnar Nordström* who discovered this solution.

It was noticed that in many examples of the solutions to Einstein's field equations, there were problematic regions where we would encounter a mathematical and a physical catastrophe. In these *places* there was a breakdown of not only *General Relativity*, but of all physics. These mysterious places were called *singularities* and in most of the cases they seemed unavoidable.

However, the physicists that were working on General Relativity in its early days did not accept these singularities as a problem. They dismissed these as mathematical artifacts due to the symmetry of the exact known solutions to Einstein's equations. Physicists like *Alexander Friedman, Georges Lemaitre, J.Rober Oppenheimer and many more* in 1950's started to consider this problem more seriously. History of how people starting considering these singularities again after decades is discussed in detail in Appendix B called *Singularity theorems : History*.

After several papers published by many physicists trying to *solve* this problem of singularities, physics finally got what it needed to clarify the nature of these singularities. These are called singularity theorems. In 1955 (just a few days after Einstein had passed away), *Amal Kumar Raychaudhuri* and *Arthur Komar* independently published what could be considered the first singularity theorem. This was the start of a new field in General relativity. Surprisingly, it took nearly half a century after the birth of General relativity to start a brand new field in it which was in no way predicted by Einstein himself.

Ten years later in 1965, *Roger Penrose* wanted to prove that singularities were not formed due to the assumption of spherical symmetry in the cosmological and astrophysical models by introducing several new ideas. As a result, he ended up proving his own singularity theorem. This theorem could be considered as the first *modern* singularity theorem in a sense that a new whole set of important concepts and developments were used in the making. This theorem inspired a lot of works particularly those by *Stephen Hawking* who also published a singularity theorem in 1967. After a few years of this, in 1970 Hawking and Penrose collected a lot of the newly developed information in a very strong theorem, which is still considered as the *main singularity theorem*. We need a precise mathematical language to express a lot of new concepts while dealing with singularity theorems. This is built up in chapter four called *Singularity theorems : Congruences and Raychaudhuri equation.* This is the how the structure of the document looks like till chapter four.

In chapter five called *General black holes and their properties*, we use all these tools that we have learned and apply them to define a *asymptotic spacetimes*, *black hole* and it's *event horizon* from a topological perspective. We will end this chapter with a detailed derivation of the *Black hole area theorem* by Hawking.

In chapter six called *Exotic topics in General relativity*, we will talk about some more bizarre predictions by General relativity like *closed timelike curves* and *warp drives*. This chapter is a bit disconnected to the first six chapters but the author has tried to make connections to the previous chapters wherever possible. The reason behind addition of this chapter was to add a few topics that are not found in standard texts easily. The author had to read papers and summarize them to include it here.

The thesis is ended with three appendices. Appendix A discusses some elementary mathematical definitions for maps between manifolds and Lie derivative. In appendix B, a detailed example of how to draw a Penrose-Carter diagram has been portrayed along with an algorithm. In Appendix C, the author discusses how singularity was dealt with before the first singularity theorem (This was also discussed in brief in the introduction).

# Notations and Conventions

## Mathematical and Physical

The author follows conventions from Wald's book [1]. He follows the metric signature to be $(-+++)$ and uses abstract index notation (Page 23, Section 2.4, [1]).

In a nutshell, indices for tensors are denoted by Latin symbols $a, b, c..., x, y, z$. An object $T^{m_1...m_k}{}_{n_1...n_l}$ would be a $(k, l)$ tensor and not the components of the tensor $T(k, l)$. Comparing to the notation used in mathematical books we would say $T \equiv T^{m_1...m_k}{}_{n_1...n_l}$. In short, an object with Latin indices would be the object itself and not it component. The components of such an object would be denoted by transforming it's Latin indices to Greek indices. Hence the components of $T^{m_1...m_k}{}_{n_1...n_l}$ would be denoted by $T^{\mu_1...\mu_k}{}_{\nu_1...\nu_l}$ and these components will hold true in a specific basis. We will follow the Einstein's summation convention where repeated indices (one up & one down) will be summed over. Einstein's field equations will be written in the following manner

$$G_{ab} = R_{ab} - \frac{1}{2} R g_{ab} = 8\pi T_{ab}$$

We will also use Geometrized units (Appendix F, [1]) $c = G = 1$ starting from section 1.2.

## Textual

The author has tried to maintain a uniform structure throughout the book where he uses *italics* **in statements** to *emphasize* on words. Other than that, *italics* are also used to highlight *keywords* in any context. Theorem statements and definitions are also in *italics*.

The author has also tried to maintain a uniform structure of important citation throughout the text which looks like the following (the font is altered)

(Page #, Section #, [1])

At the end of every chapter the author has a section called *Bibliographical notes* where all the resources are noted systematically and at times a few remarks are made whenever necessary.

## Prerequisites

The prerequisites to read the document would be an elementary understanding of General Relativity from a standard text. Chapter 1, 2, 3, 4, 6 and appendices A, B, C from [1] would be sufficient and recommended.

## A few remarks by the author

The type of this thesis can be classified as an *extensive literature review*. Different topics from various resources were bought together and stitched in such a way that someone who reads this thesis shall gain an insight and understanding on the topics discussed. The author has made several comment's and has tried to give several insights that are his own, and are not found in any of the resources. Highlighting each such comment is an astronomical task. Big chunks of derivation or intuition given by the author have been highlighted in an obvious way.

This is a Bachelor's thesis and General relativity is generally considered in most of the Universities as a part of Graduate coursework. The author hopes that the person assessing this thesis appreciates the fact that writing a Bachelor's thesis which requires General relativity as a prerequisite can be challenging. As an additional fact, author has no previous academic degrees and this is the first time he has worked on such a document which is being submitted as a thesis. The author has tried his best to keep the work lucid and understandable. The author apologies for any typos or mistakes.

The author hopes that any person who is reading this thesis is able to learn something new and are able to enjoy themselves as much as the author enjoyed typing this thesis.

# CHAPTER 1

## SCHWARZSCHILD BLACK HOLES

> The goal of this chapter is to make the reader familiar with the idea of a Schwarzschild black hole. This was one of the earliest exact solutions to the Einstein's field equations which successfully modeled a Black hole. The metric and numerous properties for the Schwarzschild solution will be explicitly derived and discussed.

We start by using basic concepts of General relativity to understand a specific type of a black hole spacetime. We will use Einstein's equations in vacuum assuming a *spherically symmetric* gravitational field and analyze the results that we get.

In particular, we are talking about the Schwarzschild spacetime. The solution of Einstein's field equations for a Black hole (non-rotating and uncharged) in a vacuum is called the *Schwarzschild metric.*

The following equation is described as the Schwarzschild equation

$$\mathrm{d}s^2 = -\left(1 - \frac{2GM}{r}\right)\mathrm{d}t^2 + \left(1 - \frac{2GM}{r}\right)^{-1}\mathrm{d}r^2 + r^2\,\mathrm{d}\theta^2 + r^2\sin^2\!\theta\,\mathrm{d}\varphi^2$$

$$= -\left(1 - \frac{2GM}{r}\right)\mathrm{d}t^2 + \left(1 - \frac{2GM}{r}\right)^{-1}\mathrm{d}r^2 + r^2\mathrm{d}\Omega^2 \qquad (1)$$

in spherical coordinates $(t, r, \theta, \phi)$ which describes a *body in a vacuum in a spherically symmetric gravitational field.* We use the notation $\mathrm{d}\Omega^2 \equiv \mathrm{d}\Omega_2^2$ to denote the metric of a two-sphere

$$\mathrm{d}\Omega^2 = \mathrm{d}\theta^2 + \sin^2\!\theta\,\mathrm{d}\varphi^2$$
$$M = \text{Mass of the gravitating object.}$$
$$G = \text{Newton's gravitation constant}$$

We want to get this metric as the solution to the *spherically symmetric vacuum Einstein's equation.*

We have our Einstein's equation as :

$$R_{mn} - \frac{1}{2}Rg_{mn} = 8\pi G T_{mn} \qquad (2)$$

With some rearrangement and substitution we can write this as :

$$R_{mn} = 8\pi G\left(T_{mn} - \frac{1}{2}T g_{mn}\right) \qquad (3)$$

We have $T_{mn} = 0$ in vacuum, which also implies $T = 0$. The equation which we want to solve and find our Schwarzschild metric is :

$$R_{mn} = 0 \qquad (4)$$

## 1.1. Derivation of the Schwarzschild metric.

In this section we derive the Schwarzschild metric. We will use Birkhoff's Theorem. A more detailed explanation is in  ([1], Chapter 6) and ([2], Chapter 5).

---

**Theorem** 1.1.1. *Birkhoff's theorem*

*Schwarzschild metric is the unique vacuum solution with a spherical symmetry and there are no time-dependent solutions of this form.*

$$\mathrm{d}s^2 = g_{mn}\mathrm{d}x^m \mathrm{d}x^n = -f(r)\mathrm{d}t^2 + h(r)\mathrm{d}r^2 + r^2(\mathrm{d}\theta^2 + \sin^2\theta\,\mathrm{d}\varphi^2) \qquad (1.1.1)$$

**Proof.** *The proof of this is quite intricate with a lot of details and background reading. One can look it up in* Page 197, Chapter 5, [2]. $\qquad \square$

---

We can make an ansatz using the *Birkhoff's theorem* for our metric in the coordinate system : $x^\mu = (x^0, x^1, x^2, x^3) = (t, r, \theta, \varphi)$

$$\mathrm{d}s^2 = g_{mn}\mathrm{d}x^m \mathrm{d}x^n = -f(r)\mathrm{d}t^2 + h(r)\mathrm{d}r^2 + r^2(\mathrm{d}\theta^2 + \sin^2\theta\,\mathrm{d}\varphi^2) \qquad (1.1.2)$$

$$g_{mn} = \begin{bmatrix} -f(r) & 0 & 0 & 0 \\ 0 & h(r) & 0 & 0 \\ 0 & 0 & r^2 & 0 \\ 0 & 0 & 0 & r^2\sin^2\theta \end{bmatrix} \qquad (1.1.3)$$

$$g^{mn} = \begin{bmatrix} -\frac{1}{f(r)} & 0 & 0 & 0 \\ 0 & \frac{1}{h(r)} & 0 & 0 \\ 0 & 0 & \frac{1}{r^2} & 0 \\ 0 & 0 & 0 & \frac{1}{r^2\sin^2\theta} \end{bmatrix} \qquad (1.1.4)$$

Now let us calculate $R_{00} = R_{tt} = 0$ and $R_{11} = R_{rr} = 0$ using the following formulae :

$$R_{mn} = R^l{}_{mln}$$
$$= \sum_l \left[ \partial_l \Gamma^l_{nm} - \partial_n \Gamma^l_{lm} + \sum_a \left( \Gamma^l_{la}\Gamma^l_{nm} - \Gamma^l_{na}\Gamma^q_{lm} \right) \right] \qquad (1.1.5)$$

$$\Gamma^q_{bc} = \frac{1}{2}g^{al}(\partial_b g_{cl} + \partial_c g_{lb} - \partial_l g_{bc}) \qquad (1.1.6)$$

We first calculate our non-vanishing Christoffel (symmetric with the lower indices) symbols. (Long and tedious exercise, but the author did it once by hand to get a feel of these calculations and confirmed using Gravipy, [9] and [5]).

The following code was used to calculate different Christoffel symbols. As an example the following bracket $(-x, y, z)$ in the computation below corresponds to $\Gamma^x_{yz}$ where the minus sign denotes an upper index. The index counting starts from 1 instead of zero.

Python 3.6.5 |Anaconda, Inc.| (default, Mar 29 2018, 13:32:41) [MSC
v.1900 64 bit (AMD64)]
Python plugin for TeXmacs.
Please see the documentation in Help -> Plugins -> Python

```
>>> from sympy import *
    init_printing(use_unicode=True)
    x, y, z, t, r = symbols('x y z t r')
    f= Function('f')(r)
    h = Function('h')(r)
    k, m, n = symbols('k m n', integer=True)
    init_printing(use_unicode=True)
    a ,b ,c, L, d = symbols('a b c L d')
```

```
>>> from gravipy import *
```

```
>>> t, r, theta, phi = symbols('t, r, theta, phi')
```

```
>>> x = Coordinates('\chi', [t, r, theta, phi])
    M = symbols('M', integer=true)
```

```
>>> x(-4)
```

phi

```
>>> Metric = diag(-f, h, r**2, r**2*sin(theta)**2)
```

```
>>> Metric
```

Matrix([[-f(r), 0, 0, 0], [0, h(r), 0, 0], [0, 0, r**2, 0], [0, 0, 0,
r**2*sin(theta)**2]])

```
>>> g = MetricTensor('g', x, Metric)
    Ga = Christoffel('Ga', g)
```

```
>>> for i in {-4,-3,-2,-1}:
        for j in range(1,3):
            for k in range(1,3):
                print((int(i),int(j),int(k),(Ga(i,j,k))))
```

```
(-4, 1, 1) 0
(-4, 1, 2) 0
(-4, 2, 1) 0
(-4, 2, 2) 0
(-3, 1, 1) 0
(-3, 1, 2) 0
(-3, 2, 1) 0
(-3, 2, 2) 0
(-2, 1, 1) Derivative(f(r), r)/(2*h(r))
(-2, 1, 2) 0
(-2, 2, 1) 0
(-2, 2, 2) Derivative(h(r), r)/(2*h(r))
(-1, 1, 1) 0
(-1, 1, 2) Derivative(f(r), r)/(2*f(r))
(-1, 2, 1) Derivative(f(r), r)/(2*f(r))
(-1, 2, 2) 0
```

```
>>>
```

```
>>> for i in {-4,-3,-2,-1}:
        for j in range(3,5):
            for k in range(3,5):
                print((int(i),int(j),int(k)),(Ga(i,j,k)))
```

```
(-4, 3, 3) 0
(-4, 3, 4) sin(2*theta)/(2*sin(theta)**2)
(-4, 4, 3) sin(2*theta)/(2*sin(theta)**2)
(-4, 4, 4) 0
(-3, 3, 3) 0
(-3, 3, 4) 0
(-3, 4, 3) 0
(-3, 4, 4) -sin(2*theta)/2
(-2, 3, 3) -r/h(r)
(-2, 3, 4) 0
(-2, 4, 3) 0
(-2, 4, 4) -r*sin(theta)**2/h(r)
(-1, 3, 3) 0
(-1, 3, 4) 0
(-1, 4, 3) 0
(-1, 4, 4) 0
```

```
>>>
```

($' \equiv \frac{\partial}{\partial x^1} = \frac{\partial}{\partial r}$ denotes derivative with respect to $x^1 = r$)

| $\Gamma^r_{tt} = \frac{1}{2}\frac{f'(r)}{h(r)}$ | $\Gamma^t_{tr} = \frac{1}{2}\frac{f(r)}{h(r)}$ | $\Gamma^r_{rr} = \frac{1}{2}\frac{h'(r)}{h(r)}$ |
|---|---|---|
| $\Gamma^r_{\theta\theta} = \frac{-r}{h(r)}$ | $\Gamma^r_{\varphi\varphi} = -\frac{r\sin^2\theta}{h(r)}$ | $\Gamma^\theta_{\theta r} = \frac{1}{r}$ |
| $\Gamma^\theta_{\varphi\varphi} = -\sin(\theta)\cos(\theta)$ | $\Gamma^\varphi_{r\varphi} = \frac{1}{r}$ | $\Gamma^\varphi_{\varphi\theta} = \frac{\cos\theta}{\sin\theta} = \cot\theta$ |

**Table 1.1.1.** Non vanishing Christoffel symbols for Schwarzschild metric.

We now calculate our non-vanishing Ricci tensor components using the above formula. (Thankfully the non-diagonal components of Ricci tensor for this metric are zero, so we just need to calculate the diagonal components).

$$R_{tt} = 0 = -\frac{f''}{2h} + \frac{f'}{4h}\left(\frac{f'}{f} + \frac{h'}{h}\right) - \frac{f'}{rh} \tag{1.1.7}$$

$$R_{rr} = 0 = \frac{f''}{2f} - \frac{f'}{4f}\left(\frac{f'}{f} + \frac{h'}{h}\right) - \frac{h'}{rh} \tag{1.1.8}$$

$$R_{\theta\theta} = 0 = \frac{1}{h} - 1 + \frac{r}{2h}\left(\frac{f'}{f} - \frac{h'}{h}\right) \tag{1.1.9}$$

$$R_{\varphi\varphi} = 0 = \left(\frac{1}{h} - 1 + \frac{r}{2h}\left(\frac{f'}{f} - \frac{h'}{h}\right)\right)\sin^2\theta = R_{\theta\theta}\sin^2\theta \tag{1.1.10}$$

The solution to (4) are the above equations obtained by setting all these diagonal Ricci tensor components to zero. We just need the first three equations as the forth one will not give us any more information.

Let us multiply $R_{tt}$ with $\frac{h}{f}$ and add it to $R_{rr}$ i.e. add the following two equations

$$-\frac{f''}{2f} + \frac{f'}{4f}\left(\frac{f'}{f} + \frac{h'}{h}\right) - \frac{f'}{rf} = 0 \tag{1.1.11}$$

$$\frac{f''}{2f} - \frac{f'}{4f}\left(\frac{f'}{f} + \frac{h'}{h}\right) - \frac{h'}{rh} = 0 \tag{1.1.12}$$

We get the result

$$f'h + fh' = 0 \tag{1.1.13}$$

We can rearrange this result as follows

$$\frac{f'}{f} = -\frac{h'}{h} \tag{1.1.14}$$

We can solve this as follows

$$\int \frac{\mathrm{d}f}{f} = -\int \frac{\mathrm{d}h}{h} \tag{1.1.15}$$

$$\ln(f) = \ln\frac{1}{h} + \ln\alpha \qquad (\alpha = \text{constant})$$

$$f = \frac{\alpha}{h}$$

$$fh = \alpha \tag{1.1.16}$$

$$h = \frac{\alpha}{f} \tag{1.1.17}$$

Let us substitute this $h$ and $h' = -\frac{\alpha}{f^2}f'$ in $R_{\theta\theta}$ (1.1.9) and we get the following equation

$$\frac{1}{h} - 1 + \frac{r}{2h}\left(\frac{f'}{f} + \frac{h'}{h}\right) = 0 \tag{1.1.18}$$

$$\frac{f}{\alpha} - 1 + \frac{rf}{2\alpha}\left(\frac{f'}{f} + \frac{f'}{f}\right) = 0$$

$$\frac{f}{\alpha} - 1 + \frac{rf'}{\alpha} = 0$$

$$f + rf' = \alpha$$

$$\frac{d}{dr}(rf) = \alpha$$

$$rf = \alpha r + \tilde{k} \tag{1.1.19}$$

$$rf = \alpha(r + k) \tag{1.1.20}$$

Using (1.1.17) and (1.1.20) we get our functions $f(r)$ and $h(r)$ from our ansatz as :

$$f(r) = \alpha\left(1 + \frac{k}{r}\right) \tag{1.1.21}$$

$$h(r) = \left(1 + \frac{k}{r}\right)^{-1} \tag{1.1.22}$$

One can easily check that these satisfy all the four equations that we get from (4).

Using the weak-field limit (Derivation: Chapter 4, [1] and [2] ), we can get $k$ and $\alpha$ to have some physical meaning using the following relation :

$$\frac{f(r)}{c^2} \to 1 + \frac{2\Phi}{c^2} \tag{1.1.23}$$

where $\Phi = -\dfrac{GM}{r}$ where $G$ is the Newton's gravitational constant, $M$ is that mass of the gravitating object and $r$ is the radial coordinate. From this we get,

$$\alpha = c^2 \tag{1.1.24}$$

$$k = -\frac{2GM}{c^2} \tag{1.1.25}$$

And thus we have derived the Schwarzschild metric for an empty spacetime outside a spherical body of mass $M$ as

$$ds^2 = -c^2\left(1 - \frac{2GM}{c^2 r}\right)dt^2 + \left(1 - \frac{2GM}{r}\right)^{-1}dr^2 + r^2d\theta^2 + r^2\sin^2\theta\,d\varphi^2 \tag{1.1.26}$$

$$g_{mn} = \begin{bmatrix} -c^2\left(1 - \dfrac{2GM}{c^2 r}\right) & 0 & 0 & 0 \\ 0 & \left(1 - \dfrac{2GM}{r}\right)^{-1} & 0 & 0 \\ 0 & 0 & r^2 & 0 \\ 0 & 0 & & r^2\sin^2\theta \end{bmatrix} \tag{1.1.27}$$

## 1.2. SINGULARITIES IN SCHWARZSCHILD METRIC

We know how our Schwarzschild metric looks in the normal coordinates $\{t, r, \theta, \varphi\}$ in (1.1.26). If you look at this line element carefully you will see that it is singular at $r = 0$ and $r_s = \dfrac{2GM}{c^2}$ . (We will use geometrized units where $c = G = 1$, for reference look into Appendix F, [1])

$r_s$ is known as **_Schwarzschild radius_**. Equation (1.1.26) is a solution to the vacuum field equations (4) and hence this solution is valid only for the surfaces of spherical matter distribution in vacuum. Let us see the Schwarzschild radius of two common objects in nature i.e. a normal star (Sun) and the proton.

Schwarzschild radius of the sun $\to (r_s)_{\text{Sun}} = 2.95\,[\text{km}]$

Schwarzschild radius of a proton $\to (r_s)_{\text{Proton}} = 2.5 \times 10^{-54}\,[\text{m}]$

Both of these values are much smaller than their respective radii. So we see that the Schwarzschild radius for both these objects lie inside them where it isn't vacuum and hence the vacuum field equations do not apply. We want to study the objects that are governed by the Schwarzschild metric for radii smaller than $2GM$ . Such objects are called Black holes. (The more precise definition of a Black hole will be in Section 5.2)

### 1.2.1. Radial null curves of Schwarzschild metric

In this section we look at the Radial null curves of the Schwarzschild metric. When $\theta$, $\varphi =$ constant and $ds^2 = 0$ we call these curves as Radial null curves.

$$ds^2 = -\left(1 - \frac{2GM}{r}\right)dt^2 + \left(1 - \frac{2GM}{r}\right)^{-1}dr^2 = 0 \tag{1.2.1}$$

Rearranging the equation we see the following equation

$$\frac{dt}{dr} = \pm\left(1 - \frac{2GM}{r}\right)^{-1} \tag{1.2.2}$$

If we take a spacetime diagram in the $t$-$r$ plane then the above equation the slope of the light cones in that plane. For $r \to \infty$ the slope is $\pm 1$ and for $r \to 2GM$ the slope is approaching $\pm\infty$. In the latter case what we interpret physically is that the light cones *close up* as they are approaching $r = 2GM$. The following figure is a pictorial depiction of what one might visualize

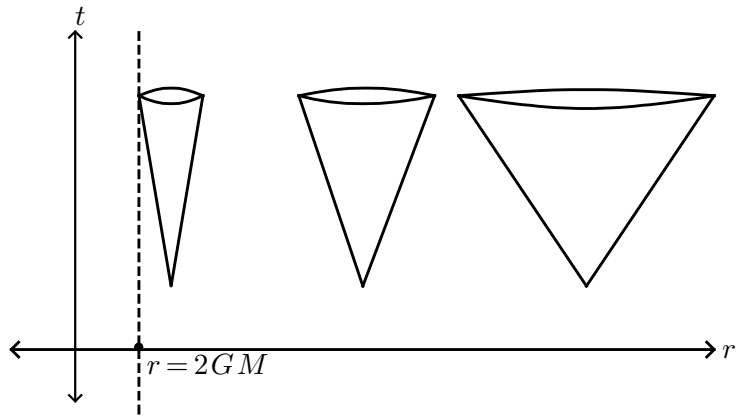

**Figure 1.2.1.** Closing of light cones near the Schwarzschild radius

(The figure above was redrawn but is inspired from Fig 5.7, Page 219, [2])

Thus a light ray that approaches this Schwarzschild radius never seems to reach there at least in this coordinate system (in this chart). So we are talking about something that is coordinate dependent and as always in physics we want something that is either invariant in all coordinate systems or we should be able to see what is the physical situation when we change the coordinate systems. In order to do this let us try to see what is the problem with the coordinate system that we are using. The problem with our current coordinates is that $\frac{dt}{dr} \to \infty$ along the radial null geodesics when approaching $r = r_s$. We can try to fix this problem by mathematically slowing down our $t$ coordinate along the $r$ direction i.e. it will move more slowly along the null geodesics.

We can solve (1.2.2) and obtain ,

$$t = \pm\left(r + 2GM\ln\left(\frac{r}{2GM} - 1\right)\right) + \text{constant} \tag{1.2.3}$$

We can define our *tortoise coordinate* (tortoise as it is moving slower along the null geodesic) by,

$$r^* = r + 2GM \ln\left(\frac{r}{2GM} - 1\right) \tag{1.2.4}$$

In terms of the tortoise coordinates with a bit of an exercise our Schwarzschild metric becomes,

$$ds^2 = \left(1 - \frac{2GM}{r}\right)(-dt^2 + dr^{*2}) + r^2\,d\Omega^2 \tag{1.2.5}$$

$r$ is considered as a function of $r^*$.

    This is some improvement as the cones do not close up like in the original coordinates.

## 1.3. Kruskal and Eddington Finkelstein coordinates. (Using Tortoise coordinates).

For the upcoming section, let's define $m = GM$ ,

### 1.3.1. Kruskal Coordinates for $r > 2m$

We consider $r > 2M$ part of the manifold and then define the Kruskal coordinates as follows :

$$u \equiv t - r^* \tag{1.3.1}$$
$$v \equiv t + r^* \tag{1.3.2}$$
$$\text{where} \quad : \quad r^* = \frac{v - u}{2} \quad \text{and}$$
$$r^* \equiv r + 2m \ln\left(\frac{r}{2m} - 1\right)$$

$r^*$ is known as the *tortoise coordinate.* The advantage of this coordinate system is that when $r \to 2M$ the tortoise coordinate $r^* \to -\infty$ and we don't get a singularity at $r_s$.

    When $r > 2m$, both the new coordinates $u$ and $v$ are defined on the whole real line $\mathbb{R}$ i.e.

$$-\infty < \ u, v \ < +\infty$$

and the limit $r \to 2M$, $t \to \infty$ corresponds to $u \to +\infty$ with some finite $v$.

    We can do the following calculation to extract some more physical intuition from this coordinate system

$$
\begin{aligned}
\frac{dr^*}{dr} &= \frac{d}{dr}\left(r + 2m \ln\left(\frac{r}{2m} - 1\right)\right) \\
&= 1 + \frac{2m}{\left(\frac{r}{2m} - 1\right)}\frac{1}{2m} \\
&= 1 + \frac{2m}{r - 2m} \\
&= \frac{r - 2m + 2m}{r - 2m} \\
&= \frac{r}{r - 2m} = \frac{1}{\left(1 - \frac{2m}{r}\right)} = \left(1 - \frac{2m}{r}\right)^{-1} \tag{1.3.3}
\end{aligned}
$$

The metric in $(u, v)$ coordinate system using the result above looks like

$$
\begin{aligned}
ds^2 &= -\left(1 - \frac{2m}{r}\right)(dt^2 - (dr^*)^2) + r^2(d\theta^2 + \sin^2\theta\, d\varphi^2) \\
&= -\left(1 - \frac{2m}{r}\right)(dt - dr^*)(dt + dr^*) + r^2(d\theta^2 + \sin^2\theta\, d\varphi^2) \quad (1.3.4)
\end{aligned}
$$

$$
ds^2 = -\left(1 - \frac{2m}{r}\right)du\, dv + r^2(d\theta^2 + \sin^2\theta\, d\varphi^2) \quad (1.3.5)
$$

Rearranging the Tortoise coordinate we get

$$
\frac{r^* - r}{2m} = \ln\left(\frac{r}{2m} - 1\right) \quad (1.3.6)
$$

$$
e^{\frac{r^* - r}{2m}} = \left(\frac{r}{2m} - 1\right)
$$

$$
e^{\frac{r^* - r}{2m}} = \left(\frac{r - 2m}{2m}\right)
$$

$$
2m\, e^{\frac{r^* - r}{2m}} = r - 2m \quad (1.3.7)
$$

$$
1 - \frac{2m}{r} = \frac{2m}{r} e^{\frac{r^* - r}{2m}} = \frac{2m}{r} e^{\frac{-r}{2m}} e^{\frac{r^*}{2m}}
$$

$$
= \frac{2m}{r} e^{\frac{-r}{2m}} e^{\frac{v - u}{4m}} \quad (1.3.8)
$$

At this point we go from $(u, v) \mapsto (U, V)$ coordinate system. The $(U, V)$ coordinate system is the Kruskal coordinate system.

Using (1.3.5) and (1.3.8) we define our variables $(U, V)$ as follows,

$$
U \equiv -e^{\frac{u}{4m}} \quad (1.3.9)
$$

$$
V \equiv e^{\frac{v}{4m}} \quad (1.3.10)
$$

This is with respect to the $(u, v)$ coordinate system. With respect to the original $(t, r)$ coordinate system $(U, V)$ looks as follows,

$$
U = -\left(\frac{r}{2m} - 1\right)^{\frac{1}{2}} e^{\frac{r + t}{4m}} \quad (1.3.11)
$$

$$
V = \left(\frac{r}{2m} - 1\right)^{\frac{1}{2}} e^{\frac{r - t}{4m}} \quad (1.3.12)
$$

Now let us calculate the differentials of $(U, V)$ with respect to $(u, v)$.

$$
\frac{dU}{du} = \frac{e^{\frac{-u}{4m}}}{4m}
$$

$$
du = -\frac{4m}{U} dU \quad (1.3.13)
$$

$$
\frac{dV}{dv} = \frac{e^{\frac{v}{4m}}}{4m}
$$

$$
dv = \frac{4m}{V} dV \quad (1.3.14)
$$

We can substitute this in (1.3.5) to get our metric in $(U, V)$ coordinate system.

$$\begin{aligned} ds^2 &= -\left(1 - \frac{2m}{r}\right) du\, dv + r^2\, d\Omega^2 \\ &= -\left(1 - \frac{2m}{r}\right) \frac{16m^2}{\left(e^{\frac{v-u}{4m}}\right)} dU\, dV + r^2\, d\Omega^2 \\ &= -\left(1 - \frac{2m}{r}\right) \frac{16m^2}{\left(e^{\frac{r^*}{2m}}\right)} dU\, dV + r^2\, d\Omega^2 \\ &= -\left(\frac{r - 2m}{r}\right)\left(\frac{r - 2m}{r}\right)^{-1} \frac{32m^3}{re^{\frac{r}{2m}}} dU\, dV + r^2\, d\Omega^2 \end{aligned}$$

(1.3.15)

$$ds^2 = \frac{-32m^3}{r} e^{\frac{-r}{2m}} dU\, dV + r^2\, d\Omega^2$$

(1.3.16)

The equation (1.3.16) is known as the metric in Kruskal Coordinates.

$$ds^2 = -\frac{32m^3}{r} e^{\frac{-r}{2m}} dU\, dV + r^2 (d\theta^2 + \sin^2\theta\, d\varphi^2)$$

(1.3.17)

$$\text{with} \quad U \in (-\infty, 0) \text{ and} \quad V \in (0, +\infty)$$

Let us introduce two more coordinates which later will be handy while making diagrams.

$$T = \frac{1}{2}(U + V)$$

(1.3.18)

$$X = \frac{1}{2}(-U + V)$$

(1.3.19)

Some algebraic rearrangement gives us $dU, dV$ in terms of $T, R, dX, dT$ as follows

$$\begin{aligned} 2T + 2X &= 2V \\ \Rightarrow \quad V &= T + X \end{aligned}$$

(1.3.20)

$$\begin{aligned} 2T - 2X &= 2U \\ \Rightarrow \quad dU &= dT - dR \end{aligned}$$

(1.3.21)

Substituting this intro (1.3.16) gives us the maximally extended Kruskal metric

$$\begin{aligned} ds^2 &= -\frac{32m^3}{r} e^{\frac{-r}{2m}} dU\, dV + r^2 (d\theta^2 + \sin^2\theta\, d\varphi^2) \\ &= -\frac{32m^3}{r} e^{\frac{-r}{2m}} (dT - dX)(dT + dX) + r^2 (d\theta^2 + \sin^2\theta\, d\varphi^2) \\ &= -\frac{32m^3}{r} e^{\frac{-r}{2m}} (dT^2 - dX^2) + r^2 (d\theta^2 + \sin^2\theta\, d\varphi^2) \end{aligned}$$

(1.3.22)

$$ds^2 = \frac{32m^3}{r} e^{\frac{-r}{2m}} (-dT^2 + dX^2) + r^2 (d\theta^2 + \sin^2\theta\, d\varphi^2)$$

(1.3.23)

This is our maximally extended Kruskal metric in the Kruskal coordinates $(T, X, \theta, \varphi)$ for the region $r > 2m$ manifold for a Schwarzschild black hole.

## 1.3.2. Kruskal Coordinates for $r < 2m$

Now consider the part of the manifold with $0 < r < 2m$ and define $u, v$ like in the previous case

$$u \equiv t - r^* \tag{1.3.24}$$

$$v \equiv t + r^* \tag{1.3.25}$$

$$r = \frac{v - u}{2} \tag{1.3.26}$$

But this time with a new tortoise coordinate (slightly different for a slightly different range of $r$).

$$r^* \equiv r + 2m \ln\left(1 - \frac{r}{2m}\right) \tag{1.3.27}$$

Here    as $r \to 0, r^* \to 0$

as $r \to 0, r^* \to -\infty$

Apart from the dependence at $r \to 0/\infty$ most of the expressions that we get are extremely similar to the case where we did this for $r > 2m$

$$\frac{\mathrm{d}r^*}{\mathrm{d}r} = \frac{1}{1 - \dfrac{2m}{r}} \tag{1.3.28}$$

then the metric expression in these new coordinates also remain the same,

$$\mathrm{d}s^2 = -\left(1 - \frac{2m}{r}\right)\mathrm{d}u\,\mathrm{d}v + r^2\mathrm{d}\Omega^2 \tag{1.3.29}$$

Following the same procedure as in the previous section

$$\frac{r^* - r}{2m} = \ln\left(1 - \frac{r}{2m}\right) \tag{1.3.30}$$

$$e^{\frac{r^* - r}{2m}} = 1 - \frac{r}{2m}$$

$$e^{\frac{r^* - r}{2m}} = \frac{2m - r}{2m}$$

$$2m\,e^{\frac{r^* - r}{2m}} = 2m - r \tag{1.3.31}$$

$$\frac{2m}{r}e^{\frac{r^* - r}{2m}} = -\left(1 - \frac{2m}{r}\right) \tag{1.3.32}$$

which gives us

$$1 - \frac{2m}{r} = -\frac{2m}{r}e^{\frac{-r}{2m}}e^{\frac{v - u}{4m}} \tag{1.3.33}$$

Defining $(u, v) \mapsto (U, V)$

$$U \equiv e^{\frac{-u}{4m}} \tag{1.3.34}$$

$$V \equiv e^{\frac{v}{4m}} \tag{1.3.35}$$

and after the same calculations as in $r > 2m$ we get

$$\mathrm{d}s^2 = -\frac{32m^3}{r}e^{-\frac{r}{2m}}\mathrm{d}U\,\mathrm{d}V + r^2\mathrm{d}\Omega^2 \tag{1.3.36}$$

which is the same as (1.3.16) but this time we have $U > 0$ and $V > 0$.

We can get the same maximally extended metric as (1.3.23) by making similar $T$, $R$ substitutions.

### 1.3.3. Metric for the exterior and interior of a Schwarzschild black hole.

The following metric

$$\mathrm{d}s^2 \;=\; -\frac{32m^3}{r}e^{-\frac{r}{2m}}\,\mathrm{d}U\mathrm{d}V + r^2\,d\Omega^2 \tag{1.3.37}$$

along with $V > 0$ and $U$ extended to $-\infty < U < \infty$ together describes the interior and exterior of the Black hole.

If we choose the metric in $(T, R, \theta, \varphi)$ coordinates (Kruskal coordinates) we get the following maximally extended solution for a Schwarzschild black hole,

$$\mathrm{d}s^2 \;=\; \frac{32m^3}{r}e^{\frac{-r}{2m}}\left(-\mathrm{d}T^2 + \mathrm{d}X^2\right) + r^2(\mathrm{d}\theta^2 + \sin^2\!\theta\,\mathrm{d}\varphi^2) \tag{1.3.38}$$

The next logical step would be to have a two-dimensional diagram which would help us understand all the properties of the metric. Such diagrams are called the *Penrose-carter diagrams* and one can read about how to make them with step by step detail in the Appendix.B. We will do this in the upcoming sections but before that, let us make a last choice of coordinates and draw a simple diagram to see a few things clearly.

## 1.4. Eddington Finkelstein coordinates

The coordinate system we are referring to in this section are $(u, r, \theta, \varphi)$ with $-\infty < u < +\infty$, $0 < r < +\infty$ and the regular domains of $\theta, \varphi$ as in spherical coordinates.

Using the two tortoise coordinates defined in the previous section, we can write down the following equations

$$r^* \;\equiv\; r + 2m\ln\left|\frac{r}{2m} - 1\right| \tag{1.4.1}$$

$$u \;=\; t + r^* \tag{1.4.2}$$

Manipulating the above equations we can do the following

$$\begin{aligned}
\mathrm{d}t^2 - (\mathrm{d}r^*)^2 \;&=\; \mathrm{d}u^2 - \mathrm{d}u\,\mathrm{d}r \\
&=\; \mathrm{d}u^2 - 2\frac{\mathrm{d}r^*}{\mathrm{d}r}\mathrm{d}u\,\mathrm{d}r \\
&=\; \mathrm{d}u^2 - \frac{2\mathrm{d}u\,\mathrm{d}r}{\left(1 - \frac{2m}{r}\right)}
\end{aligned} \tag{1.4.3}$$

Using these equations we can write down the metric in these coordinates as

$$\mathrm{d}s^2 \;=\; -\left(1 - \frac{2m}{r}\right)\mathrm{d}u^2 + (\mathrm{d}u\mathrm{d}r + \mathrm{d}r\mathrm{d}u) + r^2\,\mathrm{d}\Omega^2 \tag{1.4.4}$$

The above metric covers both, the interior and exterior of the black hole. The metric is not singular at the event horizon $r = r_s$.

### 1.4.1.  The Finkelstein diagram

The Finkelstein diagram is one of the illustrative ways to visualize the Schwarzschild spacetime. We plot the $(\tilde{t}, r)$ coordinates where

$$\begin{aligned}
\tilde{t} &= u - r \\
&= t + 2m\ln\left|\frac{r}{2m} - 1\right|
\end{aligned} \tag{1.4.5}$$

This is the diagram that we are looking for,

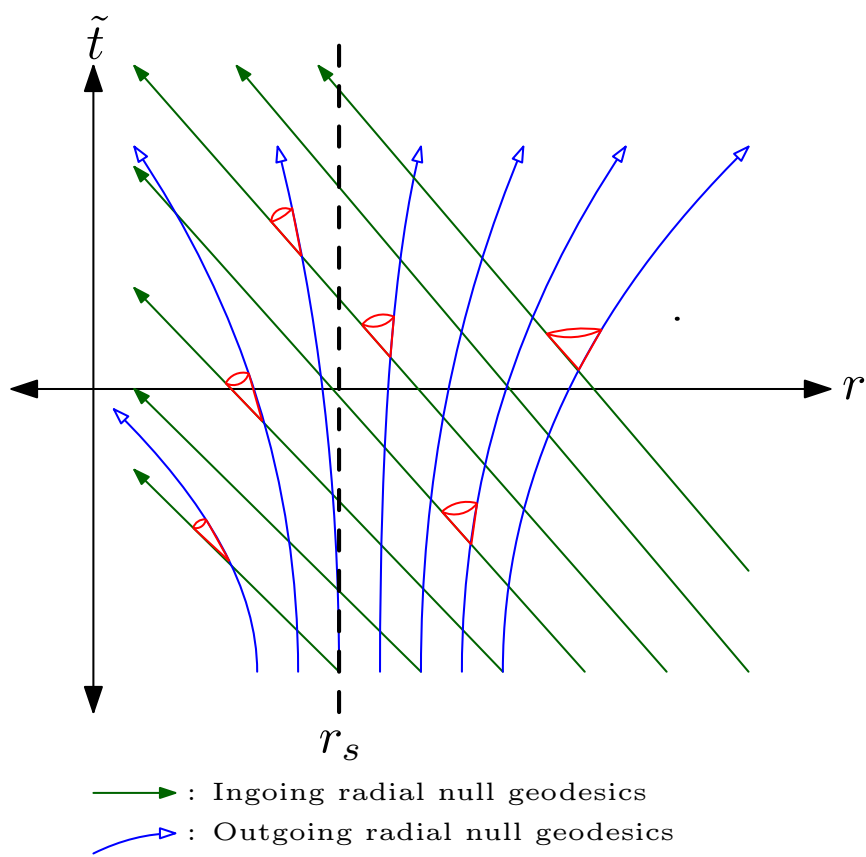

**Figure 1.4.1.** Eddington-Finkelstein diagram

(The figure above was redrawn but is inspired from **Fig 16.10, Page 220, [7]**)

The diagram has the following properties :

- The null lines $v = $ constant which correspond to the in-going massless particles are straight lines at $45°$.

- The null lines $u = $ constant which correspond to the outgoing particles are hyperbolic curves.

- These two curves together define the light cones centered at any point in the space-time giving a nice pictorial representation.

- Unlike the $(t, r)$ coordinates, for $(\tilde{t}, r)$ coordinates the light cones at the event horizon behave normally.

## 1.5. Kruskal Extension

In the previous section we have derived the full Schwarzschild metric in the form

$$\mathrm{d}s^2 \;=\; \frac{32\,m^3}{r}\,e^{\frac{-r}{2m}}\left(-\mathrm{d}T^2+\mathrm{d}X^2\right)+r^2(\mathrm{d}\theta^2+\sin^2\!\theta\,\mathrm{d}\varphi^2) \tag{1.5.1}$$

The relation between the old coordinates $(t,r)$ and the new coordinates $(T,X)$ is given by

$$\left(\frac{r}{2m}-1\right)e^{\frac{r}{2m}} \;=\; X^2-T^2 \tag{1.5.2}$$

$$\frac{t}{2m} \;=\; \ln\!\left(\frac{T+X}{X-T}\right)$$

$$\;=\; 2\tanh^{-1}\!\left(\frac{T}{X}\right) \tag{1.5.3}$$

In the equation (1.5.1), $r$ is to be interpreted as $r(X,T)$ and defined by using (1.5.2). The allowed range of coordinates for $X$ and $T$ is given by the condition $r>0$, which we can yield by taking the $\lim_{r\to 0}$ (1.5.2) giving us

$$X^2-T^2 \;>\; -1 \tag{1.5.4}$$

Let us draw a spacetime diagram for the Kruskal extension and analyze its causal structure.

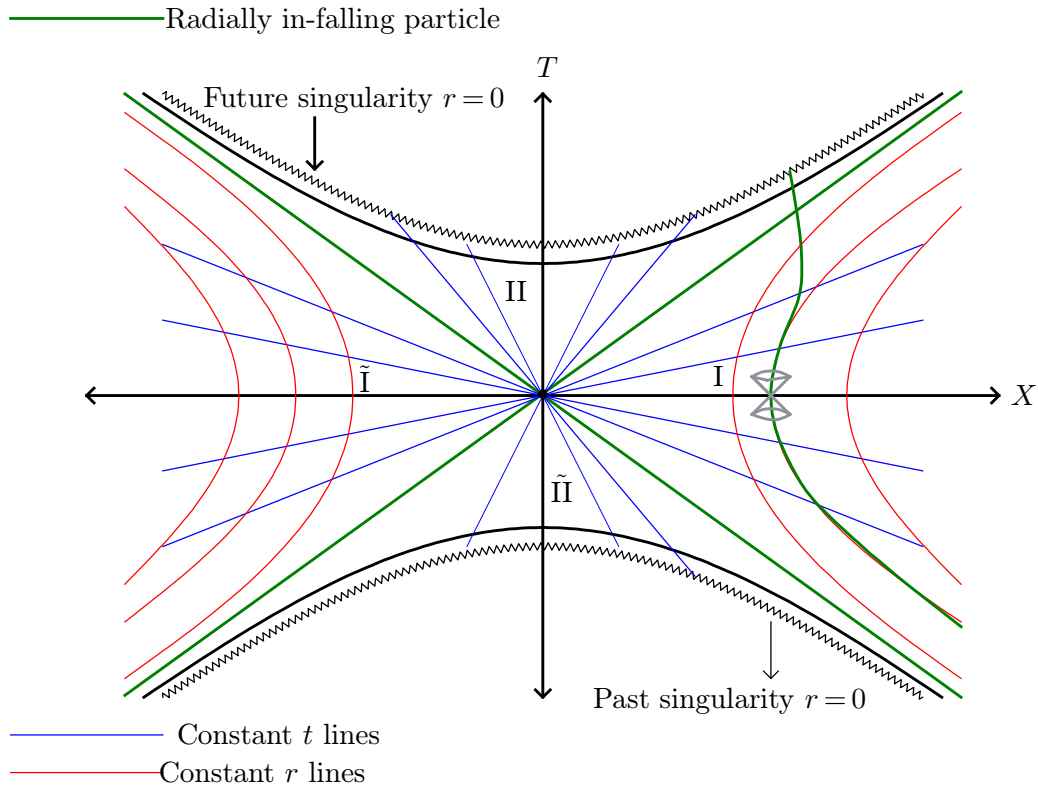

**Figure 1.5.1.** Spacetime diagram of the Kruskal coordinates.

(The figure above was redrawn but is inspired from Fig 5.12, Page 226, [2])

The diagram has the following properties :

- As expected by the construction, all the light cones are now $45°$ cones and the incoming and outgoing *radial null geodesics* are straight lines.
- Notice from equation (1.5.2), which is quadratic in $T$ and $X$, that one value of $r$ corresponds to two hypersurfaces. In two dimensions, the spacetime is bounded by two hyperbolas representing the intrinsic singularity at $r = 0$. (These are the zigzag curves in the diagram above)
    - They are termed as the *past singularity* and the *future singularity*, respectively.
- The future singularity which is spacelike and for a *globally hyperbolic spacetime* and hence it is an unavoidable in region II. (Globally hyperbolic spacetime will be defined in chapter 3).
- The asymptote to the singularity (the dark green lines in the figure above) correspond to the *event horizons* $r = 2m$ in the respective regions.
- These asymptote divide the spacetime into four regions labeled I,II,Ĩ and ĨĨ.
- The regions I and II correspond to the Eddington-Finkelstein diagram that was discussed in the previous section.
    - Region I corresponds to the Schwarzschild solution for $r > 2m$.
    - Region II corresponds to the Black hole solution.
- The regions I and ĨĨ correspond to the retarded Eddington-Finkelstein solution. (Not discussed here).
    - Region I has the same interpretation.
    - Region II has the interpretation of a **white hole.**
      (There never has had been any evidence of its existence whatsoever).

The surprising part was that there is a new region called Ĩ which is geometrically identical to the asymptotically flat exterior Schwarzschild solution.

The topology connecting I and Ĩ is complicated and we consider it in the next section.

## 1.6. Einstein-Rosen Bridge

Each point in the diagram (1.5.1) represents a 2-sphere. Let us get some intuitive idea of the overall four-dimensional structure if we consider the sub-manifold for $T = 0$. Then from equation (1.3.16) the metric induced on this hypersurface is given by

$$ds^2 = F^2 dX^2 + r^2 d\Omega^2 \tag{1.6.1}$$

As we move along the $X$-axis from $+\infty$ to $-\infty$ the value of $r$ decreases to a minimum at $2m$ at $X = 0$ and then increases again as $X$ goes to $-\infty$. We can draw a cross section of this manifold corresponding to the equatorial plane $\theta = \frac{1}{2}\pi$, in which equation (1.6.1) reduces further to

$$ds^2 = (F^2 dX^2 + r^2 d\varphi^2) \tag{1.6.2}$$

In order to interpret this we consider a two-dimensional surface possessing this line element embedded in a flat three-dimensional space. The surface appears as in the figure below where $\phi$ corresponds to our $\varphi$ and $x'$ corresponds to our $X$.

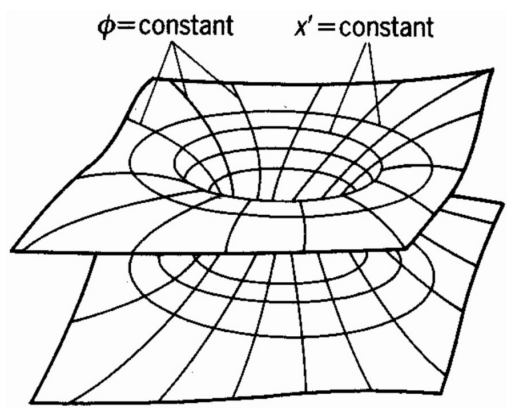

**Figure 1.6.1.** Einstein-Rosen bridge (Figure 17.2, Page 233, [7])

Therefore at $T = 0$, the Kruskal manifold can be thought as being formed by two distinct but identical asymptotically flat Schwarzschild manifolds joined at a *neck/throat* at $r = 2m$. As $T$ increases, the same qualitative picture holds but the throat narrows down, the universes join at a value $r < 2m$. At $T = 1$, the throat pinches off completely and the two universes touch at the singularity $r = 0$. For larger values of $T$, the two universes, each containing a singularity at $r = 0$, are completely separate. The Kruskal solution is time symmetric with respect $T$, and so the same thing will happen if we run time backwards from $T = 0$. The full time evolution is shown schematically in the figure below where $t'$ corresponds to our $T$.

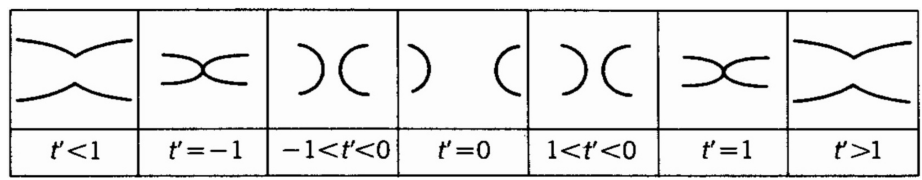

**Figure 1.6.2.** Time evolution of Einstein-Rosen bridge (Figure 17.3, Page 233, [7])

It is still an open question whether extending the solution which results in the *new universe* has any physical significance.

In the next figure we can see an embedding of the slice $T = $ constant which is geometrically identical but topologically different. This embedding leads to a Schwarzschild *wormhole* which connects two distinct regions of a single asymptotically flat universe.

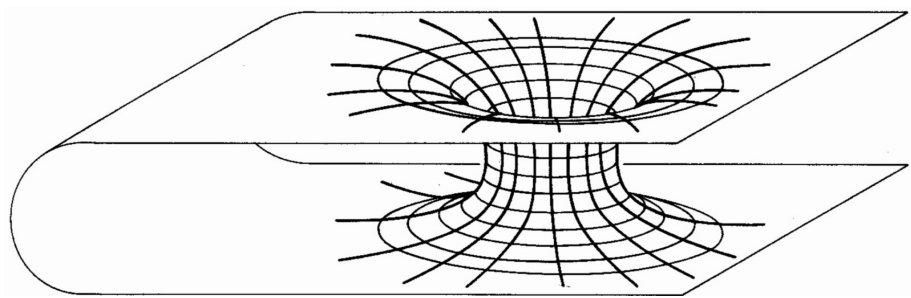

**Figure 1.6.3.** A wormhole (Figure 17.3, Page 233, [7])

## 1.7.  Penrose Diagram for the Kruskal Solution

In Appendix.B we discuss how to make a Penrose diagram for an arbitrary metric. We consider the easy Minkowski metric case over there but we go in intricate details of how to make a diagram. Using the same algorithm from there, we will form a Penrose diagram for the Kruskal metric.

Skipping the details of the calculation while summarizing it we have the following steps :

- By a careful choice of coordinates $(T', X')$ we can relabel the points of the Kruskal diagram such that light rays continue to propagate along $45°$ lines while bringing the points at infinity to a finite coordinate value. (That is indeed the whole idea of a Penrose diagram). The resulting picture of the whole slice of the Kruskal diagram (1.5.1) is called the Penrose diagram for the Schwarzschild geometry and it is an extremely useful picture for visualizing the *global* spacetime structure.

- We start with the Schwarzschild geometry in the Kruskal-Szekeres coordinates (1.5.1) and replace the coordinates $T$ and $X$ with $t$ and $x$ defined by

$$T \;=\; \frac{(t+x)}{2} \tag{1.7.1}$$

$$X \;=\; \frac{(t-x)}{2} \tag{1.7.2}$$

   − The $t$-$x$ axes are just $T$-$X$ axes rotated by $45°$ so that the light rays move on curves of constant $t$ or $x$.

- Now introduce new coordinates $(t', x')$ and $(T', X')$ defined by (For compactification purposes)

$$t' \;\equiv\; \arctan(t) \equiv T' + X' \tag{1.7.3}$$
$$x' \;\equiv\; \arctan(x) \equiv T' - X' \tag{1.7.4}$$

   − Here, light rays move on curves of constant $t'$ and $x'$ i.e. the $45°$ lines in the $T'$-$X'$ place.

   − The infinite ranges of $t$ and $x$ are each mapped to a finite range of $\left(-\frac{\pi}{2}, \frac{\pi}{2}\right)$ for $t'$ and $x'$.

- With a few calculations like in Appendix.B one can see that the hyperbola

$$r = 0,\; T > 0 \tag{1.7.5}$$

maps into the line

$$T' = \frac{\pi}{4}, \quad -\frac{\pi}{4} \leq X' \leq \frac{\pi}{4} \tag{1.7.6}$$

whereas the one with

$$r = 0,\; T < 0 \tag{1.7.7}$$

maps into

$$T' = -\frac{\pi}{4}, \quad -\frac{\pi}{4} \leq X' \leq \frac{\pi}{4} \tag{1.7.8}$$

- The horizon $T = X$ maps into the same $45°$ line as the one in the $T'$-$X'$ plane.

The results of the steps above is the following diagram,

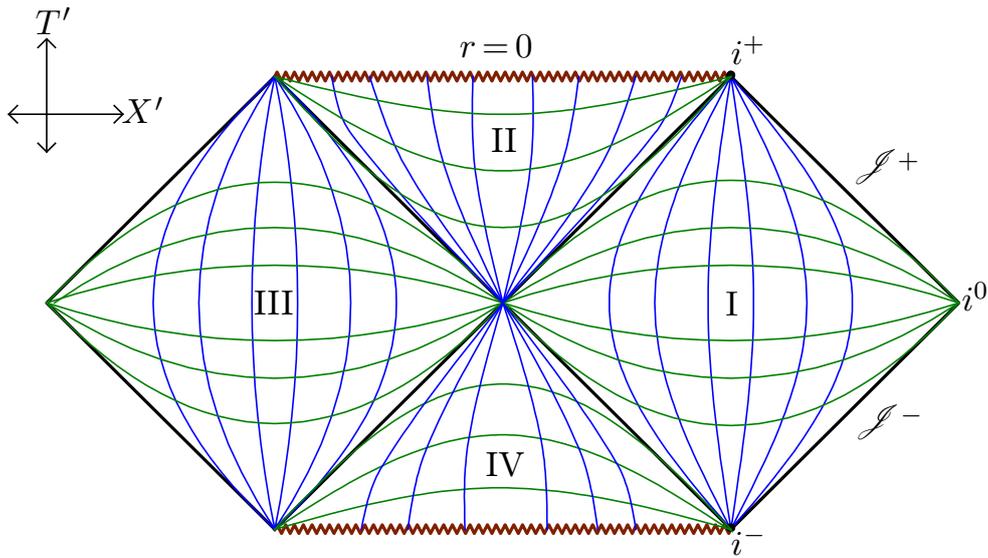

**Figure 1.7.1.** Penrose Diagram of Schwarzschild spacetime.

(The figure above was redrawn but is inspired from Fig 5.16, Page 229, [2])

Just as in the case of Minkowski space one can identify the different kinds of infinities.

* $i^0$ is called the *Spacelike infinity*
    – It is the point in spacetime where all the spacelike geodesics start and end.
* $i^+$ is called the *future timelike infinity*.
    – It is the point in spacetime where all the timelike geodesics start.
* $i^-$ is called the *past timelike infinity*.
    – It is the point in spacetime where all the timelike geodesics start.
* $\mathscr{J}^+$ is called the *future null infinity*.
    – It is the point in spacetime where all the null geodesics end.
* $\mathscr{J}^-$ is called the *past null infinity*.
    – It is the point in spacetime where all the null geodesics start.

The regions $\mathrm{I}, \mathrm{II}, \mathrm{III}, \mathrm{IV}$ have analogous meanings to the regions $I, \tilde{I}, \mathrm{II}, \tilde{\mathrm{II}}$ for the diagram (1.5.1) which were already discussed.

For a more detailed step by step derivation of this diagram one can refer to [10].

## 1.8. Bibliographical Notes

This chapter was highly influenced by the following references. Section 1.6 was more or less taken directly from the reference.

| Section | Reference | Chapter.Section | Pages |
|---------|-----------|-----------------|-------|
| 1.1 | [5] | 9.2 | 191-202 |
| 1.2 | [5] | 11.2 | 249-250 |
| 1.3 | [2] | 5.6 | 218-222 |
| 1.4 | [5] | 11.5 | 254-259 |
| 1.5 | [7] | 17.2 | 230-231 |
| 1.6 | [7] | 17.3 | 222-223 |
| 1.7 | [6] | Box 12.5 | 274 |

---***---

# Chapter 2

# Reissner-Nordström (RN) Black Holes

The RN metric is a solution to Einstein's field equations that describes the spacetime around a spherically symmetric charged non-rotating body. The only difference between this case and the Schwarzschild case is that $T_{mn} \neq 0$ anymore as we have a charge. Other than symmetry we have the assumption that the contribution to the stress-energy tensor is only due to the electromagnetic field (no matter contribution). So it should satisfy being asymptotically flat just like in the case of Schwarzschild metric. Let our body have a mass $m$ and charge $q$. Using Birkhoff's theorem we make our ansatz for the metric (spherical symmetry) (Still working where our units have $c = G = 1$).

## 2.1. Derivation of the RN metric

Derivation of the RN metric is something one does not find easily in standard texts. It is something that is left to the reader even in popular standard texts like [1],[2]. Here we will carefully go through the derivation step-by-step.

The derivation was not done **_explicitly_** in any of the references and the author takes the tiny credit for dealing it with every step carefully and deriving the final result.

We use the same coordinate system $x^\mu = (x^0, x^1, x^2, x^3) = (t, r, \theta, \varphi)$.

$$ds^2 = -f(r,t)dt^2 + h(r,t)\,dr^2 + r^2 d\theta^2 + r^2 \sin^2\theta\, d\varphi^2 \qquad (2.1.1)$$

Recalling Einstein's equations and the energy momentum tensor for electromagnetic

$$R_{ab} = 8\pi\left(T_{ab} - \frac{1}{2}T g_{ab}\right) \qquad (2.1.2)$$

$$T_{ab} = \frac{1}{\mu_0}\left(g_{bc}F_{ad}F^{cd} - \frac{1}{4}g_{ab}F_{mn}F^{mn}\right) \qquad (2.1.3)$$

$$= \frac{1}{\mu_0}\left(F_{ad}F_b^d - \frac{1}{4}g_{ab}F_{mn}F^{mn}\right)$$

$$T_a^b = g^{bb}T_{ab}$$

$$= \frac{1}{\mu_0}\left(g^{bb}g_{bc}F_{ad}F^{cd} - \frac{1}{4}g^{bb}g_{ab}F_{mn}F^{mn}\right)$$

$$= \frac{1}{\mu_0}\left(F_{ad}F^{bd} - \frac{1}{4}g_a^b F_{mn}F^{mn}\right) \qquad (2.1.4)$$

**Note 2.1.1.** $T_{ab}$ is traceless if we assume $g_{ab} = \eta_{ab}$ as we are working in vacuum (The only contribution to the energy-momentum tensor is because of the electromagnetic field).

$$T = T_a^a = \frac{1}{\mu_0}\left(F_{ad}F^{ad} - \frac{1}{4}\delta_a^a F^{mn}F_{mn}\right) \qquad (2.1.5)$$

$$= \frac{1}{\mu_0}\left(F_{ad}F^{ad} - \frac{1}{4}4 F^{ad}F_{ad}\right)$$

$$= 0 \qquad (2.1.6)$$

So we can rewrite our Einstein's equation as follows

$$R_{ab} = 8\pi T_{ab} \tag{2.1.7}$$

We need the source free Maxwell's equation in curved spacetime (basically change the partial derivative to a covariant derivative).

$$\nabla_b F^{ab} = 0 \tag{2.1.8}$$

$$\nabla_{(c} F_{ab)} = 0$$

$$= \nabla_c F_{ab} + \nabla_b F_{ca} + \nabla_a F_{bc} \tag{2.1.9}$$

We get the same equations for Christoffel symbols we got in table (1.1.1). Naturally we get the same expressions for our Ricci tensor components. (I'll copy them here so referring becomes easier). Along with that we will also need the $R_{tr} = R_{rt}$ component.

$$\left( f' \equiv \frac{\partial f}{\partial x^1} = \frac{\partial f}{\partial r}, \, g' \equiv \cdots \right)$$

$$R_{00} = R_{tt} = -\frac{f''}{2h} + \frac{f'}{4h}\left(\frac{f'}{f} + \frac{h'}{h}\right) - \frac{f'}{rh} \tag{2.1.10}$$

$$R_{11} = R_{rr} = \frac{f''}{2f} - \frac{f'}{4f}\left(\frac{f'}{f} + \frac{h'}{h}\right) - \frac{h'}{rh} \tag{2.1.11}$$

$$R_{22} = R_{\theta\theta} = \frac{1}{h} - 1 + \frac{r}{2h}\left(\frac{f'}{f} - \frac{h'}{h}\right) \tag{2.1.12}$$

$$R_{33} = R_{\varphi\varphi} = \left(\frac{1}{h} - 1 + \frac{r}{2h}\left(\frac{f'}{f} - \frac{h'}{h}\right)\right)\sin^2\theta = R_{\theta\theta}\sin^2\theta \tag{2.1.13}$$

$$R_{01} = R_{10} = R_{tr} = R_{rt} = -\frac{h'}{hr} \tag{2.1.14}$$

This is as far as we can go generalizing our spherical symmetry for our gravitational field. To find $f$ and $g$ we need to invoke Einstein's field equations for which we need $T_{ab}$. In this case knowing $F_{ab}$ will give us all the information we need about $T_{ab}$.

Having spherical symmetry, the electric field can only have a radial component. So, $E_\mu = (0, E_r, 0, 0) = F_{tr} = F_{rt}$. We can assume $B_\mu = (0, 0, 0, 0)$ as we have no currents or magnetic monopoles.

$$F_{ab} = \begin{bmatrix} 0 & E_r & 0 & 0 \\ -E_r & 0 & 0 & 0 \\ 0 & 0 & 0 & 0 \\ 0 & 0 & 0 & 0 \end{bmatrix} \tag{2.1.15}$$

Using equation (2.1.3) we can explicitly express our energy-momentum tensor, (Remember Einstein summation convention)

$$T_{ab} = \frac{1}{\mu_0}\left( g_{bc} F_{ad} F^{cd} - \frac{1}{4} g_{ab} F_{mn} F^{mn} \right) \tag{2.1.16}$$

(i) $\quad g_{bc} F_{ad} F^{cd} = g_{bc} F_{a0} F^{c0} + g_{bc} F_{a1} F^{c1} + g_{bc} F_{a2} F^{c2} + g_{bc} F_{a3} F^{c3}$

$\qquad\qquad\quad = g_{bc} F_{a0} F^{c0} + g_{bc} F_{a1} F^{c1}$

$\qquad\qquad\quad = g_{b1} F_{a0} F^{10} + g_{b0} F_{a1} F^{01} \tag{2.1.17}$

(ii) $\quad \frac{1}{4} g_{ab} F_{mn} F^{mn} = \frac{1}{4} g_{ab}(F_{m0} F^{m0} + F_{m1} F^{m1} + F_{m2} F^{m2} + F_{m3} F^{m3})$

$\qquad\qquad\qquad\quad = \frac{1}{4} g_{ab}(F_{m0} F^{m0} + F_{m1} F^{m1})$

$$
\begin{aligned}
&= \frac{1}{4} g_{ab}(F_{00}F^{00} + F_{10}F^{10} + F_{20}F^{00} + F_{30}F^{30} + \\
&\quad F_{01}F^{01} + F_{11}F^{11} + F_{21}F^{21} + F_{31}F^{31}) \\
&= \frac{1}{4} g_{ab}(F_{10}F^{10} + F_{01}F^{01}) = \frac{1}{4} g_{ab}(2F_{10}F^{10}) \\
&= \frac{1}{2} g_{ab}\, F_{10}F^{10} = \frac{1}{2} g_{ab}\, F_{01}F^{01} \quad\quad (2.1.18)
\end{aligned}
$$

$$
\begin{aligned}
T_{ab} &= \frac{1}{\mu_0}((\text{i}) - (\text{ii})) \\
&= \frac{1}{\mu_0}\left( g_{b1}F_{a0}F^{10} + g_{b0}F_{a1}F^{01} - \frac{1}{2} g_{ab}\, F_{01}F^{01} \right) \quad\quad (2.1.19)
\end{aligned}
$$

Using this we can obtain the required components $(T_{00}, T_{11}, T_{22}, T_{33})$ of the energy-momentum tensor,

$$
\begin{aligned}
T_{00} &= \frac{1}{\mu_0}\left( g_{00}\, F_{00}\, F^{10} + g_{00}\, F_{01}F^{01} - \frac{1}{2} g_{00}\, F_{01}F^{01} \right) \\
&= \frac{g_{00}}{\mu_0}\left( F_{01}F^{01} - \frac{1}{2} f_{01}F^{01} \right) \\
&= \frac{1}{2}\frac{g_{00}}{\mu_0}\, F_{01}F^{01} \\
&= -\frac{1}{2}\frac{f}{\mu_0}\, F_{01}F^{01} \quad\quad (2.1.20)
\end{aligned}
$$

$$
\begin{aligned}
T_{11} &= \frac{1}{\mu_0}\left( g_{11}\, F_{10}\, F^{10} + g_{10}F_{11}F^{01} - \frac{1}{2} g_{11}\, F_{01}\, F^{01} \right) \\
&= \frac{1}{\mu_0}\left( g_{11}F_{10}F^{10} - \frac{1}{2} g_{11}F_{01}F^{01} \right) = \frac{1}{\mu_0}\left( g_{11}F_{10}F^{10} - \frac{1}{2} g_{11}F_{10}F^{10} \right) \\
&= \frac{1}{2}\frac{g_{11}}{\mu_0}(F_{10}F^{10}) \\
&= \frac{1}{2}\frac{h}{\mu_0}(F_{10}\, F^{10}) \quad\quad (2.1.21)
\end{aligned}
$$

$$
\begin{aligned}
T_{22} &= \frac{1}{\mu_0}\left( g_{22}F_{20}F^{10} + g_{20}F_{21}F^{01} - \frac{1}{2} g_{22}F_{01}F^{01} \right) \\
&= -\frac{g_{22}}{2\mu_0}F_{01}F^{01} \\
&= -\frac{1}{2\mu_0} r^2 F_{01}F^{01} \quad\quad (2.1.22)
\end{aligned}
$$

$$
\begin{aligned}
T_{33} &= \frac{1}{\mu_0}\left( g_{33}F_{30}F^{10} + g_{30}F_{31}F^{01} - \frac{1}{2} g_{33}F_{01}F^{01} \right) \\
&= -\frac{g_{33}}{2\mu_0}F_{01}F^{01} = -\frac{r^2 \sin^2\theta}{2\mu_0}\, F_{01}\, F^{01} \\
&= T_{22}\sin^2\theta \quad\quad (2.1.23)
\end{aligned}
$$

$$
\begin{aligned}
T_{01} &= \frac{1}{\mu_0}\left( g_{11}F_{00}F^{10} + g_{10}F_{01}F^{01} - \frac{1}{2} g_{01}\, F_{01}F^{01} \right) \\
&= 0 \quad\quad (2.1.24)
\end{aligned}
$$

All the other components of $R_{ab}$ are zero so we don't need to calculate those components for $T_{ab}$.

Now we can use the Einstein's equation as we defined it at the beginning of the section :

$$R_{ab} = 8\pi(T_{ab}) \tag{2.1.25}$$

$$\begin{aligned} R_{01} = T_{01} &= 0 \\ -\frac{h'}{hr} &= 0 \\ h' &= 0 \end{aligned} \tag{2.1.26}$$

This basically means $h$ cannot depend upon $t$.

$$h(r,t) = h(r) \tag{2.1.27}$$

Looking carefully at the non-zero $T_{ab}$ components we notice,

$$\frac{T_{00}}{f} + \frac{T_{11}}{h} = 0 \tag{2.1.28}$$

$$\Rightarrow \quad 8\pi\left(\frac{R_{00}}{f} + \frac{R_{11}}{h}\right) = \frac{8\pi}{rh}\left(\frac{f'}{f} + \frac{h'}{h}\right) = 0 \tag{2.1.29}$$

$$\frac{1}{rh}\left(\frac{f'}{f} + \frac{h'}{h}\right) = 0$$

$$\left(\frac{f'}{f} + \frac{h'}{h}\right) = 0 \tag{2.1.30}$$

We know $\left(\frac{f'}{f} + \frac{h'}{h}\right) = \frac{\partial}{\partial r}\ln(fh)$ so,

$$\frac{\partial}{\partial r}\ln(fh) = 0 \tag{2.1.31}$$

That means the function $fh$ does not depend upon $r$. So we can write :

$$fh = \gamma(t) \tag{2.1.32}$$

for some function $\gamma$ that does not depend upon $r$.

We can rearrange this to get

$$f = \frac{\gamma}{h} \tag{2.1.33}$$

$$g_{00} = -\frac{\gamma}{g_{11}} \tag{2.1.34}$$

as

$$g_{00} = -f \tag{2.1.35}$$

$$g_{11} = h \tag{2.1.36}$$

We know $F_{mn} = g_{nb}g_{ma}F^{ab}$, so for $F_{01}$ we have

$$\begin{aligned} F_{01} &= g_{0a}\,g_{1b}\,F^{ab} \\ &= g_{00}\,g_{11}\,F^{01} \quad \text{(Diagonal metric)} \\ &= \frac{-\gamma}{g_{11}}g_{11}F^{01} \qquad \text{(From (2.1.34))} \end{aligned} \tag{2.1.37}$$

$$F_{01} = -\gamma F^{01} \tag{2.1.38}$$

We now use the Maxwell's equation (2.1.9) for $a=0, b=1, c=0$,

$$\begin{aligned} \nabla_0 F_{01} + \nabla_0 F_{10} + \nabla_1 F_{00} &= \nabla_0 F_{01} + \nabla_0 F_{10} \\ \rightarrow \nabla_0 F_{01} - \nabla_0 F_{01} &= 0 \end{aligned} \tag{2.1.39}$$

Similarly for other $F_{ab}$ we can easily prove the equation (2.1.9).

Now let us use equation (2.1.8) and the definition of the covariant derivative for the metric compatible connection.

$$\nabla_b F^{ab} = \partial_b F^{ab} + \Gamma^a_{mb} F^{mb} + \Gamma^b_{nb} F^{an} = 0 \tag{2.1.40}$$

We can use this with $a = 1$ to get a more explicit form of $E_r$.

$$
\begin{aligned}
\nabla_b F^{1b} &= 0 \tag{2.1.41}\\
\nabla_0 F^{10} &= \partial_b F^{1b} + \Gamma^1_{mb} F^{mb} + \Gamma^b_{nb} F^{1n}\\
&= \partial_0 F^{10} + \Gamma^1_{10} F^{10} + \Gamma^1_{01} F^{01} + \Gamma^b_{0b} F^{10}\\
&= \partial_0 F^{10} + \Gamma^1_{10} F^{10} - \Gamma^1_{10} F^{01} + \Gamma^b_{0b} F^{10}\\
&= \partial_0 F^{10} + \Gamma^b_{0b} F^{10}\\
&= \partial_0 F^{10} + F^{10}(\Gamma^0_{00} + \Gamma^1_{01} + \Gamma^2_{02} + \Gamma^3_{03})\\
&= 0 \tag{2.1.42}
\end{aligned}
$$

We had the same metric ansatz as in the Schwarzschild case. If we look up the Christoffel symbols in the previous section we see that all the Christoffel symbols in the equation above are zero. Giving us,

$$
\begin{aligned}
\nabla_0 F^{10} &= \partial_0 F^{10} = 0\\
\partial_0 F^{10} &= 0\\
\frac{\mathrm{d}}{\mathrm{d}t} E_r &= 0 \tag{2.1.43}
\end{aligned}
$$

Which gives us $E_r = E_r(r)$ i.e. it does not depend upon time. Now using the equation (2.1.40) for $a = 0$ we get,

$$
\begin{aligned}
\nabla_b F^{0b} &= 0 \tag{2.1.44}\\
\nabla_1 F^{01} &= \partial_1 F^{01} + \Gamma^0_{m1} F^{m1} + \Gamma^b_{nb} F^{0n}\\
&= \partial_1 F^{01} + \Gamma^0_{01} F^{01} + \Gamma^0_{10} F^{10} + \Gamma^b_{nb} F^{0n}\\
&= \partial_1 F^{01} + \Gamma^0_{01} F^{01} - \Gamma^0_{01} F^{10} + \Gamma^b_{nb} F^{0n}\\
&= \partial_1 F^{01} + \Gamma^b_{nb} F^{0n}\\
&= \partial_1 F^{01} + \Gamma^b_{1b} F^{01}\\
&= \partial_1 F^{01} + F^{01}(\Gamma^0_{10} + \Gamma^1_{11} + \Gamma^2_{12} + \Gamma^3_{13})\\
&= \partial_1 F^{01} + F^{01}\left(\frac{f'}{2f} + \frac{h'}{2h} + \frac{2}{r}\right) \tag{2.1.45}
\end{aligned}
$$

We have,

$$
\begin{aligned}
\left(\frac{f'}{2f} + \frac{h'}{2h} + \frac{2}{r}\right) &= \frac{1}{2}\left(\frac{\partial}{\partial r}\ln(fh)\right) + \frac{2}{r}\\
&= 0 + \frac{2}{r} \tag{2.1.46}
\end{aligned}
$$

Which gives us,

$$\nabla_1 F^{01} = \partial_1 F^{01} + F^{01}\left(\frac{2}{r}\right) = 0 \tag{2.1.47}$$

$$\partial_1 F^{01} + F^{01}\left(\frac{2}{r}\right) = 0 \tag{2.1.48}$$

We can solve this ordinary differential equation,

$$\frac{\mathrm{d}F^{01}}{\mathrm{d}r} = -\frac{2}{r}F^{01} \tag{2.1.49}$$

$$\frac{\mathrm{d}F^{01}}{F^{01}} = -\frac{2}{r}\mathrm{d}r$$

$$\ln(F^{01}) = -2\ln(r) + c$$

$$F^{01} = e^{\ln\left(\frac{1}{r^2}\right)+c} = e^{\ln\left(\frac{1}{r^2}\right)}e^c \tag{2.1.50}$$

$$= \frac{c}{r^2} \qquad c = \text{constant}$$

We know $F^{01} = E_r$ so we get,

$$E_r = \frac{c}{r^2} \tag{2.1.51}$$

Using Gauss's law for a symmetrical spherical electric field we can deduce that $E_r = \dfrac{Q}{4\pi\varepsilon_0 r^2}$ which gives us $c = \dfrac{Q}{4\pi\varepsilon_0}$. In General relativity we should always keep a track of the fact that $r$ is merely a chosen coordinate and does not necessarily measure the *actual* radial distance in the RN spacetime.

We need $f$ and $g$ in terms of $r$ and we are nearly there. Let us use the $R_{22}$ equation to get what we want,

$$R_{22} = 8\pi T_{22} \tag{2.1.52}$$

We have calculated both RHS and LHS of the above equation a few pages back. Inserting them into the equation above we get,

$$R_{22} = -\frac{r}{2h}\left(\frac{f'}{f} - \frac{h'}{h}\right) - \frac{1}{h} + 1 = 8\pi T_{22} \tag{2.1.53}$$

Using the fact that $h = \left(1 - \dfrac{2GM}{r}\right)^{-1}\dfrac{\gamma}{f}$ and $h' = \dfrac{-\gamma f'}{f^2}$ we get,

$$-\frac{r}{2h}\left(\frac{f'}{f} - \frac{-\frac{\gamma f'}{f^2}}{\frac{\gamma}{f}}\right) - \frac{1}{h} + 1 = 8\pi T_{22} \tag{2.1.54}$$

$$-\frac{r}{2h}\left(\frac{f'}{f} + \frac{f'}{f}\right) - \frac{f}{\gamma} + 1 = 8\pi\left(\frac{-1}{2\mu_0}r^2 F_{01}F^{01}\right)$$

$$-\frac{rf'}{f}\left(\frac{f}{\gamma}\right) - \frac{f}{\gamma} + 1 = 8\pi\left(\frac{-r^2}{2\mu_0}\left(\frac{-F_{01}}{\gamma}F_{01}\right)\right)$$

$$-\frac{rf'}{\gamma} - \frac{f}{\gamma} + 1 = 8\pi\left(\frac{r^2}{2\mu_0}\frac{E_r^2}{\gamma}\right)$$

$$\frac{-1}{\gamma}\frac{\partial}{\partial r}(rf) + 1 = \left(\frac{8\pi r^2}{2\mu_0\gamma}\right)\left(\frac{Q^2}{16\pi^2\varepsilon_0^2 r^4}\right)$$

$$\frac{-1}{\gamma}\frac{\partial}{\partial r}(rf) + 1 = \frac{Q^2}{4\mu_0\gamma\pi\varepsilon_0^2 r^2}$$

$$1 - \frac{Q^2}{4\mu_0\gamma\pi\varepsilon_0^2 r^2} = \frac{1}{\gamma}\frac{\partial}{\partial r}(rf)$$

$$\gamma - \frac{Q^2}{4\mu_0\pi\varepsilon_0^2 r^2} = \frac{\partial}{\partial r}(rf) \tag{2.1.55}$$

Now we use the fact that $c^2 = 1 = \dfrac{1}{\mu_0 \varepsilon_0} \Rightarrow \mu_0 = \dfrac{1}{\varepsilon_0}$. This gives us

$$\gamma - \frac{Q^2}{4\pi\varepsilon_0 r^2} = \frac{\partial}{\partial r}(rf) \tag{2.1.56}$$

We can integrate this to get an explicit expression for $f$ and then use any relation between $f$ and $h$ to get an explicit expression for $h$.

$$\gamma r + \frac{Q^2}{4\pi\varepsilon_0 r} + k(t) = rf$$

$$f = \gamma + \frac{Q^2}{4\pi\varepsilon_0 r^2} + \frac{k(t)}{r} \tag{2.1.57}$$

$k(t)$ is a function that may depend upon time. We can use $f = \dfrac{\gamma}{h}$ to get $h$.

We have seen in the previous section while deriving the Schwarzschild metric that when gravity is weak we expect the $g_{00}$ component to approach $1 - \dfrac{2m}{r}$. Basically we could get rid of $f$ by just redefining $dt \to \sqrt{\gamma(t)}\,dt$. We have defined 1 radius in the previous section as

$$r_s = \frac{2GM}{c^2} = 2M \tag{2.1.58}$$

and we can also argue that in the $Q = 0$ it should reduce to Schwarzschild metric giving us

$$k(t) = \frac{-2GM}{c^2} = -r_s \tag{2.1.59}$$

We can also define

$$r_Q^2 = \frac{GQ^2}{4\pi\varepsilon_0 c^4} = \frac{Q^2}{4\pi\varepsilon_0} \tag{2.1.60}$$

Now as we have gotten rid of $\gamma$ we have the relation $f = \dfrac{1}{g}$ . Now finally we can defined $f$ and $h$ explicitly.

$$f = 1 - \frac{r_s}{r} + \frac{r_Q^2}{r^2} \tag{2.1.61}$$

$$h = \frac{1}{f} = \left(1 - \frac{r_s}{r} + \frac{r_Q^2}{r^2}\right)^{-1} \tag{2.1.62}$$

giving us

$$ds^2 = -\left(1 - \frac{r_s}{r} + \frac{r_Q^2}{r^2}\right)dt^2 + \left(1 - \frac{r_s}{r} + \frac{r_Q^2}{r^2}\right)^{-1}dr^2 + r^2 d\theta^2 + r^2 \sin^2\theta\, d\varphi^2 \tag{2.1.63}$$

where $\quad \Delta = \left(1 - \dfrac{r_s}{r} + \dfrac{r_Q^2}{r^2}\right)$

$$g_{mn} = \begin{bmatrix} -1 - \frac{r_s}{r} + \frac{r_Q^2}{r^2} & 0 & 0 & 0 \\ 0 & \left(1 - \frac{r_s}{r} + \frac{r_Q^2}{r^2}\right)^{-1} & 0 & 0 \\ 0 & 0 & r^2 & 0 \\ 0 & 0 & 0 & r^2\sin^2\theta \end{bmatrix} \tag{2.1.64}$$

So we have finally after careful mathematical and physical consideration derived the Reissner-Nordström metric.

**Remark 2.1.1.**

We did this while assuming a spherical symmetry for our EM field. Due to no experimental evidence of magnetic monopoles we kept the $B_r$ parts of our stress energy tensor $F_{ab} = 0$. But theoretically we can still keep them in the stress energy tensor.

$$B_r \;\propto\; \frac{P}{r^2} \tag{2.1.65}$$

$$\Rightarrow\quad F_{23} \;=\; B_r r^2 \sin\theta \tag{2.1.66}$$

Basically we can introduce a theoretical magnetic mono-pole by replacing $P \to \dfrac{P^2 + Q^2}{c^2} = P^2 + Q^2$.

The RN metric has a true curvature singularity at $r = 0$ and this can be checked by computing the invariant scalar $R = R_{abcd} R^{abcd}$.

## 2.2. Few Properties of Reissner-Nordström Black holes

Without any loss of generality we will assume $Q > 0$. Let us introduce the function

$$\Delta = Q^2 - 2Mr + r^2 \;=\; (r - r_+)(r - r_-) \tag{2.2.1}$$
$$\text{where } r_\pm = M \pm \sqrt{M^2 + Q^2}$$

for convenience.

With some algebraic manipulation we can rewrite our metric (2.1.63) as

$$\mathrm{d}s^2 \;=\; -\frac{\Delta}{r^2}\mathrm{d}t^2 + \frac{r^2}{\Delta}\mathrm{d}r^2 + r^2\mathrm{d}\Omega^2 \tag{2.2.2}$$

We will have three cases to monitor because of the square root in $\Delta$. The three cases will be

1. $Q > M$ : Super-extremal RN

   In this case we won't have any real roots for $\Delta$ and is regular (non-singular) for $r > 0$. We will have a *curvature* singularity at $r = 0$ and the situation same as for the negative mass black hole (refer page 22-23,[3]).

2. $Q < M$ : Sub-extremal RN

   In this case $\Delta$ has two real roots for $r_+ > r_- > 0$ and two *coordinate* singularities. Coordinate singularities can be always removed with an appropriate choice of coordinate system. Just like in the Schwarzschild metric, let us introduce a tortoise coordinate (this time $r_*$) as

$$\frac{\Delta}{r^2}\mathrm{d}r_*^2 \;=\; \frac{r^2}{\Delta}\mathrm{d}r^2 \tag{2.2.3}$$

in terms of which the RN metric takes the form

$$\mathrm{d}s^2 \;=\; -\frac{\Delta}{r^2}(\mathrm{d}t^2 - \mathrm{d}r_*^2) + r^2\,\mathrm{d}\Omega^2 \tag{2.2.4}$$

The radial null geodesics for this are given by $t \pm r = \text{constant}$ ($\theta = \varphi = \text{constant}$).

A solution of (2.2.4) with a convenient choice of sign and an integration constant is

$$r_* \;=\; r + \frac{1}{2\kappa_+}\ln\!\left(\frac{r - r_+}{r}\right) + \frac{1}{2\kappa_-}\ln\!\left(\frac{r - r_-}{r}\right) \tag{2.2.5}$$
$$\text{where}$$
$$\kappa_+ \;=\; \frac{r_+ - r_-}{2r_+^2} > 0$$
$$\kappa_- \;=\; \frac{r_- - r_+}{2r_-^2} < 0$$

Again with similar techniques from the Schwarzschild metric, we define the null coordinates

$$u = t - r_*$$ (2.2.6)

and the Eddington-Finkelstein coordinates $(v, r, \theta, \varphi)$

$$v = t + r_*$$ (2.2.7)

In terms of the latter coordinates we get the metric

$$ds^2 = -\frac{\Delta}{r^2}dr^2 + dr\,dv + r^2\,d\Omega^2$$ (2.2.8)

which is regular for all $r > 0$, including $r = r_+$ and $r = r_-$.

To understand the spacetime structure close to $r = r_\pm$ we can use two different sets of Kruskal-type coordinates at each of the two radii

$$U^\pm = -e^{-u\kappa_\pm}$$ (2.2.9)
$$V^\pm = e^{u\kappa_\pm}$$ (2.2.10)

This gives rise to the following Penrose diagram

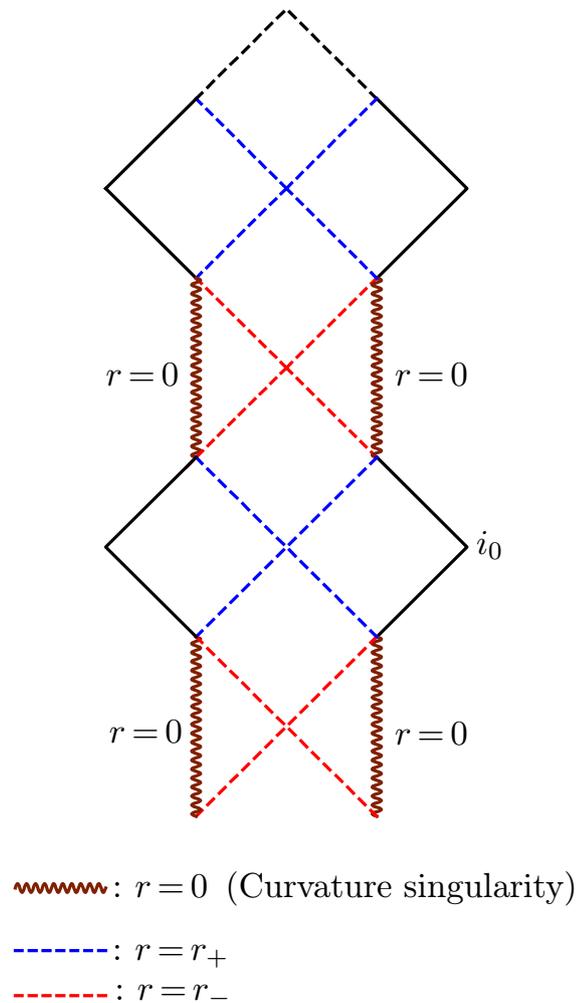

**Figure 2.2.1.** Sub-extremal RN metric diagram

(The figure above was redrawn but is inspired from Fig 6.4, Page 258, [2])

3. $Q = M$: Extremal RN

The metric of the extremal RN solution is

$$ds^2 = -\left(1 - \frac{M}{r}\right)^2 dt^2 + \left(1 - \frac{M}{r}\right)^{-2} dr^2 + r^2 d\Omega_2^2 \tag{2.2.11}$$

which has one *coordinate singularity* at $r = r_+ = r_- = M$. To get rid of this, we define the tortoise coordinate $dr_* = \left(1 - \frac{M}{r}\right)^{-2} dr$ such that

$$ds^2 = -\left(1 - \frac{M}{r}\right)^2 (dt^2 - dr_*^2) + r^2 d\Omega_2^2 \tag{2.2.12}$$

and we also change to ingoing Eddington-Finkelstein coordinates $(v, r, \theta, \phi)$, where $v = t + r_*$ label the ingoing null geodesics. This can be seen directly from (2.2.12).

After these transformations we are left with the following metric

$$ds^2 = -\left(1 - \frac{M}{r}\right) dv^2 + 2 dv dr + r^2 d\Omega_2^2 \tag{2.2.13}$$

which is regular at $r = M$. Basically the inner and outer horizons from the sub-extremal RN have been coalesced. The diagram for this can be found in **Chapter 6**, [2].

## 2.3. Bibliographical notes

This whole chapter was highly motivated by the references in the following table.

| Section | Reference | Chapter.Section | Pages |
|---------|-----------|-----------------|-------|
| 2.1 | [2] | 6.5 | 254-256 |
| 2.2 | [3] | 2.1 | 25-26 |
|     | [2] | 6.5 | 258-261 |

—***—

# CHAPTER 3
## CAUSAL STRUCTURE

> In this chapter we develop the language in which *Singularity theorems* are built. Not only that but these definitions can be used in many different areas of General relativity and also in Mathematical relativity to understand. We are going to discuss the most important definitions and concepts from *Lorentzian causality* as per our needs. We will follow Chapter 8, [1] closely but with extensive comments from the author.

In the previous sections we see that a *solution* to Einstein's field equations is some kind of a *metric tensor* $g_{ab}$ which is associated with a *4-dimensional manifold (Spacetime)*. From now on, we mathematically denote spacetime with $(M, g_{mn})$.

This metric $g_{mn}$ on any given point $p \in M$ defines a scalar product/inner product between any two vectors in the *tangent space* $T_pM$. This also defines the norm of any single vector in $T_pM$. According to the sign of the norm we have classified the vectors as (in $-+++$ notation for the Minkowski metric).

1. $g_{ab}v^a v^b < 0$ denotes a *timelike vector*.
2. $g_{ab}v^a v^b = 0$ denotes a *null vector*.
3. $g_{ab}v^a v^b > 0$ denotes a *spacelike vector*.

**Remark 3.0.1.** Time orientability

At every event $p \in M$ the tangent space is isomorphic to Minkowski spacetime. We have light cones through every event in this manifold. A light cone passing through the origin of $T_pM$ is defined as the light cone of $p$. Light cone of $p$ is a subset of $T_pM$ and not of $M$.

For each of this point $p$ we have a light cone and each of this light cone has a *future* and a *past*. If *continuous designation* of *future/past* can be made over the whole manifold $M$ then we call our spacetime $(M, g)$ to be time-orientable. The spacetimes that we will be dealing with are assumed to be time orientable.

Since a tangent vector $v_0$ at a given event $p_0 \in M$ can be thought as a velocity of a test particle passing through a point $p_0 = \gamma(t_0)$ at a given instant of time $t_0$. The curve or trajectory $\gamma(t)$ along which this point takes a path that can be identified as *spacelike,timelike* or *null*. *Timelike* and *null* can be further segregated as future or past directed.

## 3.1. DEFINITIONS OF FUTURE AND PASTS

DEFINITION 3.1.1. *Character of a point at a curve*

A curve $\gamma(t)$ is timelike (past or future directed), null (past or future directed) or space-like at $p_0 = \gamma(t)$ if it's tangent vector $\dot{\gamma}(t_0)$ at $p_0$ is timelike (past or future directed), null (past or future directed) or spacelike.

DEFINITION 3.1.2. *Global character of curve*
- A curve $\gamma$ which is timelike at every event is a timelike curve.
- A curve $\gamma$ which is spacelike at every event is a spacelike curve.
- A curve $\gamma$ which is null/lightlike at every event is a null curve.

DEFINITION 3.1.3. *Future and Past directed curve*
- A timelike or a null curve which lies in the future half of the light cone is a future directed curve.

- *A timelike or a null curve which lies in the past half of the light cone is called a past directed curve.*

## 3.2. Definitions of Causal relations between events

Definition 3.2.1. *Chronological future and chronological past of an event $p \in M$. ($I^+(p)$ and $I^-(p)$).*

*For a given event $p \in M$, the chronological future is defined by the set of all points that can be reached by a future directed timelike curve which starts at p. This set of points is denoted by $I^+(p)$.*

*For a given event $p \in M$, the chronological past is defined by the set of all points that can be reached by a past directed timelike curve which starts at p. This set of points is denoted by $I^-(p)$.*

**Remark 3.2.1.**

For a set of points $\mathcal{U}$, the chronological future/past is defined as follows:

$$I^{\pm}(\mathcal{U}) = \bigcup_{p \in \mathcal{U}} I^{\pm}(p) \qquad (3.2.1)$$

$I^{\pm}(\mathcal{U})$ is the *Chronological future/past* of $\mathcal{U}$.

Definition 3.2.2. *Causal future and causal past of an event $p \in M$. ($J^+(p)$ and $J^-(p)$).*

*For a given event $p \in M$, the causal future is defined by the set of all points that can be reached by a future directed null or timelike curve which starts at p. This set of points is denoted by $J^+(p)$.*

*For a given event $p \in M$, the causal past is defined by the set of all points that can be reached by a future directed null or timelike curve which starts at p. This set of points is denoted by $J^-(p)$.*

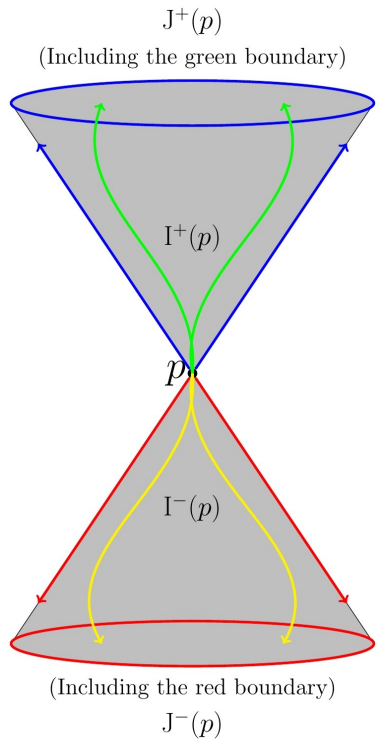

**Figure 3.2.1.** Example of a lightcone

**Definition 3.2.3.** *Future endpoint of a curve*
   *Let $\gamma$ be a future directed causal curve in M.*
*We say that $p \in M$ is a future endpoint of $\gamma$ if for every neighborhood O of p there exists $t_0 \in \mathbb{R}$ such that $\gamma(t) \in O \; \forall \, t > t_0$.*

**Definition 3.2.4.** *Future inextendible curve*
   *A causal curve $\gamma_c$ is called future inextendible if it does not have a future endpoint.*

**Definition 3.2.5.** *Past endpoint of a curve.*
   *Let $\gamma$ be a past directed causal curve in M.*
*We say that $p \in M$ is a past endpoint of $\gamma$ if for every neighborhood O of p there exists $t_0 \in \mathbb{R}$ such that $\gamma(t) \in O \; \forall \, t < t_0$.*

**Definition 3.2.6.** *Past inextendible curve*
   *A causal curve $\gamma_c$ is called past inextendible if it does not have a future endpoint.*

**Definition 3.2.7.** *Achronal sets*
   *Achronal sets $\mathcal{A}$ are subsets of spacetime M that hold the property*

$$\mathcal{A} \cap I^+(\mathcal{A}) = \emptyset \tag{3.2.2}$$

**Remark 3.2.2.**
   Intuitively what this means is that in these sets there are no such events which are in the future of another event in the set. Imagine a set of events $S$ (points in a spacetime) and any future event of these points does not belong to the set $S$. Such a set $S$ is called an *Achronal set*.
   No two events in an *achronal set* are *causally connected* to each other. What do we mean by *causally connected*? It means that no two events in an achronal set can be connected to each other by *null* or *timelike curves*.

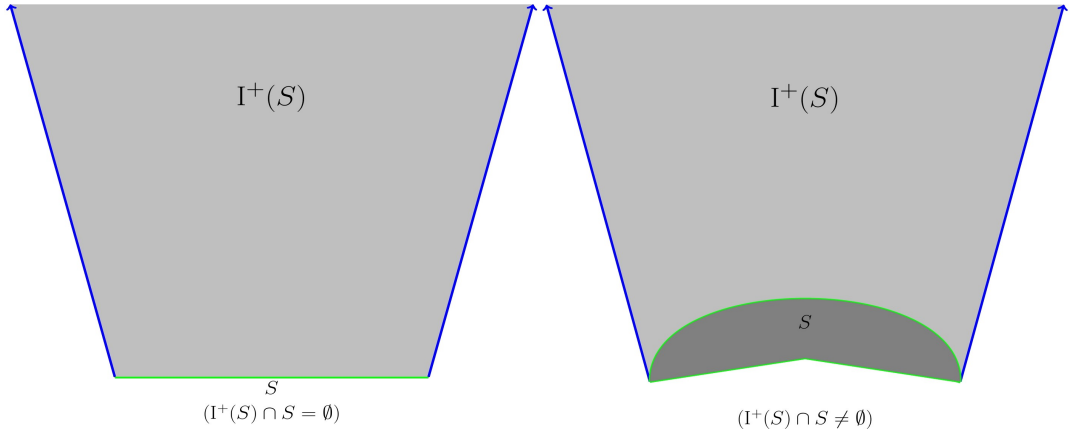

**Figure 3.2.2.** Example of an achronal set

**Definition 3.2.8.** *Edge of an achronal set*
   *Let $\mathcal{A}$ be an achronal set. Let $\partial\mathcal{A}$ be the edge of the Achronal set. Then $\partial\mathcal{A}$ is the subsets of all the events $p \in \mathcal{A}$ such that every neighborhood of p, $\mathcal{U} \in M$ contains at least a point $p_+ \in I^+(p)$ and $p_- \in I^-(p)$ and a timelike curve $\gamma_T$ connecting $p_+$ and $p_-$ where $\gamma_T \cap \mathcal{A} = \emptyset$.*

## 3.3.  Definitions of Causality, Initial conditions.

**Definition 3.3.1.** *Causal Spacetime*

    *We say that our spacetime* $(M, g)$ *is causal if it does not contain a closed causal (timelike or null) curve.*

This definition of a spacetime being *just causal* has some drawbacks. Particularly speaking, if our spacetime is arbitrarily close to being causal it could allow *timelike curves* which are not *closed*, but arbitrarily close to being *closed*. We don't like this because this allows us intuitively to *nearly* go back in time.

As an example, let's look at this diagram where the cylinder is our spacetime and the light-cones on it define our causality.

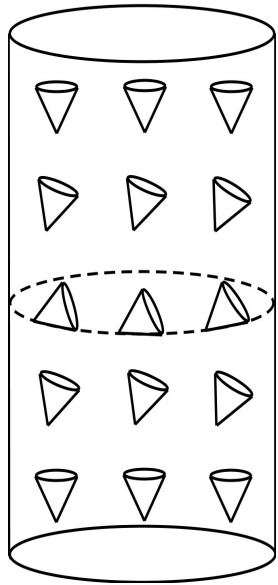

**Figure 3.3.1.** Cylindrical spacetime

(The figure above was redrawn but is inspired from Fig 8.8, Page 197, [1])

In this spacetime, you can see that if the cones become even a little more horizontal, then we will be having timelike curves which will be tending towards becoming closed timelike curves. Such a spacetime is barely causal. Such a spacetime which is barely causal is called *not strongly causal.* Now let us define what it means to be a strongly causal spacetime.

**Definition 3.3.2.** *Strongly Causal*

    $(M, g)$ *is called strongly causal if:*

    $\forall p$   *such that*   $p \in M$ *and for all neighborhoods* $\mathcal{U}$ *of p there is a neighborhood* $V \subseteq \mathcal{U}$ *such that :*

- *No causal curve $\gamma$ intersects V more than once.*
  - →  *Indeed, if $(M, g)$ is not strongly causal $\Rightarrow$ Then there exists a causal curve $\gamma$ which comes arbitrarily close to intersecting itself.*
  - →  *We require strong causality to keep causal curves at least a finite distance from intersecting themselves.*

**Remark 3.3.1.**

Spacetimes in which events from the future can influence their past i.e. spacetimes in which there are closed causal curves **do not** satisfy **strong causality.** From this point of view, strong causality seems like a sensible physical requirement.

Now it seems that the definition of *Strong causality* should be enough to do physics on our spacetime as it gives us a sensible condition that events from the future cannot effect events in their past. But there is still a small issue. We have not done anything to avoid this condition of non-existence of closed timelike curves in case of small perturbations. Lets deal with that now and we will have a causal structure of a spacetime with which we can work. For that we will define Stably Causal. But before denying that lets consider our problem with *strong causality* and then define stable causality a a result of that problem.[

---

**Problem** : Arbitrarily small perturbations in the metric could allow causal curves to self intersect. We need to find a condition to avoid this self intersection (We will define this condition as the necessary condition for Stable causality).

---

**Solution**:

Let us setup a few things,

- Consider a perturbing metric $\tilde{g}$ through :

$$g_{mn} \rightarrow \tilde{g}_{mn} = g_{mn} - \omega_m \omega_n \qquad (3.3.1)$$

  with a timelike cotangent vector field $\omega_m \omega_n$.

  - We have two metrics on the same differentiable manifold, $g_{mn}$ and $\tilde{g}_{mn}$.

  - $g_{mn}$ is a Lorentzian/ pseudo-Riemannian metric.

- Note: $\tilde{g}_{mn}$ still has the same signature but the light cones are now *wider* for $\tilde{g}_{mn}$. How do we see this?

  - Compare $g_{mn} v^m v^n$ and $\tilde{g}_{mn} v^m v^n$.
    (Calculate the length of the tangent vector $v^a$ at point $p$ with respect to the two metrics)

$$\tilde{g}_{mn} v^m v^n = g_{mn} v^m v^n - v^m \omega_m v^n \omega_n \qquad (3.3.2)$$

  Inspecting the second term on RHS carefully if $v^m \omega_m = v^n \omega_n = \alpha \in \mathbb{R}$

$$v^m \omega_m v^n \omega_n = \alpha \cdot \alpha = \alpha^2 > 0 \qquad (3.3.3)$$

  We see that in equation (3.3.2) the RHS is smaller than the LHS without the term $v^m \omega_m v^n \omega_n$. That gives us,

$$\tilde{g}_{mn} < g_{mn} \qquad (3.3.4)$$

  Thus, it is easier for vector $v^a$ to have $\|v^a\| < 0$ i.e. it is more likely to be timelike or null for $\tilde{g}_{mn}$.

- In conclusion we can say that, "Some vector that is spacelike with respect to $g_{mn}$ *maybe* timelike with respect to $\tilde{g}_{mn}$".

- $(M, \tilde{g})$ spacetime has all *causal curves* of $(M, g)$ plus more curves.

  - i.e. {Causal curves of $(M, g)$} $\subseteq$ {Causal curves of $(M, \tilde{g})$}

We have solved our problem by perturbing our metric and still maintaining the condition for causality. This condition where the perturbed metric also remains sensibly causal is called Stably causal.

Definition 3.3.3. *Stably causal*
      $(M, g)$ *is called Stably causal if there exists a covector field $\omega_a$ such that $(M, \tilde{g})$ is also causal. ($\tilde{g}_{mn} = g_{mn} - \omega_m \omega_n$)*

Now we will state some theorems on Stable causality that we might need later on. (Proofs can be found in Chapter 8, [1]).

> **Theorem 3.3.1.**
>       *If $(M, g)$ is stably causal then it implies that $(M, g)$ is strongly causal.*

> **Theorem 3.3.2.**
>       *$(M, g)$ is stably causal iff there exists a differentiable function $f \in C^\infty(M, \mathbb{R})$ such that $\nabla^a f$ is a past directed timelike vector field or $-\nabla^a f$ is a future directed timelike vector field.*

**Remark 3.3.2.**
      Intuitively, this means that $f$ can be viewed as a **cosmic clock**. (Not a unique one as we can have more than one $f$ satisfying our conditions.)

**Note 3.3.1.**
      We had defined something called *time orientability of a spacetime* at the start of this section. We said that :
$(M, g)$ is *time orientable* iff there exists a *past/future pointing smooth timelike vector field*. What separates this from the theorem above is the remark, is that there the *smooth timelike vector field* need not be a *Gradient field*.

## 3.4. Definitions/Theorems of Global dependence and Cauchy surfaces.

This section is the final buildup of the *causal structure* that we would eventually need to effectively define a black hole. (We will need some definitions from Asymptotic flatness which we will define further.)

Definition 3.4.1. *Future domain dependence of set $S$. ($D^+(S)$)*
      *Assume $S \subseteq M$ is a closed achronal set. Then the future domain of dependence of $S$ is defined as*

$$D^+(S) = \{p \in M \,|\, Every \ past \ inextendible \ causal \ curve \ through \ p \ intersects \ S\} \qquad (3.4.1)$$

Definition 3.4.2. *Past domain of dependence of set $S$. ($D^-(S)$)*

$$D^-(S) = \{p \in M \,|\, Every \ future \ inextendible \ causal \ curve \ through \ p \ intersects \ S\} \qquad (3.4.2)$$

**Note 3.4.1.** The past domain of dependence is nothing else than the set of events $p$ that affect only $S$.

Definition 3.4.3. *Domain of dependence. ($D(S)$)*
      *This is just defined as the union of the future and past domain of dependence.*

$$D(S) = D^+(S) \cup D^-(S) \qquad (3.4.3)$$

***Note* 3.4.2.** *The total domain of dependence of set $s$ i.e. both past and future events affected by $S$.*

**Example 3.4.1.**

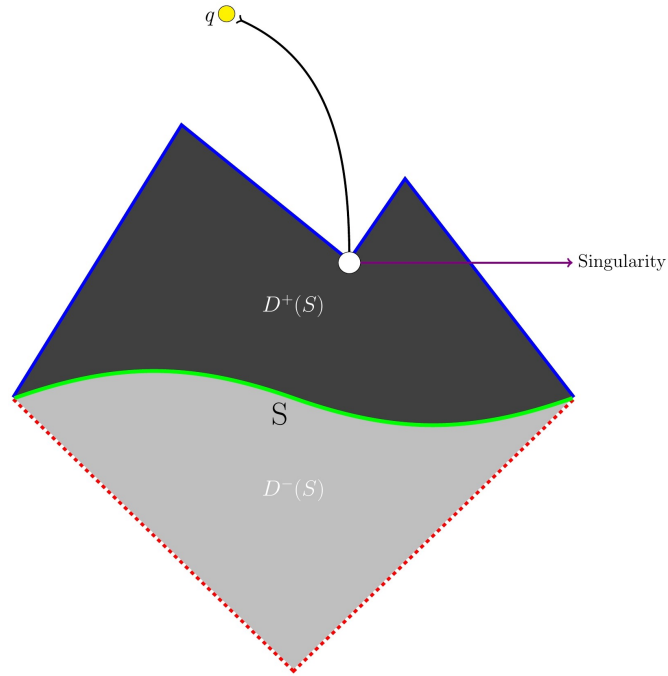

**Figure 3.4.1.** Domain of dependence

Looking at the figure above, one question we want to ask is, "Why $q \notin D^+(M)$? "
The answer is quite simple. For $q$, some *past inextendible curve* does not intersect S because it gets stuck at the *singularity*. In other words we can say, "$q$ is affected by events in the *shadow* of the singularity."

**DEFINITION 3.4.4.** *Future Cauchy horizon of S.* $(H^+(S))$

$$H^+(S) = \overline{D^+(S)} - I^-(D^+(S))$$

$\overline{D^+(S)}$: *Closed set of future domain of dependence of S.*
$I^-(D^+(S))$: *Chronological past of the future domain of dependence of S.*

**Note 3.4.3.** $H^+(S)$ is achronal. This is quite obvious as no two events in $H^+(S)$ are *causally connected* to each other.
This is the set of latest events that are affected only by $S$.

**DEFINITION 3.4.5.** *Past Cauchy horizon:* $H^-(S)$

$$H^-(S) = \overline{D^-(S)} - \overline{I^+(D^-(S))} \qquad (3.4.4)$$

**Note 3.4.4.**
This is the set of earliest events that affect only S.

**DEFINITION 3.4.6.** *Full Cauchy horizon of S.* $(H(S))$

$$H(S) = H^+(S) \cup H^-(S) \qquad (3.4.5)$$

PROPOSITION 3.4.1.
    *Full Cauchy horizon is just the boundary of Full domain of dependence.*

$$H(S) = \dot{D}(S) \qquad (3.4.6)$$

Now, let's define probably the most important definition of the whole chapter. We were building up all the definitions to get to this one.

DEFINITION 3.4.7. *Cauchy surface.*
    *A closed, achronal set S is called a Cauchy surface, if it's full Cauchy horizon vanishes. That is if,*

1. $H(S) = \emptyset$
2. $\dot{D}(S) = \emptyset$
3. $D(S) = M$

*Look at point **3** carefully, the domain of dependence of S is the entire manifold M (Spacetime). What this physically means that , if you know what happened on S (Initial conditions with suitable evolution laws) then you can predict what can happened on the entire manifold M (i.e. the entire spacetime).*

- *This is the reason why Cauchy surfaces are so important. If the conditions on a Cauchy surface are known then everything on M can be predicted.*

- *Since a Cauchy surface is Achronal, it can be viewed as an instant in time. (As no two events are connected causally on an achronal set)*

- *The term surface in Cauchy surface is motivated by the following theorem.*

    - *Theorem: Every Cauchy surface $\Sigma$, is a 3D sub-manifold $C^0$ of M.*

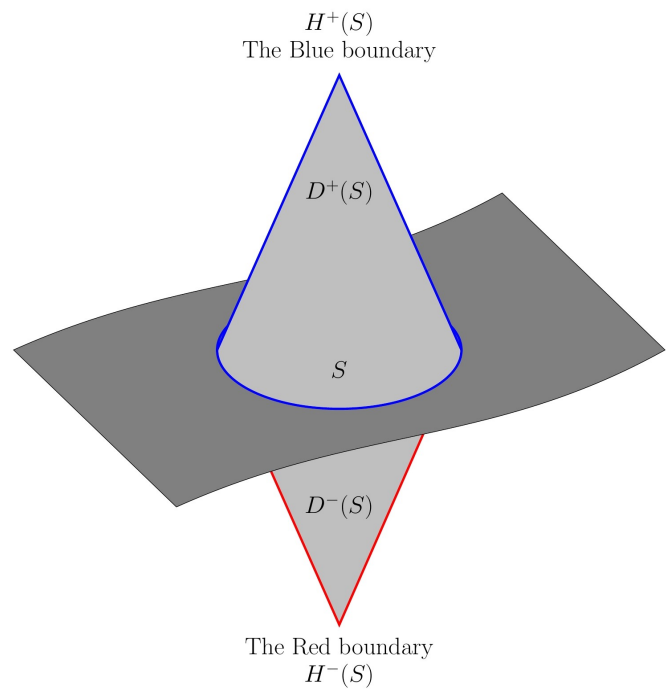

**Figure 3.4.2.** *Cauchy Horizons*

Definition 3.4.8. *Globally hyperbolic spacetime*

A spacetime $(M, g_{ab})$ is called a globally hyperbolic spacetime if it possesses a Cauchy surface $\Sigma$.

## 3.5. Bibliographical Notes

The sections in this chapter were highly motivated by the following references. Only the definitions were taken directly from the texts without any comments.

| Section | Reference | Pages |
|---------|-----------|---------|
| 3.1 | 1 | 189-190 |
| 3.2 | 1 | 190-192 |
| 3.3 | 1 | 195-206 |

—***—

# Chapter 4

## Singularity Theorems : Congruences and Raychaudhuri equation.

> The goal of this chapter is to make the reader familiar with the concept of a *singularity* from a global perspective. Not only that but also to equip the reader with some tools which are extremely effective in dealing with singularities in such a context. These tools are called the *Singularity theorems*. Instead of dealing with multiple examples of Singularity theorems, the author takes his time to build a good mathematical and intuitive grasp on frequently used mathematical tools in singularity theorems like *extrinsic curvature, geodesic congruences, Raychaudhuri's equation, etc.* The chapter is concluded by two singularity theorems by Stephen Hawking which have cosmological interpretation.
>
> A more emphasis is made on understanding the tools needed to prove these singularity theorems, in order to understand every step while the singularity theorems are being proved the reader might want to refer to numerous concepts from Chapter 8,9 from Wald's book [1].

### 4.1. What do we mean by singularity?

In all the other basic physical theories except General relativity, like Classical mechanics and Electrodynamics, the manifold and the metric structure of spacetime is already assumed. We know the *where* and *when* of all spacetime events. In these theories, the non-existence or being infinite of a certain physical quantity at certain point determines whether we have a singularity or not.

**Example.** Coulomb solution of Maxwell's equation in Special relativity has a singularity at the events labeled by $r = 0$.

This situation is different when we talk about General relativity. Here, we are trying to understand the manifold and metric structure of spacetime itself.

The notion of an event only makes sense after we have defined the spacetime $M$ and a metric $g_{ab}$ *everywhere* on $M$.

- The big bang singularity of the *Robertson-Walker (RW) metric* is not considered as a part of the spacetime manifold and hence it is not a *place* or *time*.

    (This point is meant to emphasize the fact that coordinates in General relativity may not always have a *physical meaning* like in Electrodynamics or Classical mechanics. Every time we use the word *place* in the upcoming paragraph and talk about it, we are speaking in this context.)

- The region $r > 0$ is considered to be a part of the *Schwarzschild spacetime* and hence unlike the Coulomb singularity at $r = 0$, the same singularity $r = 0$ is not a *place* in Schwarzschild spacetime.

On the basis of the above examples we might be able to think that we can still define the notion of a singular boundary. For example, by adding points $\tau = 0$ and $r = 0$ in the RW spacetime and Schwarzschild spacetime respectively we think we might achieve it. But this would allow one to talk in precise terms of a singularity as a *place* even though the metric is not defined there. While this can be done for simple cases like the above, more difficulties arise when one tries to give a meaningful general description of a singular boundary.

Singularities like the one at $r = 2\,G\,M$ i.e. the event horizon of the Schwarzschild spacetime cannot be used as an example to describe singularities as this singularity was a result of some specific coordinate chart being used. These are called *coordinate dependent singularities*. They are not really *singularities* as they can be removed by coordinate transformations. Many different attempts were made to define singular boundaries but failed due to numerous drawbacks. Hence, we must abandon the notion of singularity as being a *place*.

The failure of being able to describe singularity as a *place* does not mean that they do not exist. We know for a fact that we have singularities in Robertson-Walker and Schwarzschild spacetimes. We just need to find another way of characterizing a singularity.

One approach is the *Curvature approach* where we look at the events in the spacetime where the curvature blows up e.g. $r = 0$ in Schwarzschild spacetime and $\tau = 0$ in RW spacetime. This approach has it's own difficulties. For example :

- This approach is based on the components of the Riemann curvature tensor $R_{abc}{}^{d}$. If we use *bad* coordinates, then the components of the tensor maybe ill behaved because of the coordinate system chosen.

To avoid this :

- We instead examine the scalars formed by this curvature tensor like $R$, $R_{ab}R^{ab}$, $R_{abcd}R^{abcd}$ and similar scalars formed by polynomial expressions in derivatives of the Riemann curvature tensor. (These will be coordinate independent)

Now let us finally define how should we characterize singular spacetimes and what are the options when we go towards a singularity on a geodesic.

DEFINITION 4.1.1. *Singularity*
  *We say that $(M, g)$ possesses a singularity if it possesses an incomplete geodesic.*
  *OR*
  *A singularity in a Lorentzian manifold is an incomplete endless curve.* (Page 12, [11])

**Remark 4.1.1.**
  We distinguish singularities of null, spacelike and timelike type depending on the type of the geodesic.

DEFINITION 4.1.2. *Singular spacetime*
  *If a spacetime $M$ possesses a singularity as defined above, then $M$ is considered a singular spacetime.*

When one is traveling an *incomplete geodesic* towards a singularity, **three** things can happen. (Not necessarily independently, all of them can happen altogether)

1. **Scalar curvature singularity.**
    A scalar reconstructed from $R_{abc}{}^{d}$ diverges.

2. **Parallel-propagated curvature singularity.**
    In a parallel transported tetrad frame, a scalar component of $R_{abc}{}^{d}$ or its covarient derivatives diverge.

3. **Non curvature singularity.**
    None of the above.
    Example : Conical singularity - Take a paper and make a cone, that is not curvature because parallel transport remains trivial. That means that when you take a piece of paper and crumble it, it is not curved but it just has conical singularities.

**Fundamental problem**

As discussed before :

→   In concrete solutions, like Schwarzschild or FLRW cosmologies, curvature singularities are obviously present.

$\rightarrow$   These spacetimes are highly symmetric.

The real problem is whether more realistic spacetimes, i.e. perturbed spacetimes also show these singularities ?

**Example 4.1.1.** Spherically symmetric dust shell in-fall

- In Newtonian gravity : We use *catastrophe theory* which predicts infinite mass density but not if the spherical symmetry is perturbed.
- In Einstein's gravity : We use *Singularity theorems.*
  We predict the Black hole singularity in this case, even if the symmetry is perturbed. (This happens under assumption of some energy condition.)

**Singularity theorems** are designed in such a way that they predict singularities in spacetimes in a robust manner.

(Thus if quantum gravity is to resolve singularities then it has to overcome this robust theory).

## 4.2.  DESIGN OF SINGULARITY THEOREMS

We are now trying to look at more arbitrary spacetimes, particularly the ones which are perturbed in some or the other way from the spacetimes with which we are familiar.

### 4.2.1.  Strategy for Singularity theorems

Let us first make an algorithm for these theorems.

**Algorithm for singularity theorems**

i. Focus attention on singularities that can be identified by existence of *incomplete inextendible timelike or null geodesics.*
    Why?
    It is clear that these are the more important singularities as an observer traveling on such a geodesic will have their eigentime bounded from above and below.
    What about other singularities?
    May exist in addition, but the standard singularity theorems don't attempt to predict them.

ii. Singularities can be in the way of a geodesic. The presence of singularities interferes with the property of geodesics being extremal length curves.

iii. Let us recall two things that we will need in this point.
   1. Extremizing curve length is done by Euler-Lagrange equations which eventually lead to the geodesic equation.

$$\frac{\partial}{\partial \tau}\left(\frac{\partial L}{\partial \dot{x}^{\alpha}}\right) = \frac{\partial L}{\partial x^{\alpha}} \qquad \text{(Euler lagrange equation)}$$

$$\frac{d^2 x^{\beta}}{d\tau^2} + \Gamma^{\beta}{}_{\alpha\nu}\frac{dx^{\alpha}}{d\tau}\frac{dx^{\beta}}{d\tau} = 0 \qquad \text{(Geodesic equation)} \qquad (4.2.1)$$

   2. This point is just to state that both the equations above are differential equations and solutions to differential equations are functions.

   At least locally, geodesics are paths of extremal lengths:
   
   $\rightarrow$   Spacelike geodesics are curves of *shortest proper distance.*
   
   $\rightarrow$   Timelike geodesics are curves of *maximal proper time.*

iv. Prove that even in Generic spacetimes there exists curves of maximal length between events.

What assumptions are needed?

The spacetime being Globally hyperbolic suffices.

v. We also need to need to assume some energy condition. For example, the strong energy condition which we will define along the way to use it to prove that geodesics meet a divergence of a quantity called the *expansion* $\theta$, in finite proper time.

These extremal length curves cannot be geodesics with eigentime longer than a certain finite amount into past or future.

vi. Conclude that there are incomplete geodesics i.e. we have a singularity in the past/future.

## 4.2.2. Singularity lemma I. (Timelike geodesic congruence)

In this subsection we are gonna discuss a singularity theorem(lemma) in intricate details. We will derive most of the things that we are going need to understand this Singularity lemma.

---

**LEMMA 4.2.1. (Lemma 9.2.1, [1])**

   **Let/Suppose :**

- $\xi^a$ be a tangent field of a hypersurface orthogonal timelike geodesic congruence.

- $T_{mn}$ (Energy momentum tensor in the Einstein's field equation) obeys the **strong** energy condition $\Rightarrow R_{ab}\xi^a\xi^b \geq 0$.

- $\Theta = \Theta_\circ < 0$ at any point on a geodesic congruence.

   **Then:**

  $\Theta$ goes to $-\infty$ along that geodesic within proper time $\tau \leqslant \dfrac{3}{|\Theta_\circ|}$

---

Before we can even think about proving this theorem we need to understand a lot of concepts that will be used while proving it. We will mainly use the tools that were developed in the previous chapters and introduce some more concepts.

### 4.2.2.1. Tool I : Extrinsic curvature

The extrinsic curvature of a spacelike hypersurface describes *naively* the curvature between the spacelike hypersurface and the time dimension (Curvature between space and time). We know that the curvature between different spatial dimensions can be found using parallel transport (Take a parallelogram and then parallel transport the vector along it and you don't get the same vector back - This is how the Riemann tensor is defined).

What if one of your directions in the parallelogram is timelike?

Basically, you parallel transport your quantity to a future time and then some spatial direction. Later you parallel transport back to the time where you started it and then you parallel transport again in a spatial dimension. If the vector is now rotated then we say there is   some curvature between space and time. What intuition could we possibly have of this phenomenon? We will try to build some *naive* intuition of this in the next paragraph.

The intuition behind this could be taken as the expansion or shrinkage of spacetime. That is what shows up on these parallelograms. The condition on the extrinsic curvature is a condition on the expansion rate.

The basic idea behind this theorem is that, if certain conditions are met, such as the strong energy condition and if at a certain point in time (i.e. on a spacelike hypersurface, we can think spacelike hypersurfaces as slices of time as we are assuming our manifold to be globally hyperbolic) the spacetime was expanding at some minimum rate everywhere and nowhere slower than that. Then, the theorem concludes that it always has been expanding. Which concludes that if you go back in time, it should shrink in such a way that we meet a singularity. The same conclusion would be detected by geodesics. This idea that such a conclusion could be made by geodesics (or rather geodesic congruence which we define soon) is the sole motivation behind such singularity theorems.

If we think about it, we are not attempting to solve the Einstein's equation here for some metric. Instead, we are analyzing the geodesic equation and concluding if a bunch of geodesics collie in the past. If we successfully do conclude this, we can say that we found a singularity.

So in conclusion, extrinsic curvature tells us about the dynamics of spacetime itself. In this case it being negative, tells us about the rate of contraction.

Therefore, if we assume $\forall p \in \Sigma$

$$\kappa(p) \leq C < 0$$

this condition is true for some constant $C$, then spacetime has a finite minimum expansion rate everywhere on $\Sigma$.

### 4.2.2.2. Tool II : Strong energy condition

Before jumping straight to the strong energy condition let's look at what other energy conditions we have :

- Weak energy condition (WEC) :

$$T_{mn}v^m v^n \geq 0 \qquad \forall \text{ timelike } v: g_{mn}v^m v^n < 0$$

  For an observer with *unit tangent* $v^a$ the local energy density is $T_{mn}v^m v^n \geq 0$. As an observer you can choose the coordinate system that travels with you. That means you can choose such a frame in which you are only traveling in the time dimension and not space. That means, for your tangent vector the only non-zero component is $x^0$ component.

$$T_{mn}v^m v^n = T_{00}\,v^0\,v^0 + T_{ii}\,v^i\,v^i = T_{00}v^0 v^0 + 0 = T_{00} \geq 0$$

  where $i \in \{1, 2, 3\}$ and $v$ is a unit vector.

- Dominant energy condition (DEC) :
  Based on the weak energy condition but is slightly stronger.
  i. WEC

$$T_{mn}v^m v^n \geq 0$$

  ii. $v^a$ is any timelike vector and $K_m \equiv T_{mn}v^n$. Assuming this it satisfies

$$K_m K^m \leq 0$$

  Whenever we have $v^a$ as a *killing vector field* (It doesn't have to be for this definition, just a recall). The quantity $K_m$ is a conserved quantity (conserved 4-momentum).
  This is the condition that makes it DEC and not WEC. It basically says that the energy-momentum flow vector $K_m$ may not be conserved but it has to be *causal*, i.e. flow must be in the future.

- **Strong energy condition (SEC):**

  This is the energy condition that our theorem requires.

  We say that our matter obeys the SEC iff $\forall$ timelike $\xi^m$ :

  $$\left(T_{mn} - \frac{1}{2}T^r_{\ r}\, g_{mn}\right)\xi^m\xi^n \geq 0 \tag{4.2.2}$$

  Honestly, we will see that this energy condition is imposed because we need an equation exactly in this form to make the singularity theorem provable. So, it does not come from an intuition/observation of some condition that the actual matter in the universe is obeying. But if we look carefully, there is an interpretation.

  **Interpretation :**

  We exclude matter that causes accelerated expansion, has an equation of state parameter smaller than $-\frac{1}{3}$. That would violate the SEC.

  **Is something like this possible?**

  SEC is obeyed by known matter. It can be violated by dark energy / cosmological constant.

  **What is the relationship of SEC with WEC and DEC?**

  It is independent of WEC and DEC. SEC can be obeyed independently whether WEC or DEC are satisfied or not.

**Example 4.2.1.**

For known matter $T_{mn}$ is diagonalizable to obtain :

$$T_{mn} = \begin{bmatrix} \rho & 0 & 0 & 0 \\ 0 & p_1 & 0 & 0 \\ 0 & 0 & p_2 & 0 \\ 0 & 0 & 0 & p_3 \end{bmatrix}$$

where $\rho$ is energy density by co-moving matter and $p_i$ is principal pressure.

The energy conditions now read as follows :

- WEC (Weak) :

  $$\rho \geq 0 \quad \text{and} \quad \rho + p_i \geq 0 \quad \text{for each} \quad i \in \{1, 2, 3\}$$

- DEC (Dominant) :

  $$\text{WEC} +$$
  $$\rho \geq |p_i| \quad \text{for} \quad i \in \{1, 2, 3\}$$

- SEC (Strong) :

  $$\rho + \sum_{i=1}^{3} p_i \geq 0 \quad \text{and} \quad \rho + p_i \geq 0 \quad \text{for} \quad i \in \{1, 2, 3\}$$

Now we will be coming to our final and biggest tool/s. We will be trying to understand how to figure out the point (vi) from out algorithm for singularity theorems.

### 4.2.2.3. Tool III : Congruences and their properties.

The motivation behind this tool is the following result, *Given the SEC one can show that geodesics meet a divergence of a quantity called **expansion $\theta$**, in finite proper time.* This is called the **Focusing theorem.**

Definition 4.2.1. *Congruence*

   A congruence is defined in an open subset $O \subset M$ of a manifold $M$ as a family of curves such that through each point $p \in O$ there is exactly one curve from this family passing through it.

- *The tangents to a congruence yield a vector field.*
- *Every continuous vector field generates a congruence of curves.*
- *A congruence is smooth if the corresponding vector field is smooth.*

**Timelike geodesic congruences:**

Consider a timelike geodesic congruence with the geodesics being parameterized by proper time. For the corresponding tangent vector field $\xi^a$ :

$$\xi_a \xi^a = -1$$

Define the tensor field $B_{ab}$ :

$$B_{ab} = \nabla_b \xi_a$$

and so called the spatial metric :

$$h_{ab} = g_{ab} + \xi_a \xi_b \tag{4.2.3}$$

which projects onto the subspace of the tangent space which is perpendicular to $\xi^a$. We can easily see that $B_{ab} = \nabla_b \xi_a$ is orthogonal to the vector $\xi^a$ at any point :

$$B_{ab}\xi^a = (\nabla_b \xi^a)\xi^a = 0 = (\nabla_b \xi^a)\xi^b = B_{ab}\xi^b$$

We define the following quantities of congruence, the congruence (think of them as a spaghetti coming out of a spaghetti maker) can do some stuff geometrically.

- **Expansion** $\Theta$ :

$$\Theta = B^{ab} h_{ab} \tag{4.2.4}$$

  which is essentially the trace of $B_{ab}$.

  Intuition : It describes the average volume expansion of the infinitesimally nearby surrounding geodesics.

- **Shear** $\sigma$ :

$$\sigma_{ab} = B_{(ab)} - \frac{1}{3}\Theta h_{ab} \tag{4.2.5}$$

  this is a symmetric and traceless quantity.

  Intuition : It describes the distortion of the shape of an initial sphere in tangent space into an ellipsoid.

- **Twist (rotation)** $\omega_{ab}$ :

$$\omega_{ab} = B_{[ab]} \tag{4.2.6}$$

  which is just the anti-symmetric part of $B_{ab}$.

  Intuition : It describes the rotation of the geodesics.

This is how we can decompose a matrix and we can write down :

$$B_{ab} = \frac{1}{3}\Theta h_{ab} + \sigma_{ab} + \omega_{ab} \tag{4.2.7}$$

(A detailed decomposition of such quantities if motivated by linear algebra and can be found in Chapter 2, [4]).

### 4.2.2.4. Tool IV : Dynamics of these congruences. (Raychaudhuri's equation.)

We will now basically use the definitions of the expansion $\Theta$, shear $\sigma_{ab}$ and the twist $\omega_{ab}$ to see how does the tensor field $B_{ab}$ evolve in time. In simple words we want to see if this equation :

$$\frac{\mathrm{d}B_{ab}}{\mathrm{d}\tau} = \xi^i \nabla_i B_{ab} \tag{4.2.8}$$

gives use any useful information for this theorem.

**Derivation of the Raychaudhuri's equation :**

> The derivation was not done ***explicitly*** in any of the references and the author takes the tiny credit for dealing it with every step carefully and deriving the final result.

We need the following tools we know from differential geometry to prove Raychaudhuri's equation :

i. Definition of the Riemann tensor:

$$(\nabla_a \nabla_b - \nabla_b \nabla_a)\omega_c \;=\; R_{abc}{}^d\,\omega_d \tag{4.2.9}$$

$$= \; \nabla_a(\nabla_b\,\omega_c) - \nabla_b(\nabla_a\,\omega_c)$$

$$\nabla_a(\nabla_b\,\omega_c) \;=\; \nabla_b(\nabla_a\,\omega_c) + R_{abd}{}^c\,\omega_d \tag{4.2.10}$$

$$R_{abc}{}^d \;=\; -R_{bac}{}^d$$

ii. The Leibnitz[4.2.1] rule :

$$\nabla_d\big(A^{a_1\dots a_k}{}_{b_1\dots b_l}\,B^{c_1\dots c_{\tilde k}}{}_{d_1\dots d_{\tilde l}}\big) \;=\; \big(\nabla_d A^{a_1\dots a_k}{}_{b_1\dots b_l}\big)B^{c_1\dots c_{\tilde k}}{}_{d_1\dots d_{\tilde l}}$$

$$+ \big(\nabla_d B^{c_1\dots c_{\tilde k}}{}_{d_1\dots d_{\tilde l}}\big)A^{a_1\dots a_k}{}_{b_1\dots b_l} \tag{4.2.11}$$

$$\text{where,}$$

$$A^{a_1\dots a_k}{}_{b_1\dots b_l} \in \mathcal{T}(k,l)\text{ and}$$

$$B^{c_1\dots c_{\tilde k}}{}_{d_1\dots d_{\tilde l}} \in \mathcal{T}(\tilde k, \tilde l)$$

where $\mathcal{T}(k,l)$ is the tensor field with $k$ contravarient components and $l$ covarient components.

iii. The geodesic equation (4.2.1)

Now let us use the tools above to expand (4.2.8)

$$\frac{\mathrm{d}B_{ab}}{\mathrm{d}\tau} = \xi^i\nabla_i B_{ab} \qquad = \qquad \xi^i\nabla_i\,\nabla_b\xi_a = \xi^i\nabla_i\left(\nabla_b\xi_a\right)$$

$$\overset{(4.2.10)}{=\joinrel=\joinrel=} \quad \xi^i(\nabla_b(\nabla_i\xi_a) + R_{ibd}{}^a\xi_d)$$

$$= \quad \xi^i\nabla_b\nabla_i\xi_a + R_{ibd}{}^d\,\xi^i\xi_d$$

$$\overset{(4.2.11)}{=\joinrel=\joinrel=} \quad \nabla_b(\xi^i\nabla_i\xi_a) - (\nabla_b\xi^i)(\nabla_i\xi_a) + R_{ibd}{}^a\,\xi^i\xi_d$$

$$\overset{(4.2.1)}{=\joinrel=\joinrel=} \quad -B^i{}_b B_{ai} + R_{ibd}{}^a\,\xi^i\xi_d \tag{4.2.12}$$

$$\frac{\mathrm{d}B_{ab}}{\mathrm{d}\tau} = \xi^i\nabla_i B_{ab} \qquad = \qquad -B^i{}_b B_{ai} + R_{ibd}{}^a\,\xi^i\xi_d \tag{4.2.13}$$

Now we take the trace of (4.2.13) :

First lets take the trace on the left hand side (LHS) of (4.2.13):

$$\text{Tr}(\text{LHS of }(4.2.13)) = \text{Tr}\left(\frac{\mathrm{d}B_{ab}}{\mathrm{d}\tau}\right) \;=\; \frac{\mathrm{d}}{\mathrm{d}\tau}(\text{Tr}(B_{ab}))$$

$$(\text{Trace is a linear operator.})$$

$$= \; \frac{\mathrm{d}}{\mathrm{d}\tau}(B_{ab}\,h^{ab}) = \frac{\mathrm{d}\Theta}{\mathrm{d}\tau} \tag{4.2.14}$$

$$\text{Tr}\left(\frac{\mathrm{d}B_{ab}}{\mathrm{d}\tau}\right) \;=\; \frac{\mathrm{d}\Theta}{\mathrm{d}\tau} \tag{4.2.15}$$

---

4.2.1. [1] uses "Leibnitz".

Let's take the trace on first term on the right hand side (RHS) of (4.2.13) and expand it using (4.2.8):

$$
\begin{aligned}
B^{ab}B_{ba} &= \left(\frac{1}{3}\Theta h^{ab}+\sigma^{ab}+\omega^{ab}\right)\left(\frac{1}{3}\Theta h_{ba}+\sigma_{ba}+\omega_{ba}\right) \\
&= \frac{1}{3}\Theta h^{ab}\left(\frac{1}{3}\Theta h_{ba}+\sigma_{ba}+\omega_{ba}\right)+ \\
&\quad +\sigma^{ab}\left(\frac{1}{3}\Theta h_{ba}+\sigma_{ba}+\omega_{ba}\right)+\omega_{ba}\left(\frac{1}{3}\Theta h_{ba}+\sigma_{ba}+\omega_{ba}\right)
\end{aligned}
\tag{4.2.16}
$$

All the mixed terms of $\Theta h^{ab},\sigma^{ab},\omega^{ab}$ are zero because each of the individual terms are irreducible and they do not interact with each other. So basically ,

$$
\frac{\Theta}{3}h^{ab}\sigma_{ba},\frac{\Theta}{3}h^{ab}\omega_{ba},......,\frac{\Theta}{3}\omega^{ab}h_{ba},\omega^{ab}\sigma_{ba}=0
\tag{4.2.17}
$$

Using this fact we get the following equation,

$$
B^{ab}B_{ba} = \frac{\Theta}{9}h^{ab}h_{ba}+\sigma^{ab}\sigma_{ba}+\omega^{ab}\omega_{ba}
\tag{4.2.18}
$$

In the equation above $h_{ba},\sigma_{ba}$ are symmetric. The way we defined $h_{ba}$ is such that it is the diagonal part of $B_{ab}$, and $\sigma_{ba}$ is the non-diagonal symmetric part. $\omega_{ba}$ is anti-symmetric. Swapping the indices which are downstairs for the equation above we get,

$$
B^{ab}B_{ab} = \frac{\Theta^2}{9}h^{ab}h_{ab}+\sigma^{ab}\sigma_{ab}-\omega^{ab}\omega_{ab}
\tag{4.2.19}
$$

$h^{ab}$ is the spatial metric as defined in (4.2.3). By spatial metric we mean that it is the projection onto the spacelike hypersurface that has its normal vector $\xi^a$ as we defined before. We can check this by taking the inner product of any vector $x^a$ with $\xi^a$ using the spatial metric. The result should be zero if it projects it to the hypersurface defined by $\xi^a$.

$$
\begin{aligned}
h_{ab}\xi^a x^b &= (g_{ab}+\xi_a\xi_b)\xi^a x^b \\
&= g_{ab}\xi^a x^b+\xi_a\xi_b\xi^a x^b \\
&= \xi^a x_a+(\xi_a\xi^a)\xi_b x^b \\
&= \xi^a x_a+(-1)\xi_a x^a \\
&= \xi^a x_a-\xi_a x^a
\end{aligned}
\tag{4.2.20}
$$
$$
h_{ab}\xi^a x^b = 0
\tag{4.2.21}
$$

Hence, we have established that $h_{ab}$ projects vectors from our 4-dimensional manifold to a 3-dimensional hypersurface. The trace of $h_{ab}$ is 3 using a local Lorentz frame. We have

$$
\mathrm{Tr}(h^{ab})=h_{ab}h^{ab} = 3
\tag{4.2.22}
$$

We can plug this into (4.2.19) to get,

$$
B^{ab}B_{ab} = \frac{\Theta}{3}+\sigma^{ab}\sigma_{ab}-\omega^{ab}\omega_{ab}
\tag{4.2.23}
$$

We now plug this as the first term for the right hand side of Tr(4.2.13) and take the taking the trace of the second term ($R_{mn}$) we get (Just replacing some indices) :

$$
\begin{aligned}
\mathrm{Tr}\left(\frac{\mathrm{d}B_{ab}}{\mathrm{d}\tau}\right) &= -\frac{\Theta^2}{3}-\sigma^{ab}\sigma_{ab}+\omega^{ab}\omega_{ab}-R_{cd}\,\xi^c\xi^d \\
\frac{\mathrm{d}\Theta}{\mathrm{d}\tau} &= -\frac{\Theta^2}{3}-\sigma^{ab}\sigma_{ab}+\omega^{ab}\omega_{ab}-R_{cd}\,\xi^c\xi^d
\end{aligned}
\tag{4.2.24}
$$

This equation (4.2.24) is known as the **Raychaudhuri's equation** for a timelike geodesic congruence.

### 4.2.2.5. Proof

After four tools and several pages we finally have the machinery to prove Theorem (4.2.1).

Let us assume that in we have the strong energy condition satisfied i.e. equation (4.2.2). Let's impose that condition on the term containing the Ricci tensor in equation (4.2.24). Before that let us express that term using Einstein's equation.

$$R_{cd}\xi^c\xi^d \;=\; 8\pi\left(T_{cd}-\frac{1}{2}Tg_{cd}\right)\xi^c\xi^d \tag{4.2.25}$$

$$=\; 8\pi\left(T_{cd}\xi^c\xi^d+\frac{1}{2}T\right) \tag{4.2.26}$$

Basically the term $T_{cd}\xi^c\xi^d$ represents the energy density of matter as being measured by an observer with 4-velocity $\xi^c$. If we impose the strong energy condition which precisely makes the right hand side of the equation above non-negative we get :

$$T_{cd}\xi^c\xi^d+\frac{1}{2}T \;\geq\; 0 \tag{4.2.27}$$

$$T_{cd}\xi^c\xi^d \;\geq\; -\frac{1}{2}T \tag{4.2.28}$$

We can use this to prove our theorem as it was one of the assumption we made while stating the theorem. $\sigma_{ab}$ and $\omega_{ab}$ both being spatial vectors we can know that the inner product of the terms with themselves will be positive. We can even get rid of $\omega_{ab}$ by the Frobenius theorem (Appendix B.3, Page 434, [1]) if we assume the congruence is hypersurface orthogonal. Under all these assumptions we have the equation (4.2.24) as,

$$\frac{\mathrm{d}\Theta}{\mathrm{d}\tau}+\frac{\Theta^2}{3} \;\leq\; 0 \tag{4.2.29}$$

$$-\frac{1}{\Theta^2}\frac{\mathrm{d}\Theta}{\mathrm{d}\tau}-\frac{1}{3} \;\geq\; 0$$

$$\frac{\mathrm{d}}{\mathrm{d}\tau}\Theta^{-1}-\frac{1}{3} \;\geq\; 0$$

$$\frac{\mathrm{d}}{\mathrm{d}\tau}\Theta^{-1} \;\geq\; \frac{1}{3} \tag{4.2.30}$$

Integrating this we get,

$$\Theta^{-1}(\tau) \;\geq\; \Theta_0^{-1}+\frac{1}{3}\tau \tag{4.2.31}$$

Where $\Theta_0$ is the initial value of $\Theta$.

Let's consider a hypersurface-orthogonal congruence. Let it be initially converging ($\Theta<0$). Then (4.2.30) tells us that convergence will continue and eventually we will hit a point where geodesics will cross (such points is also known as *caustic*) in a *finite* proper time $\tau \leq -3\Theta_0^{-1}$.

Basically what we are saying is that, matter obeying SEC cannot push geodesics apart, it can only increase the rate at which they are converging. Obviously this result only applies to some arbitrarily-chosen congruence and the appearance of caustics doesn't indicate any singularity in the spacetime. (Geodesics cross all the time, even in flat spacetime. Focal points of lenses is one example where null geodesics in flat spacetime cross each other, this does not indicate the existence of a singularity). Many proofs for singularity theorems take advantage of this property of the Raychaudhuri equation to show that spacetime must be geodesically incomplete in some way.

### 4.2.3. Singularity lemma II (Null geodesic congruence)

LEMMA 4.2.2. (Lemma 9.2.2, Page 223, [1])

**Let/Suppose :**
- $k^a$ be the tangent field of a hypersurface orthogonal null geodesic congruence.
- $R_{ab}k^a k^b \geqslant 0$ is satisfied. (Einstein's equation holds in the spacetime and strong or weak energy condition is satisfied by matter).
- The expansion $\Theta = \Theta_\circ < 0$ at any point on a geodesic in the congruence.

**Then :**
$\rightarrow$    $\Theta$ goes to $-\infty$ along that geodesic within affine length $\lambda \leqslant \dfrac{2}{|\Theta_\circ|}$.

**Proof.** The computation for this proof is extremely similar to the one for timelike geodesic congruence case i.e. the previous lemma.

    The main differences are as follows:
- The $\frac{1}{3}$ factor in front of $\Theta$ in $B_{ab}$ becomes a $\frac{1}{2}$ giving us
  (For a detailed explanation of how this happens one can refer to Page 221-222:[1] or Page 461-465:[2])
  $$\hat{B}_{ab} \;=\; \frac{1}{2}\Theta\,\hat{h}_{ab} + \hat{\sigma}_{ab} + \hat{\omega}_{ab}$$

- The Raychaudhuri equation looks like the following
  $$\frac{\mathrm{d}\Theta}{\mathrm{d}\lambda} \;=\; -\frac{1}{2}\Theta^2 - \hat{\sigma}_{ab}\hat{\sigma}^{ab} + \hat{\omega}_{ab}\hat{\omega}^{ab} - R_{cd}\,k^c\,k^d$$

- After similar computations and reasoning from the previous lemma we get the final equation to prove this statement as follows :
  $$\Theta^{-1}(\tau) \;\geq\; \Theta_0^{-1} + \frac{1}{2}\tau \tag{4.2.32}$$

The details of the proof can be found on (Page 221-223,[1])        $\square$

## 4.3. CONJUGATE POINTS

DEFINITION 4.3.1. (Page 223, [1]) *Jacobi Fields*

    On a manifold M, with a connection and for a geodesic $\gamma$ with tangent vector $\xi^a$. A solution $\eta^a$ of the geodesic deviation equation ([1],p. 46)
$$\xi^a\,\nabla_a(\xi^b\,\nabla_b\,\eta^c) = -R_{abd}{}^c\,\eta^b\,\xi^a\,\xi^d \tag{4.3.1}$$
is called a **Jacobi field** on $\gamma$.

**Remark 4.3.1.** Two points $p, q \in \gamma$ are called **conjugate points** if there exists a non-zero Jacobi field which vanishes at $p, q$.

**Example 4.3.1.**
    In Riemannian geometry if you consider a sphere to be your manifold and its longitudinal geodesics, the north and south poles are conjugate points on this Sphere.

### 4.3.1. Why are we interested in conjugate points?

- In spacetimes they tell us the stage at which a timelike geodesic fails to be a local maximum of the proper time (as it should be) between two points.

- In case of a null geodesic they tell us when does it fail to remain on the boundary of a point.
- In case of Riemannian geometry - They tell us the stage at which a geodesic fails to be the minimum length curve (locally) connecting points.

### 4.3.2. Existence of Conjugate points.

Let us consider the conjugate points on timelike geodesics and analyze them. We will use results from the previous section.

---

PROPOSITION 4.3.1. (Proposition 9.3.1, Page 226, [1])

**Let/Suppose :**

- $(M, g_{ab})$ be a spacetime satisfying $R_{ab}\xi^a\xi^b \geqslant 0$ for all timelike $\xi^a$.

- $\gamma$ be a timelike geodesic and $p \in \gamma$.

- The convergence of the timelike geodesics propagating into the future from $p$ attains a negative value for $\Theta$ at $r \in \gamma$.

   **Then :**

   Within proper time $\tau \leqslant \dfrac{3}{|\Theta_0|}$ from $r$ along $\gamma$ there exists a point $q$ conjugate to $p$ , assuming that $\gamma$ extends that far.

**Proof.**

   *We say that the proposition above is an obvious result from lemma (4.2.1) and the following lemma.*

LEMMA 4.3.1. (Proof on page 225-226, [1])

**Let/Suppose :**

- $\gamma$ be a timelike geodesic $\xi^a$ and $p \in \gamma$.

- A congruence of timelike geodesics passing through $p$. $p$ itself is excluded from $O \in M$. (This congruence is singular at $p$ itself)

   **Then:**

   A point $q \in \gamma$ where $q$ lies to the future of $p$ is conjugate to $p$ iff the expansion $\Theta$ tends to $-\infty$ at $q$.

   Considering all our assumptions from lemma (4.3.1) and theorem (4.2.1) we can effectively deduce this proof *obvious*.          □

---

## 4.4. TOPOLOGY OF CAUSAL CURVES

In order to study causal curves and identify the condition in which some causal curves are incomplete, it is convenient to consider the set of causal curves between two points as a *Topological space*. Let $(M, g_{ab})$ be a strongly causal spacetime and $p, q \in M$.

DEFINITION 4.4.1. *Set of continuous future directed causal curves*
   *We define $C(p, q)$ to be the set of continuous, future directed causal curves from $p$ to $q$. If there are two curves in $C$ that differ only by reparametrization then they are considered to be the same curve.*

   In order to understand this definition we have to expand our definition of causal curves to continuous curves.

Definition 4.4.2. *Continuous future directed causal curves*

A continuous future directed causal curve $\lambda$ is such that for any $p \in \lambda$ there is a convex normal neighborhood $U$ of $p$ such that if $\lambda(t_1), \lambda(t_2) \in U$ with $t_1 < t_2$ then there is a future directed differentiable causal curves from $\lambda(t_1)$ to $\lambda(t_2)$ which lies entirely in $U$.

To understand this definition we need to understand a *convex normal neighborhood*.

Definition 4.4.3. *Convex normal neighborhood*

A convex normal neighborhood is an open set $U$ with $p \in U$ such that $\forall q, r \in U$ there exists a unique geodesic $\gamma$ connecting $q$ and $r$ which stays entirely in $U$.

**Note 4.4.1.** For an arbitrary spacetime $(M, g_{ab})$ one can show that a convex normal neighborhood exists for any point $p \in M$.

**Note 4.4.2.** $C(p, q)$ will be an empty set if $q \notin J^+(q)$

We can define a topology $\mathscr{T}$ on $C(p, q)$ making $(C(p, q), \mathscr{T})$ a topological space as follows :

Let $U \in M$ be open, and define $O(U) \subset C(p, q)$ by

$$O(U) \;=\; \{\lambda \in C(p, q) | \lambda \subset U\} \tag{4.4.1}$$

Basically, $O(U)$ consists of all causal curves from $p$ to $q$, which lie entirely within $U$. We define our topology $\mathscr{T}$ by calling a subset $O$ of $C(p, q)$ as *open* if it can be expressed as

$$O \;=\; \bigcup O(U) \tag{4.4.2}$$

where each $O(U)$ is of the form (4.4.1).

It can be proved that $\mathscr{T}$ is *Hausdorff* and *second countable.* We define convergence as follows : a sequence of curves $\lambda_n \to \lambda$ if for every open set $U \in M$ with $\lambda \subset U$, there exists $N$ such that $\lambda_n \subset U$ for all $n > N$.

We will now state one of the most important theorems for $C(p, q)$.

---

**Theorem** 4.4.1. *Let $(M, g_{ab})$ be a globally hyperbolic spacetime and let $p, q \in M$, then $C(p, q)$ is compact.*

---

**Sketch of proof** : (Technical details of this can be found in Section 4.5.2)

We need to prove that every sequence $\{\lambda_n\}$ in $C(p, q)$ has a limit curve. This proof makes use of a fairly technical lemma which states that, if $\{\lambda_n\}$ is a sequence of future inextendible causal curves passing through $p$, then there exists a future inextendible causal curve $\lambda$ passing through $p$ which is a limit curve of $\lambda_n$.

Now suppose $p, q \in D^-(\Sigma)$ where $\Sigma$ is a Cauchy surface, then if we remove the point $q$ from the spacetime, then all curves in $C(p, q)$ become future inextendible, and from the lemma we see that they have a limit curve $\lambda$. If we add the point $q$ back then it has to be a future end point of $\lambda$, so we found a limit curve in $C(p, q)$. Similar for $p, q \in D^+(\Sigma)$. Now if $p$ and $q$ are on different sides of $\Sigma$, then we can find a point $r$ in $I^+(\Sigma)$ and a future directed causal curve $\lambda$ from $p$ to $r$ such that a sub-sequence $\{\tilde{\lambda}_n\}$ converges to $\lambda$ point-wise. Reversing the process we can find a limit curve $\tilde{\lambda}$ of $\{\tilde{\lambda}_n\}$ from $q$ to $r$. Joining the two curves we get a limit curve for the sequence $\{\lambda_n\}$.

## 4.5. Singularity Theorems

In this section we will be closely following (Section 9.5, Pages 237-242, [1]).

### 4.5.1. First Theorem / Timelike case

The first theorem that we will be proving will let us interpret that the universe is globally hyperbolic. We also can establish that the universe at one instant of time is expanding everywhere at a rate bounded away from zero and hence the universe must have begun in a singular state a finite time ago.

---

**Theorem** 4.5.1. **First Theorem** (Theorem 9.5.1, Page 237, [1])

    **Let / Suppose :**

1. $(M, g)$ is a globally hyperbolic spacetime.

2. $R_{ab}\xi^a\xi^b \geq 0$ for all timelike $\xi^a$

    — *This is satisfied if the Einstein's equation is satisfied with the strong energy condition holding for matter.*

    — *This is the point that makes it the timelike case.*

3. There exists a smooth (at least $C^2$) spacelike Cauchy surface $\Sigma$ for which the trace of the extrinsic curvature satisfies $K \leq \tilde{C} < 0$ everywhere, $\tilde{C}$ being a constant.

    **Then** :

→    *No past directed timelike curve from $\Sigma$ can have length greater than $\dfrac{3}{|\tilde{C}|}$. (Basically, all past directed timelike geodesics are incomplete.)*

---

**Proof.** We will prove this using contradiction.

    Claim : Suppose we have a past directed timelike curve $\Gamma$, from $\Sigma$ with length greater than $\dfrac{3}{|\tilde{C}|}$.

- Let $p$ be a point on $\Gamma$ lying beyond the length $\dfrac{3}{|\tilde{C}|}$ from $\Sigma$.

- From the theorem stated below this point, we can say there exists a maximum length curve $\gamma$ ($\gamma$ not the same as $\Gamma$) from $p$ to $\Sigma$, which definitely also must have length greater than $\dfrac{3}{|\tilde{C}|}$

---

**Theorem.** (Theorem 9.4.5, Page 237, [1])
Let $(M, g_{ab})$ be a globally hyperbolic spacetime.
Let $p \in M$ and let $\Sigma$ be a Cauchy surface.
Then there exists a curve $\gamma \in C(\Sigma, p)$ for which $\tau$ attains its maximum value $C(\Sigma, p)$.
($C(\Sigma, p)$ is defined in definition (4.4.1), $\tau$ is the length function defined on $C(\Sigma, p)$)

---

- From the theorem stated below this point, we say that $\gamma$ must be a geodesic with no conjugate point between $\Sigma$ and $p$.

---

**Theorem.** (Theorem 9.4.3, Page 236, [1])
Let $(M, g_{ab})$ be a strongly causal spacetime.
Let $p \in M$, let $\Sigma$ be an achronal, smooth spacelike hypersurface and consider the length function $\tau$ defined on $C(\Sigma, p)$.
A necessary condition for $\tau$ to attain its maximum value at $\gamma \in C(\Sigma, p)$ is that $\gamma$ must be geodesic orthogonal to $\Sigma$ with no point conjugate to $\Sigma$ between $\Sigma$ and $p$.

- We use the proposition which is stated below this point, we impose contradiction as the proposition says that $\gamma$ must have a conjugate point between $\Sigma$ and $p$.

---

PROPOSITION. (Proposition 9.3.4, Page 230, [1])

Let $(M, g_{ab})$ be a spacetime satisfying $R_{ab}\xi^a\xi^b \geq 0$ for all timelike $\xi^a$.

Let $\Sigma$ be a spacelike hypersurface with $K = \Theta < 0$ at a point $q \in \Sigma$.

Then within proper time $t \leq \dfrac{3}{K}$ there exists a point $p$ conjugate to $\Sigma$ along the geodesic $\gamma$ orthogonal to $\Sigma$ and passing through $q$, assuming $\gamma$ can be extended that far.

---

Reading every point and the theorem/proposition corresponding to it, we can conclude that the original curve $\Gamma$ cannot exist.

<div align="right">□</div>

## 4.5.2. Hawking's Theorem

---

THEOREM 4.5.2. (Theorem 9.5.2, Page 238,239, [1])

**Let/Suppose :**

1. $(M, g_{ab})$ be strongly causal.

2. $R_{ab}\xi^a\xi^b \geq 0$ for all $\xi^a$ timelike.
   - This is satisfied if the Einstein's equation is satisfied with the strong energy condition holding for matter.

3. There exists a compact, edge-less, achronal, smooth spacelike hypersurface $\Sigma$ such that for past directed normal geodesic congruence from $\Sigma$ we have $K < 0$ everywhere on $\Sigma$.

4. $C$ denote the maximum value of $K$, so $K \leq C < 0$ everywhere on $\Sigma$.

   **Then :**

$\rightarrow$   At least one inextendible past directed timelike geodesic from $\Sigma$ has length no greater than $\dfrac{3}{|C|}$.

---

**Proof.** We will prove this using contradiction.

**Assumption** : Suppose all past directed inextendible timelike geodesics from $\Sigma$ had length greater then $\dfrac{3}{|C|}$. Let us denote these geodesics by $\sigma$.

- Since the spacetime $(\text{int}[D(\Sigma)], g_{ab})$ satisfies all the assumptions from theorem 4.5.1, all $\sigma$ must least $\text{int}[D(\Sigma)]$.

- $H(\Sigma)$ is the boundary of $D(\Sigma)$

$$H(\Sigma) \;=\; \dot{D}(\Sigma)$$

due to (Proposition 8.3.6, Page 204, [1]). Because of this, all $\sigma$ must intersect $H^-(\Sigma)$ before their length becomes greater than $\dfrac{3}{|C|}$. This obviously implies that $H^-(\Sigma) \neq \emptyset$.

We will now prove that $H^-(\Sigma)$ must be compact and then show that this leads to a contradiction.

Lemma 4.5.1. $H^-(\Sigma)$ is compact.

We will present the proof here :

The main step in proving compactness for $H^-(\Sigma)$ is the demonstration that for each $p \in H^-(S)$ there exists a maximum length orthogonal geodesic from $S$ to $p$.

We can start the proof by saying

- The length of any causal curve from $\Sigma$ to $p$ is less than equal to $\dfrac{3}{|C|}$. Using this fact we can say that there exists a least upper bound $\tau_\circ$ of length for all causal curves from $\Sigma$ to $p$.

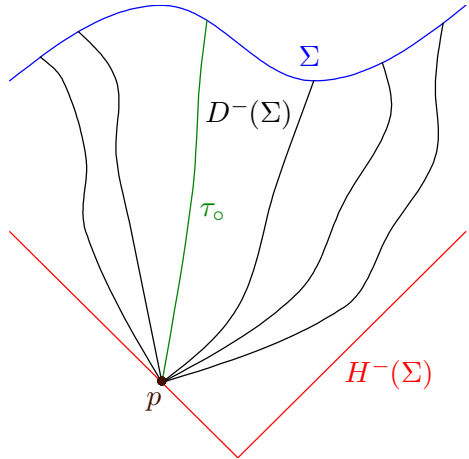

**Figure 4.5.1.** Diagram showing the idea of the least upper bound $\tau_\circ$

(The figure above was redrawn but is inspired from **Fig 9.6, Page 239, [1]**)

We wish to find an orthogonal geodesic from $\Sigma$ to $p$.

- Let $\{\lambda_i\}$ be a *sequence* of timelike curves from $\Sigma$ to $p$ satisfying

$$\lim_{i \to \infty} \tau[\lambda_i] \;=\; \tau_\circ \tag{4.5.1}$$

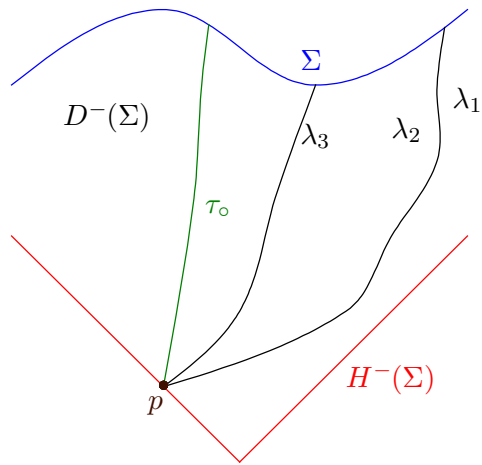

**Figure.** *Step 1:* $\tau(\lambda_i) \to \tau_\circ$

- Choose $q_n \in \lambda_n$, $q_n \neq p$, $q_n \in I^+(p)$ such that

$$\lim_{n \to \infty} q_n = p \tag{4.5.2}$$

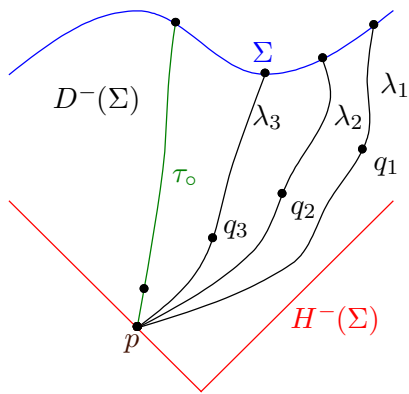

***Figure.*** *Step 2*

- Now we have $q_n \in \text{int}[D^-(\Sigma)]$. This implies that there exists a normal geodesic $\gamma_n$ from $\Sigma$ to $q_n$ which maximizes the length of all causal curves from $\Sigma$ to $q_n$. Basically we have

$$\lim_{n \to \infty} \tau[\gamma_n] = \tau_\circ \tag{4.5.3}$$

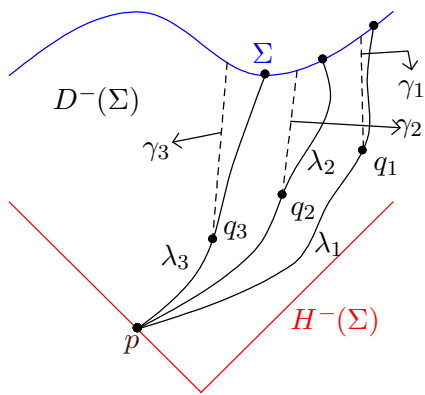

***Figure.*** *Step 3*

We have this result because of (Theorem 9.4.5, [1])

- Now we define $r_n$ to be the intersection point of $\gamma_n$ with $\Sigma$. We know that $\Sigma$ is compact and hence there exists an accumulation point $r$ of the sequence $\{r_n\}$.

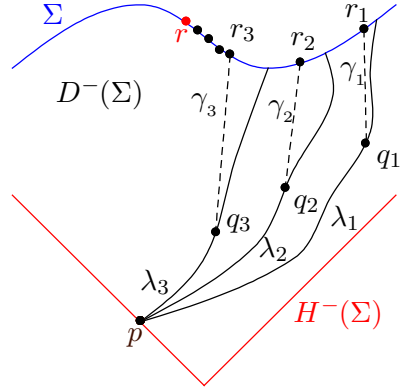

**Figure.** *Step 4*

- We denote $\gamma$ to be the geodesic normal to $\Sigma$ originating from $r$. Due to continuity, $\gamma$ intersects $H^-(\Sigma)$ at $p$. We get

$$\tau[\gamma] = \lim_{n \to \infty} \tau[\gamma_n] = \tau_\circ \qquad (4.5.4)$$

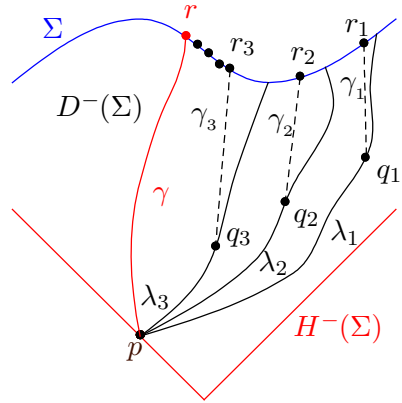

**Figure 4.5.2.** *Step 5*

Hence, we have found the timelike geodesic orthogonal to $\Sigma$ which maximizes the length from $\Sigma$ to $p$.

We prove the compactness of $H^-(\Sigma)$ by showing that every sequence $\{p_n\}$ in $H^-(\Sigma)$ has an accumulation point $p \in H^-(\Sigma)$.

$$\{p_n\} \xrightarrow[n \to \infty]{} p \qquad (4.5.5)$$

Details of the proof for the above statement :

- Let $\{\tilde{\gamma}_n\}$ be a sequence of maximum length orthogonal geodesics from $\Sigma$ to $p_n$.
- Let $\tilde{r}_n$ be the intersection point of $\tilde{\gamma}_n$ with $\Sigma$.
- Let $\tilde{r}$ be an accumulation point of $\{\tilde{r}_n\}$
- Let $\tilde{\gamma}$ be the geodesic starting from $\tilde{r}$ orthogonal to $\Sigma$.

- Let $p$ be the intersection point of $\tilde{\gamma}$ with $H^-(\Sigma)$.
  Then,
  $\rightarrow$   $p$ is an accumulation point of $\{p_n\}$.
  $\rightarrow$   Thus, $H^-(\Sigma)$ is compact.

Hence, we have successfully shown that $H^-(\Sigma)$ is compact.

We know edge$(\Sigma) = \emptyset$ because of the following theorem

---

**Theorem.** (Page 203, Theorem 8.3.5, [1])

   *Every point $p \in H^+$ lies on a null geodesic $\lambda$ contained entirely within $H^+(\Sigma)$ which is either past inextendible or has a past endpoint on the edge of $\Sigma$.*

---

Hence, $H^-(\Sigma)$ contains future inextendible null geodesic.

Since $(M, g_{ab})$ is strongly causal, we can use the following lemma

---

**Lemma.** (Page 197, Lemma 8.2.1, [1])

   *Let $(M, g_{ab})$ be strongly causal and let $K \subset M$ be compact. Then every causal curve in $\lambda$ confined within $K$ must have past and future endpoints in $K$.*

---

to say that our previous claim is impossible if $H^-(\Sigma)$ is compact, which we just proved.

Thus, our assumption that all past directed inextendible timelike geodesics from $\Sigma$ have greater length than $\dfrac{3}{|C|}$ has led to a contradiction.

$\square$

Both the singularity theorems that we proved were timelike geodesic incompleteness in a cosmological context.

Two more significant theorems having a different context can be found in **Page 239-242, [1]**. We didn't include them here because they require a slightly extra set of mathematical toolkit which we didn't include in the previous chapters.

## 4.6.  Bibliographical Notes

This whole chapter was highly influenced by **Chapter 9, [1]**. Whenever statements of theorems and proofs were taken, they were cited accordingly. A lot of comments were made by the author during this whole chapter which cannot be found in any texts.

— ✳✳✳ —

# Chapter 5
## General Black Holes and their properties

In the few of the previous chapters we set up some *causal structure* language using which we can understand the structure of spacetime as defined by the causal relations among the events. An asymptotically flat spacetime in simple terms is a spacetime which behaves like the Minkowski spacetime far away from the source. At the start of this section we will define the concept of asymptotically flat and simple space in the language we developed in the previous sections. This will give us the possibility of defining a *no-escape* region. To get an early intuitive grasp of this concept we can make a proposition.

### 5.1. Asymptotic Structure needed for defining Black hole

PROPOSITION 5.1.1. *Starting from a given event p, we can find only* **causal curves** *connecting it with events in a spatially bounded proper subset of space, then we will say that we cannot escape from some region containing p.*
*Otherwise, we will be free to have a causal connection with events that are outside any bounded proper subset of space, i.e. even with events which are at infinity.*

Infinity in the above proposition is considered in a broad sense as a part of the boundary of our spacetime $(M, g)$.

**Note 5.1.1.** In the following definitions whenever we will be referring to $\mathscr{I}^+$ of a spacetime we mean it as the *future null infinity*. What it means is that all the *future directed null curves* of the spacetime will end on this surface (hypersurface).

DEFINITION 5.1.1. *Asymptotically empty and simple spacetime*
Let us consider a strongly causal spacetime $(M, g)$. It is said to be **asymptotically empty and simple** if there exists a spacetime $(M, g)$ which is called the **associated unphysical space** and a chart :

$$\lambda \colon M \to \tilde{M}$$

*an embedding of M as the manifold with smooth boundary $\partial M$ in $\tilde{M}$ such that*

1. There is a smooth (at least $C^3$) function $\Omega$ on $M$ such that on $\lambda(M)$

$$\Omega > 0 \qquad and \qquad \Omega^2(g) = \lambda_*(\tilde{g}) \quad , (* \ denotes \ pullback, \ see \ A.1)$$

   *i.e $\tilde{g}$ is conformal to g on $\lambda(M)$.*
2. On $\partial M$, $\Omega = 0$ and $d\Omega \neq 0$.
3. Every *null geodesic* on $M$ has **two** endpoints on $\partial M$.
4. $R_{mn} = 0$, on an open neighborhood of $\partial M$ in $M \cup \partial M = \bar{M}$.

So here we can see that an *asymptotically empty and simple spacetime* has a boundary that resembles the properties of infinity in Minkowski space. Moreover from the definition above, we can see that in an asymptotically empty and simple spacetime the boundary $\partial M$ is a null surface. Spacetime is then in its past or in it's future.

Definition 5.1.2. *Weakly asymptotically empty and simple spacetime.*

A spacetime $(M, g)$ is called **weakly asymptotically and simple spacetime** *if there exists an asymptotically empty and simple spacetime $(M', g')$ and a neighborhood $N'$ of $\partial M'$ in $M'$ such that $N' \cap M'$ is isometric to an open set $N \subseteq M$: in this case the region $N$ is asymptotically empty and simple.*

As we see no restriction from the definition, a *weakly asymptotically and simple spacetime* can have many (even infinite) number of *asymptotically empty and simple regions*.

Definition 5.1.3. *Strongly asymptotically predictable spacetime.* (SAP)

Let us consider $(M, g)$ as a weakly asymptotically empty and simple spacetime and a partial Cauchy surface $\mathscr{J}$ in M. $(M, g)$ is called a **strongly future asymptotically predictable spacetime** from the partial Cauchy surface $\mathscr{S}$ if $\mathscr{J}^+$ is contained in $D^+(\mathscr{S})$.

What this physically means is, starting from $\mathscr{S}$ we can predict all the future until $\mathscr{J}^+$ and there will be no singular points in the spacetime which can be seen from the future null infinity and the same is true for a neighborhood of $\partial J^-(\mathscr{J}^+)$

Moreover a strongly future asymptotically predictable space has the nice property that given a partial Cauchy surface $\mathscr{S}$ there exists a family $\mathscr{S}(\tau)$ of spacelike surfaces, homeomorphic to $\mathscr{S}$, which cover $D^+(\mathscr{S}) - \mathscr{S}$ and intersect $\mathscr{J}^+$. Each $\mathscr{S}(\tau)$ can be interpreted as a constant-time surface (Time slices). Moreover the family $\mathscr{S}(\tau)$ is such that:

1. $I^+(\mathscr{S}(\tau)) \supseteq \mathscr{S}(\tau_2)$ if $\tau_1 < \tau_2$ so that, as one would expect, a spacelike surface of the family is always in the (chronological) future of surfaces at previous instants of time.

2. The edge of each spacelike surface $\mathscr{S}(\tau)$ in $\tilde{M}$ is a 2-sphere $S^2(\tau)$ in $\mathscr{J}^+$.

3. At all instants of time $\mathscr{S}(\tau) \cup (\mathscr{J}^+ \cap J^-(S^2(\tau)))$ is a *Cauchy surface* for $D(\mathscr{S})$.

**Remark 5.1.1.** In this situation if $\partial J^-(\mathscr{J}^+) \neq \emptyset$ then it has a nonempty intersection with $\mathscr{S}(\tau)$ for sufficiently large $\tau$. There then will be some events on $\mathscr{S}(\tau)$ at least for sufficiently large $\tau$, which are not causally connected to future null infinity. We will use this remark in the definition of a black hole.

## 5.2. Definition of a Black hole and event horizon.

In this section we will formulate a precise notion of a Black hole in the most abstract setting of a spacetime $(M, g_{ab})$. Unlike the *Schwarzschild Black hole* or the *RN black hole* from the earlier chapters where we defined Black holes using a particular metric, here we will define it from a topological perspective.

This is the first time that we are talking about a general Black hole so maybe it would be fair to state the basic idea of a Black hole in layman terms.

A Black hole is a point-like super-heavy mass which has a *region of no escape.* In physical terms, the gravity is so strong that any particle of light entering it cannot escape it.

In a naive way, we could define a Black hole topologically as $A \subset M$ such that $\forall p \in A$ with $J^+(p) \in A$. We do define a *no escape region* in spacetime but if we stick to this definition then the causal future of any set in any spacetime will be a Black hole. We need to take a little bit more case while specifying what portion of spacetime would be a Black hole.

For an asymptotically flat spacetime, the impossibility of escaping to $\mathscr{J}^+$ provides an appropriate characterization of Black hole.

Now we are close to defining a Black hole precisely. The idea that $J^-(\mathscr{J}^+)$ is *well behaved* but does not include the entire spacetime leads us to define the definition.

**Remark 5.2.1.** One line summary of Strongly asymptotically predictable spacetime.

Let $(M, g_{ab})$ be a asymptotically flat spacetime with associated unphysical spacetime $(\tilde{M}, \tilde{g}_{ab})$. We say that $(M, g_{ab})$ is SAP if in the unphysical spacetime there exists an open region with $\overline{M \cap J^-(\mathscr{J}^+)} \subset \tilde{V}$, such that $(\tilde{V}, \tilde{g}_{ab})$ is *globally hyperbolic*.

DEFINITION 5.2.1. **Black hole** ($\mathscr{B}$)

*A strongly asymptotically predictable spacetime is said to have a black hole if M is not contained in $J^-(\mathscr{J}^+)$. Black hole of such a region is defined by*

$$\mathscr{B} = [M - J^-(\mathscr{J}^+)] \tag{5.2.1}$$

DEFINITION 5.2.2. **Event Horizon** of $\mathscr{B}$

$$\mathscr{H} = \dot{J}^-(\mathscr{J}^+) \cap M \tag{5.2.2}$$

**Note 5.2.1.** Summary/Remarks of the idea of a *Black hole*

- The definition of a *Black hole* above translates the idea of the presence of a *no-escape* region in an strongly predictable spacetime a proper geometric language.

- A *Black hole* is defined as a region from which particles and light rays cannot escape to $\mathscr{J}^+$.

- If we let $\mathscr{S}(\tau)$ denote the family of spacelike surfaces in terms of which an asymptotically empty and simple region in it can be foliated. A Black hole on $\mathscr{S}(\tau)$ is a connected component of the set

$$\mathscr{B}(\tau) = \mathscr{S}(\tau) - J^-(\mathscr{J}^+) \tag{5.2.3}$$

we can see here the idea of a *hole* in $\mathscr{B}$ in the spatial sense as it is contained in the constant time surface $\mathscr{S}(\tau)$.

- The event horizon $\mathscr{H}$ is an *achronal set* and the boundary of the region from which particles and light rays can escape to infinity is generated by null geodesic segments with possibly past endpoints but no future endpoints.

## 5.3. HAWKING'S AREA THEOREM

The proof for the following theorem was taken from [1]. The author deals with it by trying to provide insightful comments to every step. This was not a particularly easy task as it requires a decent amount of sophistication in the mathematical abstraction used to state and prove the theorem.

**THEOREM** 5.3.1. *Hawking's Black hole area theorem.* (`Theorem 12.2.6,`[1])

**Let/Suppose :**

1. *$(M, g_{ab})$ be a strongly asymptotically predictable spacetime satisfying $R_{ab}\chi^a\chi^b \geqslant 0$ for all null $\chi^a$.*

2. *$\Sigma_1$ and $\Sigma_2$ be spacelike Cauchy surfaces for the globally hyperbolic region $\tilde{V}$ with $\Sigma_2 \subset I^+(\Sigma_1)$. (Where $\tilde{V}$ is the unphysical spacetime)*

3. *H is the event horizon (boundary of the black hole region of $(M, g_{ab})$ and let $\mathscr{H}_1 = \mathscr{H} \cap \Sigma_1$, $\mathscr{H}_2 = \mathscr{H} \cap \Sigma_2$.*

**Then :**

→ *The **area** of $\mathscr{H}_2$ is greater than or equal to the area of $\mathscr{H}_1$.*

**Proof.**

Few comments on the assumptions :

- From point 1,
  Using the definition of SAP we get a good mathematical setting that every spacelike surface can be interpreted as time slices.

- In point 2,
  $\Sigma_1, \Sigma_2$ can be thought as time slices in our spacetime. The equation $\Sigma_2 \subset I^+(\Sigma_1)$ means that $\Sigma_2$ is a time slice after $\Sigma_1$. All points of $\Sigma_2$ are accessible from $\Sigma_1$ using appropriate timelike geodesics.

- In point 3,
  Using the argument above, $\mathscr{H}_1, \mathscr{H}_2$ can be presumed as the event horizons at a particular time on the manifold.

For the start of the proof we use contradiction of the following statement/fact and build up on it to get some properties formally :

The null geodesic generators of $\mathscr{H}$ are non-negative i.e. non-converging i.e. their expansion $\Theta \geqslant 0$ everywhere on $H$.

Contradiction :

- Suppose $\Theta < 0$ for some $p \in \mathscr{H}$.

  *Buildup using this contradiction :*

- Let $\Sigma$ be spacelike Cauchy surface for $\tilde{V}$ passing through $p \in \Sigma$, i.e. $p \in \Sigma$.

  $\rightarrow$ Spacelike Cauchy surfaces act like slices of time as we discussed earlier in SAP. So a point $p \in \Sigma$ will have the same $\tau$ (some time parameter) as for all $p' \in \Sigma$.

- Let $\mathscr{H}_\oplus = H \cap \Sigma$

  $\rightarrow$ Deducing using the remark for the previous point, $\mathscr{H}$ is the event horizon of our black hole at some instance in time (that's what the intersection of $H$ with $\Sigma$ does).

- As $\Theta < 0$ at $p$ (from assumption), we can deform $\mathscr{H}$ in an outward neighborhood of $p$ to obtain a surface $\mathscr{H}_\oplus$ on $\Sigma$ such that $\mathscr{H}_\oplus$ enters $J^-(\mathscr{J}+)$ and has $\Theta < 0$ everywhere in $J^-(\mathscr{J}+)$.

  $\rightarrow$ In simple words,

$$\mathscr{H}_\oplus \ \subset \ \Sigma \tag{5.3.1}$$

$$\mathscr{H}_\oplus \cap J^-(\mathscr{J}+) \ \neq \ \emptyset \tag{5.3.2}$$

$$\forall p \in \mathscr{H}_\oplus \ \rightarrow \ \Theta < 0 \tag{5.3.3}$$

  Using all the three properties we can deduce

$$\forall p \in \mathscr{H}_\oplus \cap J^-(\mathscr{J}+) \ \rightarrow \ \Theta < 0 \tag{5.3.4}$$

  $\rightarrow$ This statement here is basically saying that we can construct a neighborhood around the event horizon in a particular time slice such that a few points from $\mathscr{H}_\oplus$ can end up at $\mathscr{J}+$ (future null infinity) giving us $\mathscr{H}_\oplus \cap J^-(\mathscr{J}+) \neq \emptyset$.

At this point we can show that we reach a contradiction.

- Let $K \subset \Sigma$ be the closed region lying between $\mathscr{H}$ and $\mathscr{H}_\oplus$.

  $\rightarrow$ In Minkowski spacetime iff we think about this situation we will have a ring between $\mathscr{H}$ and $\mathscr{H}_\oplus$. Analogous to this we can think of this closed region in an arbitrary spacetimee.

- Let $q \in \mathscr{J}+$ with $q \in \dot{J}^+(K)$.

- The null geodesic generator of $\dot{J}^+(K)$ on which $q$ lies must meet $\mathscr{H}_\oplus$ orthogonally.
  - $\rightarrow$ It is a null geodesic generator (it generates null geodesics for all $p \in K$) which explains it's orthogonality with $\mathscr{H}_\oplus$.
- Having a null geodesic through a point $q \in \dot{J}^+(K)$ would mean that $\Theta \geqslant 0$.

*This is a contradiction from our assumption that $\Theta < 0$ and hence our generator will have a conjugate point before reaching $q$.*

Hence, we can claim that $\Theta > 0$ everywhere on $\mathscr{H}$.

*We still haven't proved our main statement yet. The fact that $\Theta > 0$ on $\mathscr{H}$ is going to help us prove this with a good precision. We will do this now,*

- Each $p \in \mathscr{H}_1$ lies on a future inextendible null geodesic $\gamma$, contained in $H$.
  - In simple words, as the geodesic through any point inside the event horizon will be inextendible as it won't have a future endpoint on the manifold, or we can say we don't know what happens to it once its inside the Black hole.
- As we know $\Sigma_2$ is a Cauchy surface and by the properties of Cauchy surfaces, $\gamma$ (which starts at $\Sigma_1$-also a Cauchy surface) must intersect at some point $q \in \mathscr{H}_2$.
  - $\Sigma_2$ is a Cauchy surface which implies $\exists q \in \mathscr{H}_2$ such that $\gamma \cap \Sigma_2 = \{q\}$.
- Hence, all the points from $\mathscr{H}_1$ can be mapped to a region of $\mathscr{H}_2$ and we obtain a natural map from $\mathscr{H}_1$ to $\mathscr{H}_2$.
  - $f \colon \mathscr{H}_1 \to \mathscr{H}_2 \colon p \mapsto q$ where $f$ is the natural map we talked about in the point.
- We can denote the area of a set $\mathcal{U} \in \Sigma$ by $\mathcal{A}(\mathcal{U})$. We have $\Theta \geqslant 0$ which implies $\mathcal{A}(\mathscr{H}_1) \leqslant \mathcal{A}(\mathscr{H}_2)$.
  - f is injective (all points from $\mathscr{H}_1$ will be mapped to $\mathscr{H}_2$) but it need not be surjective (all points from $\mathscr{H}_2$ need not be mapped back to $\mathscr{H}_1$) as $\Theta \geqslant 0$.

$$f(\mathscr{H}_1) \subseteq \mathscr{H}_2 \;\Rightarrow\; \mathcal{A}(\mathscr{H}_1) \leqslant \mathcal{A}(f(\mathscr{H}_1)) < \mathcal{A}(\mathscr{H}_2) \tag{5.3.5}$$

*Therefore, the area of $\mathscr{H}_1$ has to be smaller than area of $\mathscr{H}_2$.*

$\square$

## 5.4. Bibliographical Notes

This whole chapter was highly influenced by **Chapter 11,12 in** [1] and [8]. Whenever statements of theorems and proofs were taken, they were cited accordingly. A lot of comments were made by the author during this whole chapter which cannot be found in any texts.

———————— *∗∗∗* ————————

# CHAPTER 6

# EXOTIC TOPICS IN GENERAL RELATIVITY

Up to this point since chapter 3, we have focused on spacetimes which do not allow closed timelike curves (CTCs) and obey some kind of a energy condition. This does not mean that spacetimes permitting CTCs or violating all three energy conditions cannot exist. The reason this chapter is named *exotic* has to do with these facts. We start by analyzing exact solutions to the Einstein's equations which permit CTCs. We do this in the so called, *Misner space* and the *Gödel Universe*. We then give a conclusion to why CTCs give problems in physics. Then we discuss the *Alcubierre warp drive* which not only violates all three energy conditions but provides a nice idea for what we know as *warp drives* from science fiction. The metric for such a warp drive can also have CTCs. We do not discuss this artifact here but provide a link where interested readers could look it up.

## 6.1. CLOSED TIMELIKE CURVES (CTCS) IN MISNER SPACE

Misner space is a 2-dimensional spacetime metric with

$$ds^2 = -2dt^2 \, d\psi - t \, d\psi^2 \tag{6.1.1}$$

where $t \in (-\infty, \infty)$ and $\psi$ is periodic i.e. each $\psi$ is identified with $\psi + \psi_0$ where $\psi_0 > 0$.

All the curves with $T = $ constant are closed due to the periodicity of $\psi$. $T < 0$ curves are spacelike and $T > 0$ curves are timelike. It follows from the previous point that all points with $T > 0$ *can* lie on *closed timelike curves*(CTCs) whereas for points with $T < 0$ cannot. We will indeed show that all points $T > 0$ do lie on CTCs. The curve corresponding to $T = 0$ serves as the *chronology horizon*.

DEFINITION 6.1.1. *Chronology Horizon*

*The hypersurface separating the causal and non-causal parts of any spacetime is caled-Chronology horizon.*

The metric (6.1.1) is flat and hence in a local sense it is equivalent to the Minkowski metric. But globally speaking it is staggeringly different than the Minkowski metric due to the identification (periodicity in this case) of $\psi$.

### 6.1.1. Misner process

We will call the procedure which transforms the Minkowski spacetime into Misner as *Misner process.*

Let us start with the two dimensional Minkowski metric

$$ds^2 = -dt^2 + dx^2 \tag{6.1.2}$$

Misner's space occupies only the portion $x < t$ of Minkowski, namely the shaded region I and II of the figure below

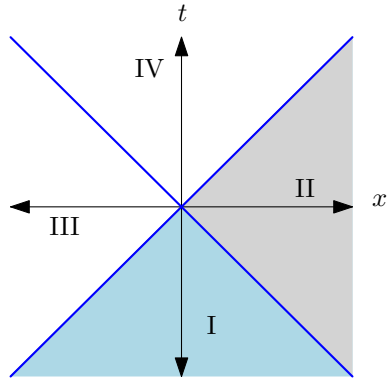

**Figure 6.1.1.** Misner Space

(The figure above was redrawn but is inspired from Pg.3 of [12])

We first make our changes on region I. Let's consider the following coordinate transformation :

$$
\begin{aligned}
x &\equiv -2\sqrt{-T}\sinh\!\left(\frac{z}{2}\right), \\
t &\equiv -2\sqrt{-T}\cosh\!\left(\frac{z}{2}\right)
\end{aligned}
\tag{6.1.3}
$$

where $T \in (-\infty, 0)$ and $z \in (-\infty, \infty)$.

Using these coordinate transformations for the Minkowski metric we get the metric

$$dx = -2\left(-\frac{1}{2\sqrt{-T}}\sinh\!\left(\frac{z}{2}\right)dT + \sqrt{-T}\cosh\!\left(\frac{z}{2}\right)\frac{dz}{2}\right) \tag{6.1.4}$$

$$dt = -2\left(-\frac{1}{2\sqrt{-T}}\cosh\!\left(\frac{z}{2}\right)dT + \sqrt{-T}\sinh\!\left(\frac{z}{2}\right)\frac{dz}{2}\right) \tag{6.1.5}$$

$$
\begin{aligned}
\mathrm{d}x^2 \;=\; 4\Bigg(&-\frac{1}{4T}\sinh^2\!\Big(\frac{z}{2}\Big)\,\mathrm{d}T^2 - T\cosh^2\!\Big(\frac{z}{2}\Big)\frac{\mathrm{d}z^2}{4} - \frac{1}{4}\sinh\!\Big(\frac{z}{2}\Big)\cosh\!\Big(\frac{z}{2}\Big)\mathrm{d}T\,\mathrm{d}z \\
&-\frac{1}{4}\sinh\!\Big(\frac{z}{2}\Big)\cosh\!\Big(\frac{z}{2}\Big)\mathrm{d}z\,\mathrm{d}T\Bigg)
\end{aligned}
\tag{6.1.6}
$$

$$
\begin{aligned}
\mathrm{d}t^2 \;=\; 4\Bigg(&-\frac{1}{4T}\cosh^2\!\Big(\frac{z}{2}\Big)\,\mathrm{d}T^2 - T\sinh^2\!\Big(\frac{z}{2}\Big)\frac{\mathrm{d}z^2}{4} - \frac{1}{4}\cosh\!\Big(\frac{z}{2}\Big)\sinh\!\Big(\frac{z}{2}\Big)\mathrm{d}T\,\mathrm{d}z \\
&-\frac{1}{4}\cosh\!\Big(\frac{z}{2}\Big)\sinh\!\Big(\frac{z}{2}\Big)\mathrm{d}z\,\mathrm{d}T\Bigg)
\end{aligned}
\tag{6.1.7}
$$

Now we calculate

$$
-\mathrm{d}t^2 + \mathrm{d}x^2 \;=\; 4\Bigg(\frac{1}{4T}\mathrm{d}T^2 - \frac{T}{4}\mathrm{d}z^2\Bigg)
\tag{6.1.8}
$$

$$
\mathrm{d}s^2 = \frac{\mathrm{d}T^2}{T} - T\,\mathrm{d}z
\tag{6.1.9}
$$

Now we introduce the coordinate $\psi$

$$
\psi = z - \ln|T|
\tag{6.1.10}
$$

Now imposing this coordinate on (6.1.9) we get (6.1.1).

However, the transformation (6.1.3) applies only to region I of figure:(6.1.1). In order to transform region II from $(t,x) \mapsto (T,z)$ we use the following modified transformation

$$
\begin{aligned}
x &\equiv 2\sqrt{T}\sinh\!\Big(\frac{z}{2}\Big), \\
t &\equiv 2\sqrt{T}\cosh\!\Big(\frac{z}{2}\Big)
\end{aligned}
\tag{6.1.11}
$$

The transformation (6.1.10) applies to $T > 0$ without any modification. So (6.1.10) along with (6.1.11) gives us (6.1.9) in region II as well.

**Note 6.1.1.**

- Curves with $T = $ constant are *spacelike* in region I and *timelike* in region II.

- Curves with $z = $ constant are *timelike* in I and *spacelike* in region II

- Curves with $\psi = $ constant are everywhere *null*.

Until now in this section, we have constructed the Misner's metric (6.1.1) on the half-Minkowski manifold i.e. regions I and II in Fig. (6.1.1).

Now, we choose a parameter $\psi = \psi_\circ > 0$ and we fold the $\psi$ coordinate by identifying $\psi$ with $\psi + \psi_\circ$ (keeping $T$ constant). The coordinate $T$ still has the range $-\infty < T < \infty$.

A pair of such identified constant-$\psi$ lines embedded in the half-Minkowski space is shown in the figure below

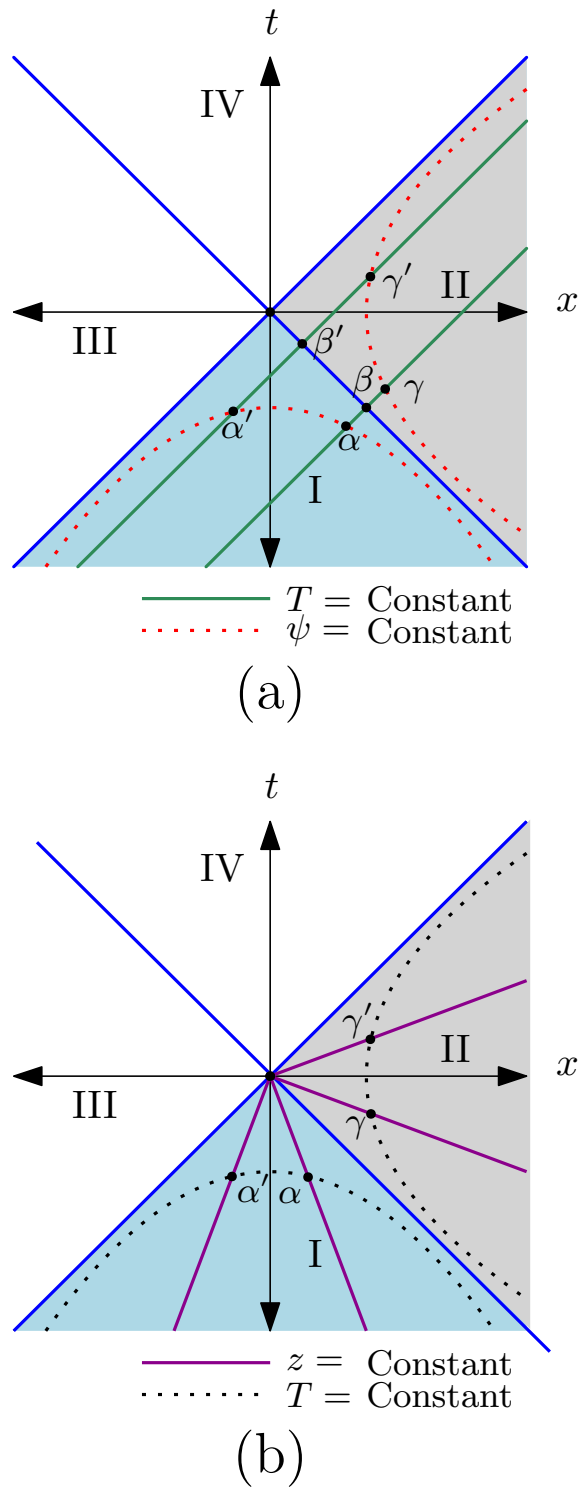

**Figure 6.1.2.** Identified points in Misner space (Denoted by Greek symbols ∼ Greek symbols primed)

(The figure above was redrawn but is inspired from **Page 3**, [12]).

**Note 6.1.2.**

Identifying $\psi$ coordinate at the same $T$ value is equivalent to the identification (by identification we mean a topological identification, for example when we identify two sides of a square/rectangle we get a Torus) of the $z$ coordinate at the same $T$. We can see this from the equation (6.1.10). We can see the lines of constant $z$ along with lines of with constant $T$ in the Fig. (6.1.2).

The transformation from *Minkowski* to *Misner* altogether is

$$\begin{aligned} t &= Te^{\frac{\psi}{2}} - e^{-\frac{\psi}{2}}, \\ x &= Te^{\frac{\psi}{2}} + e^{-\frac{\psi}{2}} \end{aligned} \qquad (6.1.12)$$

and the inverse transformation is

$$\begin{aligned} \psi &= -2\ln\!\left(\frac{x-t}{2}\right), \\ T &= \frac{x^2 - t^2}{4} \end{aligned} \qquad (6.1.13)$$

the transformations above hold in both regions I and II.

**Remark 6.1.1.**

As we talked in the note above, Misner's identification can be imposed by identifying two lines of constant $z$ (at same $T$ and with $z$ values separated by $\psi_\circ$). The velocity $\frac{\mathrm{d}x}{\mathrm{d}t}$ along each such line of constant $z$ is fixed and hence the relative velocity between a pair of identified $z =$ constant lines is well defined.

The relative velocity is (after some calculations) is given by

$$u = \tanh\!\left(\frac{\psi_\circ}{2}\right) \qquad (6.1.14)$$

This is a boost we know on Minkowski space from special relativity. Thus, Misner's *folding* can be viewed as an identification under the action of a boost with velocity $u$.

## 6.1.2. Geodesics in Misner Space

We want to see have a body with mass behaves in Misner space. For simplicity purposes, we will use the boost symmetry mentioned above. Along with that, we will also choose a Lorentz frame in which the object is at rest ($x(t) =$ Constant). We will analyze the properties of a single such geodesic.

As the $\psi$ coordinate is null, it increases monotonically along any timelike geodesic and is suitable to use as a parameter. Hence, we conveniently express our geodesics using the function $T(\psi)$.

Therefore, a single static geodesic satisfies (in covering space i.e. region I and II) $x =$ constant $\equiv x_0$. Due to (6.1.12) we get

$$T(\psi) = -e^{-\psi} + x_0\, e^{-\frac{\psi}{2}} \qquad (6.1.15)$$

Now we try to analyze how does this geodesic behave when we propagate it from some $T < 0$ to $T = 0$.

Looking at the relation (6.1.15) we can see that there are two different classes of such geodesics:

- The ones with $x_0 < 0$ only approach $T = 0$ at $\psi \to \infty$.
- The ones with $x_0 > 0$ will all reach $T = 0$ at a *finite* $\psi$ and continue their journey in the region $T > 0$.

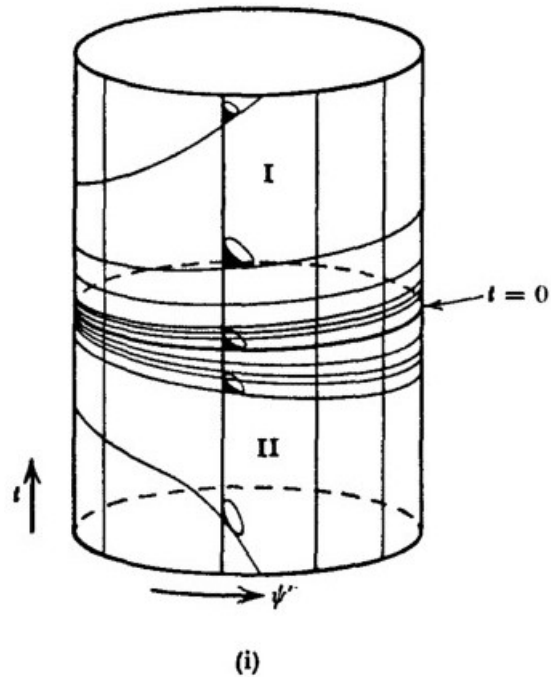

**Figure 6.1.3.** Geodesics in Misner space. Taken from Pg.172,[8]

The goal of this section is to understand the motion of objects in the region where CTCs exist. So for the remainder of this section (which is a brief discussion) we will restrict our attention to the second class of geodesics that we mentioned above, namely the $x_0 > 0$ geodesics.

Let us consider the behavior of these geodesics at $T > 0$. For each of such geodesics the function $T(\psi)$ will reach its maximum at its intersection point with $t = 0$.

The following figure will demonstrate the behavior described above

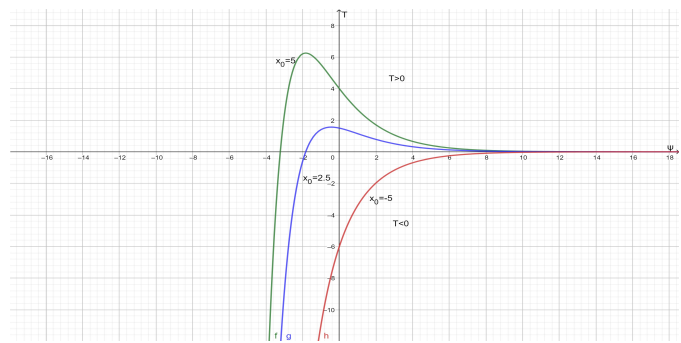

**Figure 6.1.4.** Constant $x$ geodesics plotted in $(T, \psi)$ coordinates.

The figure above displays two different $x = x_0 > 0$ geodesics and the $t = 0$ line.

These static geodesics exhibit a simple symmetry when displayed in the $(T, z)$ coordinates. We can easily see that in the $T > 0$ region, the relation $x = x_0$ yields

$$T(z) = \left( \frac{x_0}{2\cosh\left(\frac{z}{2}\right)} \right)^2 \tag{6.1.16}$$

This function is symmetric around $z = 0$ and hence, it's maximum is obtained at $z = 0$. This is illustrated in the figure below

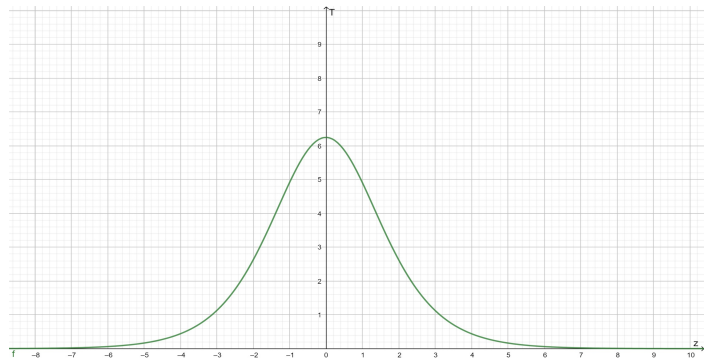

**Figure 6.1.5.** Symmetry of a single $x_0 > 0$ geodesic around $z = 0$

The figure above displays a single $x = x_0$ geodesic and the line $t = 0$ coincides the line $z = 0$ in $(T, z)$ coordinates, again demonstrating that the geodesics $x = x_0$ reach their maximal $T$ value at a point where $t$ vanishes.

### 6.1.3. Bibliographical Notes

The section above was taken from [12] (A few simple calculations which were unclear were done explicitly by the author). Some background reading was done using the following resources : [3],[8] and hence a few concepts might be motivated from explanations based in these resources. The figures drawn were also inspired from the same paper [12].

## 6.2. CTCs in the Gödel Universe

Derived in 1935 by Kurt Gödel the Gödel metric is an exact solution to Einstein's field equations(EFEs). The Gödel metric has some unique characteristics which make us want to dig in more. One of the most unique characteristic is that it allows the existence of *closed timelike curves* (CTCs). In this section we are not going to derive the metric but instead analyze the interesting scenarios that could come up in this universe.

### 6.2.1. Gödel metric

The Gödel universe is a solution to EFEs for an infinitely long rotating dust cylinder. In technical terms it is a solution to EFEs for a pressure-less perfect fluid source rotating around an axis.

The metric in $(t, x, y, z)$ system is the following equation (along with it's matrix form and the inverse matrix) :

$$\mathrm{d}s^2 \;=\; -\mathrm{d}t^2 + \mathrm{d}x^2 - \frac{1}{2}e^{2x}\mathrm{d}y^2 + \mathrm{d}z^2 - e^x(\mathrm{d}t\,\mathrm{d}y + \mathrm{d}y\,\mathrm{d}t) \tag{6.2.1}$$

$$g_{mn} \;=\; \begin{bmatrix} -1 & 0 & -e^x & 0 \\ 0 & 1 & 0 & 0 \\ -e^x & 0 & -\dfrac{1}{2}e^{2x} & 0 \\ 0 & 0 & 0 & 1 \end{bmatrix} \tag{6.2.2}$$

$$g^{mn} \;=\; \begin{bmatrix} 1 & 0 & -2e^{-x} & 0 \\ 0 & 1 & 0 & 0 \\ -2e^{-x} & 0 & 2e^{-2x} & 0 \\ 0 & 0 & 0 & 1 \end{bmatrix} \tag{6.2.3}$$

DEFINITION 6.2.1. *Gödel-type metrics*

*These are metrics for spacetimes where the metric $g_{mn}$ can be written as the difference between a degenerate background metric $b_{mn}$ which has one dimension lower than the entire spacetime and the tensor product of two unit timelike vectors $v_m$.*

$$g_{mn} \;=\; b_{mn} - v_m v_n$$

*Basically $b_{mn}$ is a matrix of $d \times d$ dimension with a zero row (Matrix with rank $d-1$).*

The metric 6.2.1 is one of the most trivial case of Gödel-type metric giving us the most trivial Gödel-type spacetime.

We should be able to write down the metric (6.2.1) according to the definition of Gödel-type metrics above. We can achieve this by writing down the following background metric and a timelike vector field,

- Background metric $b_{mn}$ :

$$b_{mn} \;=\; \begin{bmatrix} 0 & 0 & 0 & 0 \\ 0 & 1 & 0 & 0 \\ 0 & 0 & \dfrac{1}{2}e^{2x} & 0 \\ 0 & 0 & 0 & 1 \end{bmatrix} \tag{6.2.4}$$

- Timelike vector field (lower-indexed four velocity) $v_m$:

$$v_m = (1, 0, e^x, 0) \tag{6.2.5}$$

$$v_m v_n = \begin{bmatrix} 1 \\ 0 \\ e^x \\ 0 \end{bmatrix} \begin{bmatrix} 1 & 0 & e^x & 0 \end{bmatrix}$$

$$= \begin{bmatrix} 1 & 0 & e^x & 0 \\ 0 & 0 & 0 & 0 \\ e^x & 0 & e^{2x} & 0 \\ 0 & 0 & 0 & 0 \end{bmatrix} \tag{6.2.6}$$

We can see now that :

$$b_{mn} - v_m v_n = \begin{bmatrix} 0 & 0 & 0 & 0 \\ 0 & 1 & 0 & 0 \\ 0 & 0 & \dfrac{1}{2}e^{2x} & 0 \\ 0 & 0 & 0 & 1 \end{bmatrix} - \begin{bmatrix} 1 & 0 & e^x & 0 \\ 0 & 0 & 0 & 0 \\ e^x & 0 & e^{2x} & 0 \\ 0 & 0 & 0 & 0 \end{bmatrix} \tag{6.2.7}$$

$$= \begin{bmatrix} -1 & 0 & -e^x & 0 \\ 0 & 1 & 0 & 0 \\ -e^x & 0 & -\dfrac{1}{2}e^{2x} & 0 \\ 0 & 0 & 0 & 1 \end{bmatrix} = g_{mn} \tag{6.2.8}$$

Therefore we indeed can write the Gödel metric as ,

$$g_{mn} = b_{mn} - v_m m_n \tag{6.2.9}$$

**Remark 6.2.1.**

- We can see that $b_{mn}$ is a degenerate metric of dimension $d = 4$ (our spacetime has dimension 4) and rank $d - 1 = 3$.

- The norm of our 4-velocity vector $v_n$ can be checked to be $-1$.

$$v_n v^n = g^{mn} v_m v_n \tag{6.2.10}$$
$$= g^{00} v_0 v_0 + 2 g^{20} v_2 v_0 + g^{22} v_2 v_2 + g^{33} v_3 v_3 \tag{6.2.11}$$
$$= (1)^2 + 2(-2e^{-x})e^x + 2e^{-2x}(e^x)^2 \tag{6.2.12}$$
$$= 1 - 4 + 2 \tag{6.2.13}$$
$$= -1 \tag{6.2.14}$$

$v_m$ is indeed a timelike unit vector with respect to our metric.

## 6.2.2. CTCs in the Gödel universe

Let us make a coordinate transformation from our original coordinates $(t, x, y, z)$ to cylindrical coordinates $(t, r, \varphi, z)$ where

$$t \in (-\infty, \infty) \qquad r \in (0, \infty) \qquad \varphi \in [0, 2\pi] \qquad z \in (-\infty, \infty)$$

After a few calculations we get the following metric :

$$ds^2 \;=\; -dt^2 + dr^2 + dz^2 - \sinh^2 r(\sinh^2 r - 1)\,d\varphi^2 + \sqrt{2}\,\sinh^2 r(d\varphi\,dt + dt\,d\varphi) \qquad (6.2.15)$$

where $\varphi = 0$ and $\varphi = 2\pi$ are identified (that means we consider that we came back to the same point in that coordinate). Hence, $\varphi = 0, \varphi = 2\pi$ being the point where the curve *closes*.

Let us consider a parameterized curve $\gamma(s)$ where all coordinates except $\varphi$ are constant. Then the tangent vector to this curve looks as follows:

$$v^\mu \;=\; \frac{d\gamma^\mu}{ds} = \left( \frac{d\gamma^0}{ds}, \frac{d\gamma^1}{ds}, \frac{d\gamma^2}{ds}, \frac{d\gamma^3}{ds} \right) \qquad (6.2.16)$$

$$=\; \left( 0, 0, \frac{d\varphi}{ds}, 0 \right) \qquad (6.2.17)$$

If $\gamma$ is a timelike curve then its tangent must be a timelike vector for all points $s$ on this curve. We can see this by writing $dt = dr = dz = 0$ in  (6.2.2).

$$ds^2 \;=\; -\sinh^2 r_c(\sinh^2 r_c - 1)d\varphi^2 \qquad (6.2.18)$$

where $r_c$ is the constant $r$ coordinate along this curve.

$$g_{mn}v^m v^n \;=\; g_{\varphi\varphi}\,v^\varphi v^\varphi$$

$$=\; -\sinh^2 r_c(\sinh^2 r_c - 1)\left( \frac{d\varphi}{ds} \right)^2 < 0 \qquad (6.2.19)$$

$$-\sin^2 r_c - 1 \;<\; 0 \qquad (6.2.20)$$
$$\sinh^2 r_c - 1 \;>\; 0$$
$$\sinh^2 r_c \;>\; 1$$
$$\sinh r_c \;>\; 1 \qquad (6.2.21)$$

Solving the above equation we get,

$$\frac{e^{2x}-1}{2e^x} = \frac{e^x e^x - 1}{2e^x} \;>\; 1 \qquad (6.2.22)$$

$$\text{Let} \;\; : \;\; y = e^x \qquad (6.2.23)$$

$$\frac{y^2-1}{2y} \;>\; 1$$

$$y^2 - 1 - 2y \;>\; 0$$

$$y^2 - 2y - 1 \;>\; 0 \qquad (6.2.24)$$

Solving this we get :

$$y \;>\; 1 + \sqrt{2} \qquad (6.2.25)$$
$$e^{r_c} \;>\; 1 + \sqrt{2}$$
$$r_c \;>\; \log(1 + \sqrt{2}) \qquad (6.2.26)$$

What we calculated was the radius outside which lines of constant time will be closed timelike curves. Basically, whenever we are outside the radius $r_c$, we will be traveling on a closed timelike curve which is constant in coordinate time. This is illustrated in the diagram below

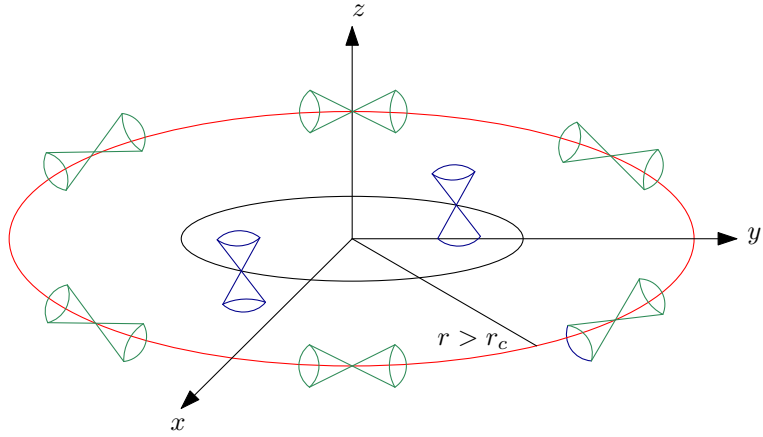

**Figure 6.2.1.** Closed timelike curves in Gödel Universe

During our travel some proper time elapses, and when returning to the initial point $\varphi = 0$, we will have traveled back to the point where it started, which for us will be a time travel backwards in time.

The amount of proper time $\Delta \tau$ passing for the traveler will be given from the line element on the curve (6.2.18) and by using the definition of proper time we get

$$
\begin{aligned}
\mathrm{d}\tau^2 \;&=\; -\mathrm{d}s^2 && (6.2.27)\\
&=\; \sinh^2(r_c)(\sinh^2(r_c)-1)\mathrm{d}\varphi^2 && (6.2.28)
\end{aligned}
$$

$$
\begin{aligned}
\Delta \tau \;&=\; \int_0^{2\pi} \sinh{(r_c)}\sqrt{(\sinh^2 r_c - 1)}\,\mathrm{d}\varphi \\
&=\; 2\pi \sinh{(r_c)}\sqrt{(\sinh^2 r_c - 1)} && (6.2.29)
\end{aligned}
$$

which will be defined and positive once we have the condition (6.2.26) for a timelike curve.

Let's return to general Gödel type metrics. We can always do a cylindrical coordinate change in these metrics, so that the line elements become

$$
\begin{aligned}
\mathrm{d}s^2 \;&=\; -(\mathrm{d}t + s(r,\varphi)\mathrm{d}z)^2 + \mathrm{d}r^2 + r^2\mathrm{d}\varphi^2 + \mathrm{d}z^2 && (6.2.30)\\
g_{\mu\nu} \;&=\; \begin{bmatrix} -1 & 0 & 0 & -s \\ 0 & 1 & 0 & 0 \\ 0 & 0 & r^2 & 0 \\ -s & 0 & 0 & 1-s^2 \end{bmatrix} && (6.2.31)
\end{aligned}
$$

where $s(r,\varphi)$ is an arbitrary function of the coordinates $r$ and $\varphi$. Let us now consider a general curve $\gamma(\lambda)$, with all coordinates having a general dependence on the parameter, where our parameter $\lambda \in [0, 2\pi]$. The tangent vector $T^a$ to the curve is

$$
\begin{aligned}
T^m \;&=\; \frac{\mathrm{d}\gamma}{\mathrm{d}\lambda} && (6.2.32)\\
&=\; \left(\frac{\mathrm{d}t}{\mathrm{d}\lambda}, \frac{\mathrm{d}r}{\mathrm{d}\lambda}, \frac{\mathrm{d}\varphi}{\mathrm{d}\lambda}, \frac{\mathrm{d}z}{\mathrm{d}\lambda}\right) && (6.2.33)
\end{aligned}
$$

We can impose that this curve should be timelike ($V_a V^a$) and use the metric (6.2.30) to explicitly calculate it

$$g_{ab}V^a V^b = -1 \qquad (6.2.34)$$

$$g_{00}V^0 V^0 + (g_{03}V^0 V^3 + g_{30}V^3 V^0) + g_{11}V^1 V^1 + g_{22}V^2 V^2 + g_{33}V^3 V^3 = -1 \qquad (6.2.35)$$

$$-\left(\frac{\mathrm{d}t}{\mathrm{d}\lambda}\right)^2 - 2\frac{\mathrm{d}t}{\mathrm{d}\lambda}\frac{\mathrm{d}z}{\mathrm{d}\lambda} + \left(\frac{\mathrm{d}r}{\mathrm{d}\lambda}\right)^2 + r^2\left(\frac{\mathrm{d}\varphi}{\mathrm{d}\lambda}\right)^2 + (1-s^2)\left(\frac{\mathrm{d}z}{\mathrm{d}\lambda}\right)^2 = -1 \qquad (6.2.36)$$

$$-s\frac{\mathrm{d}z}{\mathrm{d}\lambda} \pm \sqrt{\left(\frac{\mathrm{d}r}{\mathrm{d}\lambda}\right)^2 + r^2\left(\frac{\mathrm{d}\phi}{\mathrm{d}\lambda}\right)^2 + \left(\frac{\mathrm{d}z}{\mathrm{d}\lambda}\right)^2 + 1} = \frac{\mathrm{d}t}{\mathrm{d}\lambda} \qquad (6.2.37)$$

Let us now Fourier-series expand this expression conventionally, having $\lambda$ in the entire defined interval $[0, 2\pi]$. The functions $r(\lambda), \phi(\lambda), z(\lambda)$ are assumed to be periodic in $\lambda$ at this stage, in order to have CTCs.

Fourier expansion is

$$\frac{\mathrm{d}t}{\mathrm{d}\lambda} = \sum_{n=-\infty}^{\infty} c_n e^{in\lambda} \qquad (6.2.38)$$

with the coefficients given by

$$c_n = \frac{1}{2\pi}\int_0^{2\pi} \frac{\mathrm{d}t}{\mathrm{d}\lambda} e^{-in\lambda}\, \mathrm{d}\lambda \qquad (6.2.39)$$

Now, look at the first term in the expansion i.e. $n=0$ with coefficient $c_0$. This will be the constant term in our expansion. We can write it as

$$
\begin{aligned}
c_0 &= \frac{1}{2\pi}\int_0^{2\pi}\left(-s\frac{\mathrm{d}z}{\mathrm{d}\lambda} \pm \sqrt{\left(\frac{\mathrm{d}r}{\mathrm{d}\lambda}\right)^2 + r^2\left(\frac{\mathrm{d}\phi}{\mathrm{d}\lambda}\right)^2 + \left(\frac{\mathrm{d}z}{\mathrm{d}\lambda}\right)^2 + 1}\right)\mathrm{d}\lambda \\
&= \frac{1}{2\pi}\int_0^{2\pi}\left(-s\frac{\mathrm{d}z}{\mathrm{d}\lambda}\right)\mathrm{d}\lambda \pm \frac{1}{2\pi}\int_0^{2\pi}\left(\sqrt{\left(\frac{\mathrm{d}r}{\mathrm{d}\lambda}\right)^2 + r^2\left(\frac{\mathrm{d}\phi}{\mathrm{d}\lambda}\right)^2 + \left(\frac{\mathrm{d}z}{\mathrm{d}\lambda}\right)^2 + 1}\right)\mathrm{d}\lambda \\
&\equiv c_{0a} \pm c_{0b} \qquad (6.2.40)
\end{aligned}
$$

Notice that the integrand of the second integral is a positive definite i.e. $c_{0b} > 0$. After plugging this in (6.2.37) and integrating, these constants will give a linear term

$$t(\lambda) = (c_{0a} \pm c_{0b})(\lambda - \lambda_0) + \frac{1}{2\pi}\int_{\lambda_0}^{\lambda}\sum_{n\neq 0} c_n\, e^{in\lambda}\, \mathrm{d}\lambda \qquad (6.2.41)$$

Assuming, as stated earlier, if all other components are periodic in $\lambda$ (except for $t$ now), then we can find solutions such that the integral term in this is periodic in $\lambda$. However for $t$ to be periodic, since $\lambda \neq \lambda_0$, we must have that $c_{0a} + c_{0b}$ vanishes. But we already know that $c_{0b}$ is strictly positive definite. Therefore we can conclude that we can not have CTCs if $c_{0a} = 0$, for then the linear term in $\lambda$ has no chance of vanishing. Now let us look at what this condition $c_{0a} = 0$ is,

$$\frac{1}{2\pi}\int_0^{2\pi}\left(-s(r,\phi)\frac{\mathrm{d}z}{\mathrm{d}\lambda}\right)\mathrm{d}\lambda = 0 \qquad (6.2.42)$$

where we now added the explicit dependence of the function $s(r, \phi)$. Since we are assuming $r(\lambda)$ and $\phi(\lambda)$ to be periodic in order to find CTCs, we also have that $s(r, \phi)$ is periodic, which lets us Fourier-series expand it

$$s(r(\lambda), \phi(\lambda)) = d_0 + f(\lambda) = d_0 + \sum_{n \neq 0} d_n e^{in\lambda} \tag{6.2.43}$$

where $f(\lambda)$ is guaranteed to be a periodic function. Now

$$-\frac{1}{2\pi} d_0 \int_0^{2\pi} \frac{\mathrm{d}z}{\mathrm{d}\lambda} \mathrm{d}\lambda - \frac{1}{2\pi} \int_0^{2\pi} f(\lambda) \frac{\mathrm{d}z}{\mathrm{d}\lambda} \mathrm{d}\lambda = 0 \tag{6.2.44}$$

Since we are assuming $z(\lambda)$ to be a periodic function, the first integral will vanish, leaving us with the demand of

$$\int_0^{2\pi} f(\lambda) \frac{\mathrm{d}z}{\mathrm{d}\lambda} \mathrm{d}\lambda = 0 \tag{6.2.45}$$

But in a search of CTCs e have the freedom to choose the periodic function $z(\lambda)$ as we wish, say

$$\int_0^{2\pi} f(\lambda)^2 \mathrm{d}\lambda = 0 \tag{6.2.46}$$

This is only possible if $f(\lambda)$ is a constant and thus also $s(r, \phi)$ is a constant. However, suppose we exclude the case of constant $s(r, \phi)$, then we can always find periodic functions $s(r, \phi), z(\lambda), r(\lambda), v(\lambda)$ such that indeed the demand (6.2.37) is not satisfied, and then we can choose coefficients $c_{0a}$ such that it cancels $c_{0a}$ and so $t(\lambda)$ can also be periodic.

We have rephrased our question about whether CTCs can exist in this metric to the question whether $s(r, \phi)$ is constant or not. If this function is *not* constant, then there always exist closed timelike curves in the Gödel-type metrics.

### 6.2.3. Problems for Gödel-type metrics

When we derive the geodesics of Gödel-type metrics, we will note that these CTCs are *not geodesics* of the spacetime. This implies that a massive particle will need an external force acting on it and so forcing it to move on the given closed timelike curve. The interaction connecting the particle and this forcing machine will need to essentially also travel on a closed timelike curve to be able to complete the entire trajectory. But then this forcing machines, also being a massive particle by assumption, needs also something to force it on a CTC, and would that be the particle it is initially forcing? This would mean that they are producing an internal force which pushes the two together, but we need an external force acting on the entire system as a whole. What will push the system of the particle + force machine? We need a third object pushing the two and then we are back in the beginning and we can go on forever. To solve this paradox, we need a force machine which can act over time and space and needs not the interaction to follow the same type of path as the forced particle, however this is an unphysical interaction which needs more study.

In our derivations of the CTCs in the Gödel-type metrics, we were assuming all the time the identification of the angular coordinate $\theta = 0$ with $\theta = 2\pi$, even when the angular coordinate $\phi$ becomes a timelike coordinate. There is no mathematical or physical demand which allows us to still identify the angular coordinate as we do in the case when it is spacelike as in the region where it becomes timelike.

### 6.2.4. Bibliographical notes

This section and the next section (Conclusion for CTCs) was taken from http://www.diva-portal.org/smash/get/diva2:1111006/FULLTEXT01.pdf.

## 6.3.  Conclusion for CTCs

Why is are physicists trying to avoid the CTCs in the solutions for (2)? CTCs provide a possibility for time travel to the past. As an example, if an observer enters a closed timelike curve at his birth and stays in it during his entire life, then at some point the person would realize that he came back to the moment of his birth, this will be the point where the CTC closes into its start point. This way, the observer travels back in time. This possibility gives rise to some questions.

First and foremost, consider that the entire world goes into a CTC at an early stage of one's ancestor's life. The world then develops as usual until the person in question is born. Then suddenly, the CTC closes into its initial point by returning to the past moment of the early stage of the persons ancestors life. The person can cause the death of his ancestor, assuming he is free to take any action. Eventually, this person in question, will he exist or not? This is called the *Grandfather paradox* and evokes questions about how affecting events in the past can and will cause changes in the future. Think of this situation as a non-linear differential equation, small changes in initial conditions can cause a lot of chaos.

Another paradox arises when we assume there is an absolute free will for the traveler on a CTC. Suppose an observer A meets an older version of himself, let us call him B. B has been traveling on a CTC and thus has been aging, but could get back to the initial point of his departure. B tells A how to travel on the CTC, allowing thus for A to depart on the CTC. Now, when A arrives to the same point as B was earlier, he meets the older version of himself A (in short, A becomes B along the CTC). However, this time, assuming he can do some he decides not to share the information about the secret of the CTC. Therefore, A will not be able to travel on the CTC. But then who will have told the initial A how to travel on the CTC in the first place? Did the CTC never even exist? This is referred to as the *Free will paradox.*

## 6.4. Alcubierre Warp Drive

### 6.4.1. Introduction

In this section we will be discussing a paper on warp drives in General relativity. Particularly summarizing the paper called *The warp drive: hyper-fast travel within general relativity* by *Miguel Alcubierre* [13]. While summarizing the paper we will focus on the key aspects without leaving out the mathematical intricacies.

### 6.4.2. Idea behind such a Warp Drive

Wormholes or more precisely mathematical formulations such as the *Einstein-Rosen bridge* is considered to be one of well known ideas for warp drives. A warp drive is something that can take you from point $A \to B$ faster than what light would take from $A \to B$. In this model we won't be needing a wormhole but we will be needing exotic matter to create a distortion of spacetime as needed.

The idea:

- We modify the spacetime in such a way that it allows a spaceship to travel with an arbitrarily large speed.

How?

- We make a purely local expansion of spacetime behind the spaceship and an opposite contraction in front of it.
- As a result of this, motion faster than the speed of light as seen by observers outside the disturbed region is achieved mathematically.

Any violations of known physics?

- Faster than speed of light travel?
  - → Apparently no. We will obtain a value for speed that is much larger than the speed of light. Nevertheless, we will always stay inside the local light cones.
  - → The enormous speed of separation will come from the expansion of spacetime itself.

### 6.4.3. Physics behind such a Warp drive

We will use the 3+1 formulation of General relativity with a metric. Any metric described in the language of 3+1 formalism will have no causal curves. We can also guarantee our spacetime to be Globally hyperbolic i.e. spacetime will be described by spacelike hypersurfaces which will define the chronology (constant coordinate time $t$ spacelike hypersurfaces).

The geometry of spacetime will be given by the following quantities ($G = c = 1$):

1. The 3-metric of the hypersurfaces denoted by $\gamma_{ij}$.
   - It is positive definite for all values of $t$ just like any other spatial metric.
2. The lapse function $\alpha$.
   - This gives the interval of proper time between nearby hypersurfaces as measured by observers whose four velocity is normal to the hypersurface (also known as "Eulerian observers").
3. The shift vector $\beta^i$.
   - This relates the shift vector spatial coordinate systems on different hypersurfaces.

Using the above quantities we can write the metric of spacetime as :

$$\mathrm{d}s^2 = -\mathrm{d}\tau^2 = g_{\alpha\beta}\,\mathrm{d}x^\alpha\,\mathrm{d}x^\beta = -(\alpha^2 - \beta_i\beta^i)\mathrm{d}t^2 + 2\beta_i\,\mathrm{d}x^i\,\mathrm{d}t + \gamma_{ij}\,\mathrm{d}x^i\,\mathrm{d}x^j \qquad (6.4.1)$$

where the Greek indices take values from the set $\{0, 1, 2, 3\}$ and Latin indices take values from $\{1, 2, 3\}$.

Let our spaceship move along the $x$ axis of a Cartesian coordinate system. Our goal is to find the metric that will *push* the spaceship along a trajectory described by $x_s(t)$ which is an arbitrary function of time. We can get our metric (6.4.1) to have the properties mentioned above by setting the 3 parameters that we discussed to be the following :

$$\alpha \;=\; 1 \tag{6.4.2}$$

$$\beta^x \;=\; -v_s(t)\, f(r_s(t)) \tag{6.4.3}$$

$$\beta^y = \beta^z \;=\; 0 \tag{6.4.4}$$

$$\gamma_{ij} \;=\; \delta_{ij} \tag{6.4.5}$$

where:

$$v_s(t) \;=\; \frac{\mathrm{d}x_s(t)}{\mathrm{d}t} \tag{6.4.6}$$

$$r_s(t) \;=\; [(x - x_s(t))^2 + y^2 + z^2]^{\frac{1}{2}} \tag{6.4.7}$$

$$f(r_s) \;=\; \frac{\tanh(\sigma(r_s + R)) - \tanh(\sigma(r_s - R))}{2\tanh(\sigma R)} \tag{6.4.8}$$

$$R > 0 \quad \text{and} \quad \sigma > 0 \qquad \text{are arbitrary parameters}$$

If one is wondering where are these $\tanh(x)$ functions coming from you might want to recall the *Lorentz boosts* in *special relativity* and just assume the function is doing some analogous work here.

For $\sigma \gg 0$ the function $f$ approaches a top hat function rapidly :

$$\lim_{\sigma \to \infty} f(r_s) \;=\; \left\{ \begin{array}{ll} 1 & \text{for } r_s \in [-R, R] \\ 0 & \text{otherwise} \end{array} \right.$$

With all this framework we can define our metric (6.4.1) as

$$\mathrm{d}s^2 \;=\; -\mathrm{d}t^2 + (\mathrm{d}x - v_s f(r_s)\mathrm{d}t)^2 + \mathrm{d}y^2 + \mathrm{d}z^2 \tag{6.4.9}$$

**Note 6.4.1.**

- We saw in equation (6.4.5) that the 3D geometry of our hypersurfaces is flat.
- Equation (6.4.2) refers to the fact that timelike curves normal to the hypersurfaces are in a free fall i.e. they are geodesics.
- The shift vector $\beta^i$ vanishes for the case $r_s \gg R$.
- In conclusion we can say that the spacetime will be essentially flat everywhere except within a radius of order $R$, centered at $(x_s(t), 0, 0)$.

Because our hypersurfaces are flat, the information about the curvature of the spacetime will be encoded in the *extrinsic curvature tensor* $K_{ij}$.

- Extrinsic curvature is defined as how 3D hypersurfaces are embedded in a 4D spacetime and is defined by the following formula:

$$K_{ij} \;=\; \frac{1}{2\alpha}\left( \nabla_i \beta_j + \nabla_j \beta_i - \frac{\partial g_{ij}}{\partial t} \right) \tag{6.4.10}$$

$$\nabla_i \;:\; \text{Covariant derivative with respect to 3 metric } \gamma_{ij}.$$

We can compute this using our metric and the shift vector $\beta^i$ to give out the following form of the extrinsic curvature:

$$K_{ij} = \frac{1}{2}(\partial_i \beta_j + \partial_j \beta_i) \tag{6.4.11}$$

- The expansion $\theta$ of the volume elements associated with the Eulerian observers us given in terms of $K_{ij}$ as

$$\theta = -\alpha \operatorname{Tr}(K_{ij}) \tag{6.4.12}$$

Using the above expression one can show that

$$\theta = v_s \frac{x_s}{r_s} \frac{\mathrm{d}f}{\mathrm{d}r_s} \tag{6.4.13}$$

The following figure shows a graph of $\theta$ as a function of $x$ and $\rho = \sqrt{y^2 + z^2}$ for the particular case when $\sigma = 8$ and $R = v_s = 1$. The center of the perturbation corresponds to the spaceship position $x_s(t)$.

One can clearly see how the volume elements are expanding behind the spaceship and contracting in front of it in the following figure :

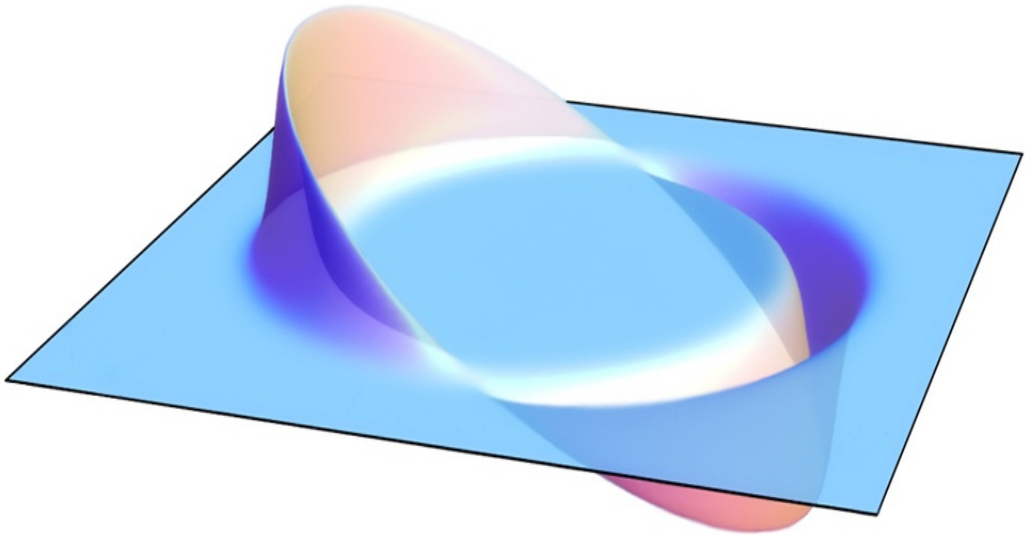

**Figure 6.4.1.** Warping of Spacetime in the Alcubierre drive
https://de.wikipedia.org/wiki/Datei:Alcubierre.png

We can prove that the trajectory of the Spaceship is like a timelike curve regardless of the value of $v_s(t)$, we substitute $x = x_s(t)$ in the metric (6.4.9). We can easily notice that the trajectory we will have is

$$\mathrm{d}\tau = \mathrm{d}t \tag{6.4.14}$$

This not only implies that the spaceship moves on a timelike curve, but also that its proper time is equal to its coordinate time. Since the coordinate time is also equal to the proper time of distant observers in the flat region, we conclude that the spaceship suffers no time dilation as it moves!

It can also be shown that the observers with four-velocity

$$U^m = (1, 0, 0, vf) \Rightarrow U_m = (1, 0, 0, 0) \tag{6.4.15}$$

moves on a geodesic as their acceleration is zero i.e. $a^m = U^m \nabla_m U_n = 0$. These are the observers denoted as *Eulerian observers*.

Moreover, it is not difficult to convince one that when the parameter $\sigma$ is large, the tidal forces in the immediate vicinity of the spaceship are very small (provided that $R$ is larger than the size of the spaceship). The region where $r_s \cong R$, the tidal forces can be very large indeed.

## 6.4.4. Using the Warp drive

To see how one can use this metric to make a round trip to a distant star in an arbitrary small time, let us consider the following situation : Two stars $A$ and $B$ are separated by a distance $D$ in flat spacetime. At time $t_0$, a spaceship starts to move away from $A$ at a speed $v < 1$ using its rocket engines. Then the spaceship stops at a distance $d$ away from $A$. We assume that $d$ is such that

$$R \ll d \ll D \tag{6.4.16}$$

At this point we feel a disturbance of the spacetime as described in the previous sections. The disturbance is such that the spaceship is pushed away from $A$ with a rapidly changing coordinate acceleration from 0 to $a$ where $a = $ constant. We had assumed that the spaceship is initially at rest ($v_s$=0), this disturbance will develop smoothly from a flat spacetime before using the equation which we mentioned before

$$ds^2 = -dt^2 + (dx - v_s f(r_s) dt)^2 + dy^2 + dz^2$$

When we are halfway between $A$ and $B$, the disturbance is modified in such a way that the coordinate acceleration in the second part of the trip is exactly opposite to the one we had in first part of the trip. As a result of this the spaceship will eventually find itself at rest at a distance $d$ away from $B$. This is the instance when the disturbance will disappear (as $v_s = 0$) again. We complete the journey now by moving through flat spacetime at $v$.

If we assume each of the changes in acceleration are extremely rapid, then the total coordinate time $T$ , elapsed in the one way trip will be essentially given by

$$T = 2\left[\frac{d}{v} + \sqrt{\frac{D - 2d}{a}}\right] \tag{6.4.17}$$

Since both stars remain in flat space, their proper time is equal to coordinate time. On the other hand, the proper time measured on the spaceship will be

$$T = 2\left[\frac{d}{\gamma v} + \sqrt{\frac{D - 2d}{a}}\right] \tag{6.4.18}$$

with $\gamma = (1 - v^2)^{\frac{1}{2}}$.

We now see that the time dilation comes only from the initial and final stages of the trip, when the spaceship moves through flat spacetime. Now, if (6.4.16) hols true, we will have

$$\tau \simeq T \simeq 2\sqrt{\frac{D}{a}} \tag{6.4.19}$$

Now, it is clear that $T$ can be made as small as we want by increasing the value of $a$. Since a round trip will only take twice as long, we find that we can be back at star $A$ after an arbitrarily small proper time, both from the point of view of the spaceship and from the point of view of the star. The spaceship will then be able to travel much faster than the speed of light. However, as we have seen, it will always remain on a timelike trajectory, that is, inside its local light cone: light itself is also being pushed by the distortion of spacetime. A propulsion mechanism based on such a local distortion of spacetime just begs to be given the familiar name of the warp drive in science fiction.

### 6.4.5. Drawback of the Warp drive

The metric we just described that does all these fascinating things has a big drawback, it violates all three energy conditions (WEC, SEC and DEC). Both weak and dominant energy conditions require energy density to be positive for all observers. If one calculates the Einstein tensor from the metric, and uses the fact that the four-velocity of the Eulerian observers is given by :

$$n^a = \frac{1}{\alpha}(1, -\beta^i), \quad n_a = (-\alpha, 0) \tag{6.4.20}$$

then one can show that all these observers will see an energy density given by :

$$T^{ab} n_a n_b = \alpha^2 T^{00} = \frac{1}{8\pi} G^{00} = -\frac{1}{8\pi} \frac{v_s^2 \rho^2}{4 r_c^2} \left( \frac{\mathrm{d}f}{\mathrm{d}r_s} \right)^2 \tag{6.4.21}$$

The fact that this expression is everywhere negative implies that the weak and dominant energy conditions are violated. In a similar way one can show that the strong energy condition is also violated.

### 6.4.6. Conclusion for the Warp drive

We see that, just like in the case of wormholes, one need exotic matter to travel faster than the speed of light. However, if one believes that exotic matter is forbidden, this is true classically. Quantum field theory permits the existence of regions with negative energy densities in some special circumstances (as, for example, in the Casimir effect).

The need of exotic matter therefore doesn't necessarily eliminate the possibility of using a spacetime distortion like thee one described above for hyper fast interstellar travel.

The spacetime given by the metric (6.4.9) is indeed Globally hyperbolic and hence it contains no closed causal curves. Although it is not difficult to construct such a spacetime which contain such curves. The author wanted to include this as a comment to the previous section and it would also add on to the discussion of the causal structure of a spacetime but the details of such a calculation are a bit intricate and time consuming. The details about such curves can be read from the following source : Page 15, [14].

---***---

# Appendix A

# Symmetries and Lie Derivative

## A.1. Maps between Manifolds

**Definition A.1.1.** *Pullback of a function.*

*Consider two Manifolds $M$ and $N$. Let $x^\alpha$ and $y^\alpha$ be the coordinate systems/charts on them respectively. Let $\varphi$ be a map from $M$ to $N$ and $f$ be a function from $N$ to $\mathbb{R}$.*

$$\varphi \colon M \;\to\; N$$
$$f \colon N \;\to\; \mathbb{R}$$

*Using all the things that we defined above we can construct a schematic like this,*

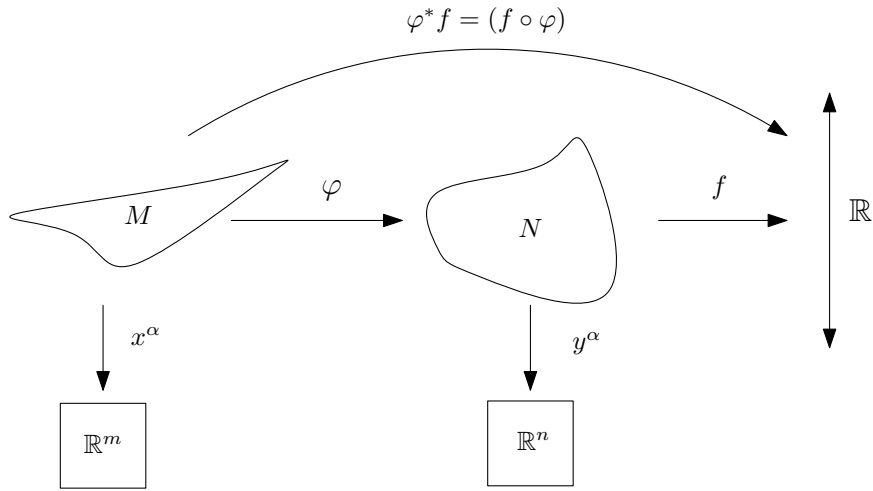

**Figure A.1.1.**
Pullback of a function

*The construction $f \circ \varphi$ is quite obvious looking at the schematic. It basically "pulls back" the function $f$ from $N$ to $M$. It is used so often that it gets its own name and notation.*

$$\varphi^* f \;=\; (f \circ \varphi) \tag{A.1.1}$$
$$\varphi^* f = (f \circ \varphi) \colon M \;\to\; \mathbb{R}$$

*Equation (A.1.1) is called the **pullback** of $f$ by $\varphi$ from $N$ to $M$.*

**Remark A.1.1.** We can pull functions back but we cannot push them forward. Basically if we have $\gamma \colon M \to \mathbb{R}$ there is no way that we can compose $\gamma$ with $\varphi$ to create a function on $N$.

**Definition A.1.2.** *Push forward of a vector*

*Given two manifolds $M$ and $N$ let $\varphi$ be a smooth map from $M$ to $N$. For some point $p$ in $M$ and $q$ in $N$ let $T_pM$ and $T_qN$ be the tangent space at $p$ and $q$ in $M$ and $N$ respectively. Let $\varphi(p) = q$. Then the **pushforward** of a vector $v^a \in T_pM$ is a vector $\varphi_* v^a \in T_qN$ defined by,*

$$\varphi_* v^a(f) = v^a(f \circ \varphi) \tag{A.1.2}$$

*for all smooth functions $f\colon N \to \mathbb{R}$*

We can rewrite this using the definition of the pullback of a function.

$$\varphi_* v^a(f) = v^a(\varphi^* f) \tag{A.1.3}$$

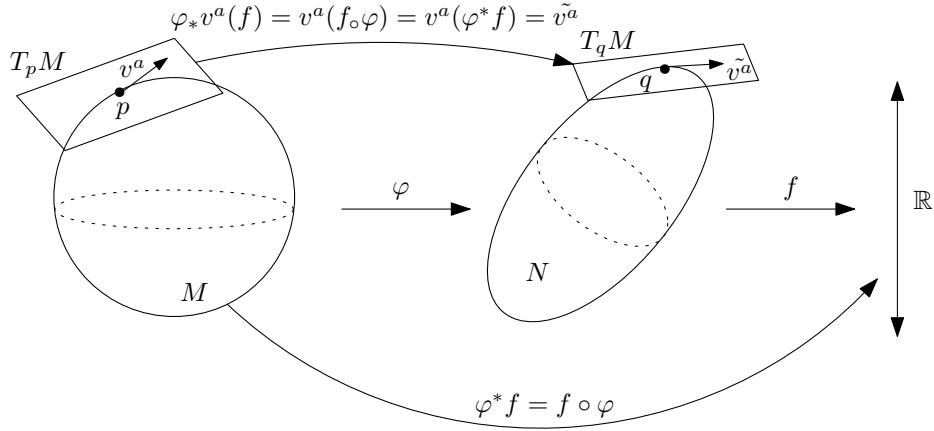

**Figure A.1.2.** *Pushforward of a vector*

# A.2. Lie Derivative

**Definition A.2.1.** *Integral curves*

Let $(M, g)$ be a Lorentzian manifold. Given a smooth vector field $\xi^a$ on M, an integral curve of the vector field $\xi^a$ is the curve

$$\gamma\colon \mathbb{R} \to M$$

*whose tangent vector is equal to $\xi^a$ at every point $p \in \gamma$. Let $\gamma$ be parameterized by $\lambda$ giving us $\gamma(\lambda)$ as our curve.*

We basically require this condition mathematically assuming we have smooth functions $f\colon M \to \mathbb{R}$, (The second bracket $(p)$ on both LHS and RHS in the upcoming parts means "at a point $p$")

$$(\xi^a(f))(p) \;=\; \left(\frac{\mathrm{d}}{\mathrm{d}\lambda}(f \circ \gamma(\lambda))\right)(p)$$

*If we have a coordinate system $x^\alpha$ on M, the components $\xi^\alpha$ of $\xi^a$ must satisfy*

$$\xi^\alpha(p) \;=\; \left(\frac{\mathrm{d}}{\mathrm{d}\lambda}x^\alpha(\gamma(\lambda))\right)(p)$$

**Remark A.2.1.** If $\xi^a$ is smooth and non-zero everywhere then the integral set of curves on $\xi^a$ form a congruence. Basically a congruence means that every point $p \in M$ lies on a unique integral curve.

**Note A.2.1.** Given any congruence we can associate it with a one-parameter family of diffeomorphisms from $M \to M$. They are defined as follows ,

For each $s \in \mathbb{R}$ define $h_s\colon M \to M$ where $h_s(p)$ is a point away from $p$ at a parameter distance $s$ along $\xi^a$.

$$\text{If } p \;=\; \gamma(\lambda_0)$$
$$\text{then}$$
$$h_s(p) \;=\; \gamma(\lambda_0 + s)$$

These transformations form an abelian group. Composition law in abelian group is defined as

$$h_s \circ h_t = h_{s+t}$$

Identity map is $h_0$ and the inverse is defined as

$$(h_s)^{-1} = h_{-s}$$

Using the definitions and remarks above we can define a really important concept in differential geometry known as the *Lie derivative*.

DEFINITION A.2.2. *Lie Derivative along a vector field*

*The lie derivative along a vector field $\xi^a$ denoted by $\mathcal{L}_{\xi^a}$. Applied to a vector $v^a$ at point $p$ it is defined as*

$$(\mathcal{L}_{\xi^a} v^b)(p) \;=\; \lim_{\delta\lambda \to 0} \frac{(v^b)(p) - (h_{\delta\lambda})_*(v^b(h_{-\delta\lambda}(p)))}{\delta\lambda}$$

Where the term $(h_{\delta\lambda})_*(v^b(h_{-\delta\lambda}(p)))$ might seem a bit confusing but it makes sense after breaking it down.

- $(h_{\delta\lambda})_*$ is pushing forward a vector at a distance $\delta\lambda$ along $\xi^a$ *(Integral curve)*
- $p$ is the point at which we have $v^b$ and $h_{-\delta\lambda}(p)$ is a point $-\delta\lambda$ away from $p$ along $\xi^a$.

## A.3.  BIBLIOGRAPHICAL NOTES

The author used [3],[1],[2] as reference while writing this appendix. It was kept short and only the definitions used from such topics were illustrated here.

—***—

# APPENDIX B
## PENROSE-CARTER DIAGRAMS

The main purpose of this appendix is to cover the idea with an example of how to draw Penrose diagrams. Penrose diagrams are an essentially helpful tool to visualize the entire spacetime on a finite sheet of paper.

## B.1. ALGORITHM FOR CONSTRUCTING PENROSE DIAGRAMS

**The Algorithm.**

1. Write down the given spacetime metric in some coordinate system and keep a track of the coordinate ranges.
2. Now redefine the original non-compact[B.1.1] coordinates by two new (still non-compact) but *null* coordinates $v$ and $w$.
3. Introduce new coordinates $p, q$ to compactify $v$ and $w$ individually.

$$p \equiv \tan^{-1}(v)$$
$$q \equiv \tan^{-1}(w)$$

for which we get the following ranges for the coordinates,

$$p \in \left(-\frac{\pi}{2}, \frac{\pi}{2}\right)$$
$$q \in \left(-\frac{\pi}{2}, \frac{\pi}{2}\right)$$

4. Define a temporal $(T)$ and spatial $(X)$ coordinate.

$$T = p + q$$
$$X = p - q$$

5. Express the metric $g_{mn}$ in the coordinates $(T, X)$ and denote it as $\tilde{g}$.

   If we have $g = \tilde{g}\,\Omega^{-2}(T, X)$ then we can obtain the non-physical diagram metric by dropping the $\Omega$ factor.

6. Draw the diagram using $(T, X, ...)$ coordinates.

**Remark.**

- We choose null coordinates in step(2) and compactifies them in step(3) above because this maintains the core structure.
- In step(5) we talked about dropping the $\Omega$ factor. This works if $\Omega \neq 0$ everywhere. $\Omega$ is called the conformal factor.

   Dropping the conformal factor does change the shape of the timelike and spacelike geodesics but not of the null geodesics.

   Precisely,

---

B.1.1. By non-compact we mean that the range of at least one of the coordinate goes to infinity and so we cannot draw it on a finite sheet of paper.

$\gamma$ is a null geodesic of the metric $g_{mn}$ iff $\gamma$ is a null geodesic of the metric $\Omega^2 g$ where $\Omega$ is non-vanishing smooth function everywhere on $M$.

## B.2.   Example : Penrose diagram for Minkowski metric

### B.2.1.   Drawing the diagram.

The simplest vacuum solution for the Einstein's equations is the Minkowski spacetime in the coordinates $(t, r, \theta, \varphi)$

Let's use the algorithm we defined above in B.1 as follow it one step at a time.

1. We have the metric :

$$ds^2 \;=\; -dt^2 + dr^2 + r^2(d\theta^2 + \sin^2\theta \, d\varphi^2)$$
$$\text{in the coordinates} \;\rightarrow\; (t \in (-\infty, +\infty), r \in (-\infty, +\infty),$$
$$\theta \in (0, \pi), \varphi \in (0, 2\pi))$$

Now before continuing further, we only consider $(t, r)$ coordinates and suppress the unit sphere shell $r^2(d\theta^2 + \sin^2\theta \, d\varphi^2)$. So every point in our $(t, r)$ diagram will be a like a suppressed sphere.

2. Introduce the null coordinates :

$$v \;=\; t + r$$
$$w \;=\; t - r$$

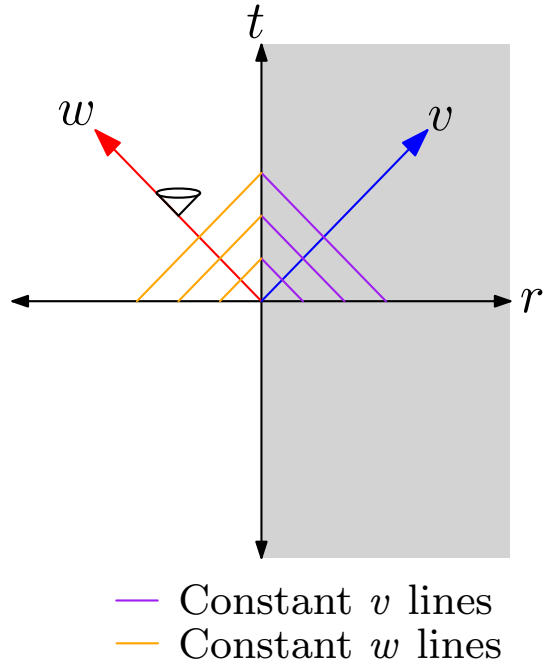

— Constant $v$ lines
— Constant $w$ lines

**Figure B.2.1.**
Minkowski spacetime $(v, w)$

We also get the following equation by rearranging the ones above,

$$r = \frac{1}{2}(v - w) > 0$$

for $v, w \in \mathbb{R}$. So the shaded part of the diagram can be covered by $v > w$ coordinates.

3. Compactify $(v, w)$ :

$$p \equiv \tan^{-1}(v)$$
$$q \equiv \tan^{-1}(w)$$

We have $p > q$ and $p, q \in \left(-\dfrac{\pi}{2}, \dfrac{\pi}{2}\right)$ as the constraints on this $(p, q)$ coordinate system.

4. Construct the spatial and temporal coordinates for convenience.

$$T \equiv p + q$$
$$X \equiv p - q$$

The constraints here are,

$$T > X \implies \qquad -\frac{\pi}{2} < \frac{1}{2}(T+X) < \frac{\pi}{2}$$
$$-\frac{\pi}{2} < \frac{1}{2}(T-X) < \frac{\pi}{2}$$
$$\implies \qquad -\pi < (T+X) < \pi$$
$$-\pi < (T-X) < \pi$$

which gives us,

$$(T+X) > (T-X)$$

We also get the result after some basic algebra :

$$2X > 0 \implies X > 0$$

5. Now we express the metric in $(T, X, \theta, \varphi)$.

How do we do that?

Use the metric $g_{\mu\nu}$ in the coordinates $x^\mu = (t, r, \theta, \varphi)$ as a starting point. Then use transform this metric to be the metric $g'_{\mu\nu}$ in the coordinates $y^\mu = (v, w, \theta, \varphi)$ using the following transformation formula :

$$g'_{\alpha\beta} = \sum_{\mu,\nu} \frac{\partial x^\mu}{\partial y^\alpha} \frac{\partial x^\nu}{\partial y^\beta} g_{\mu\nu} \tag{B.2.1}$$

and we get the following term :

$$\frac{\partial x^\mu}{\partial y^\nu} = \begin{bmatrix} \dfrac{1}{2} & \dfrac{1}{2} & 0 & 0 \\ \dfrac{1}{2} & -\dfrac{1}{2} & 0 & 0 \\ 0 & 0 & 1 & 0 \\ 0 & 0 & 0 & 1 \end{bmatrix}$$

Using this and B.2.1 we get

$$g'_{\mu\nu}(v, w, \theta, \varphi) = \begin{bmatrix} 0 & \dfrac{1}{2} & 0 & 0 \\ \dfrac{1}{2} & 0 & 0 & 0 \\ 0 & 0 & \left(\dfrac{1}{2}(v-w)\right)^2 & 0 \\ 0 & 0 & 0 & \left(\dfrac{1}{2}(v-w)\sin\theta\right)^2 \end{bmatrix} \tag{B.2.2}$$

Therefore getting the metric :

$$g'_{\mu\nu} \;=\; \frac{1}{2}\mathrm{d}v\,\mathrm{d}w + \frac{1}{2}\,\mathrm{d}w\,\mathrm{d}v + \left(\frac{1}{2}(v-w)\right)^2 \mathrm{d}\theta^2 + \left(\frac{1}{2}(v-w)\sin\theta\right)^2 \mathrm{d}\varphi^2$$

We keep on doing this till be get the metric in the coordinates $(T, X, \theta, \varphi)$. Doing the same exercise as we did from $g_{\mu\nu} \to g'_{\mu\nu}$ we go from $g'_{\mu\nu} \to \cdots \to \tilde{g}_{\mu\nu}(T, X, \theta, \varphi)$ and get the following equation,

$$\tilde{g}_{\mu\nu} \;=\; \frac{1}{4}\frac{1}{\tan^2\!\left(\frac{1}{2}(T+X)\right)}\frac{1}{\tan^2\!\left(\frac{1}{2}(T-X)\right)}\,g_{\mu\nu}$$

where

$$\Omega \;=\; \frac{1}{4}\frac{1}{\tan^2\!\left(\frac{1}{2}(T+X)\right)}\frac{1}{\tan^2\!\left(\frac{1}{2}(T-X)\right)}$$

is non zero everywhere. (Looking at the ranges of $T$ and $X$)

So this indeed is a spacetime for which we can draw a Penrose diagram.

6. Draw the diagram.

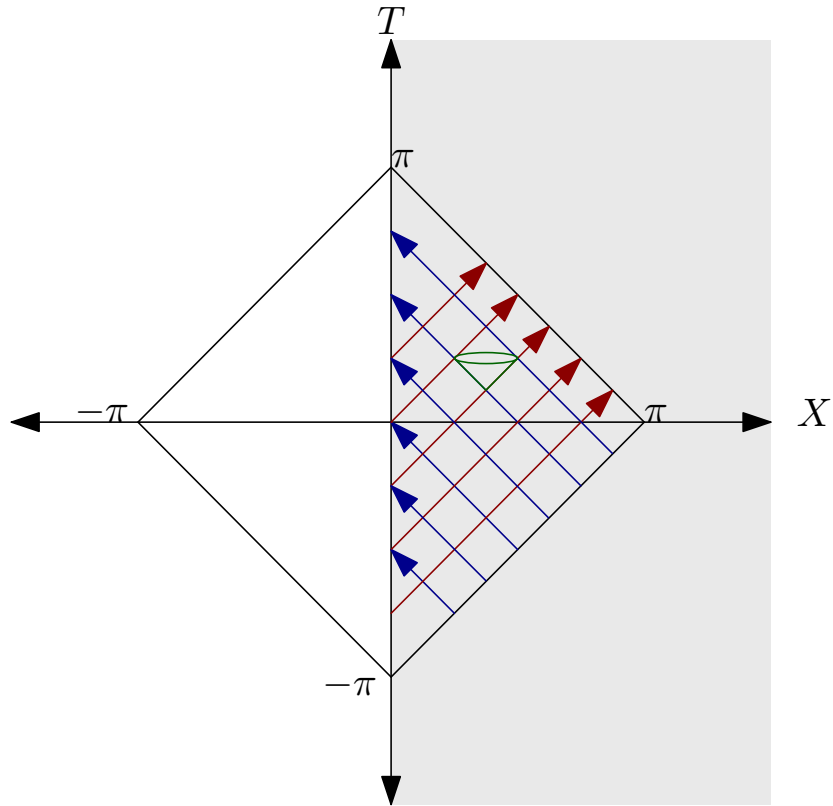

**Figure B.2.2.** Minkowski spacetime $(T, X)$

In the figure above, the red and the blue lines are the null geodesics.

## B.2.2. Analyzing the diagram.

In this section let us redraw the diagram with only the shaded region and label it with some technical details.

The null geodesics remain the same as in diagram B.2.2.

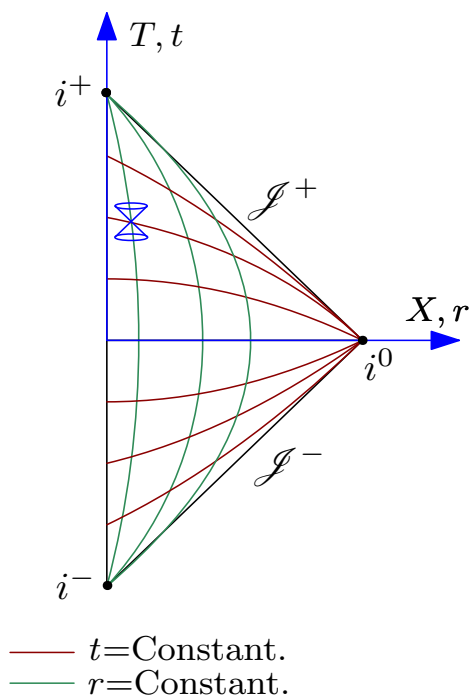

**Figure B.2.3.** Timelike and Spacelike geodesics in Minkowski spacetime.

Let us now discuss the importance/meaning of these specific points :

- $i^0$ is called the *Spacelike infinity*
    - It is the point in spacetime where all the spacelike geodesics start and end.
- $i^+$ is called the *future timelike infinity*.
    - It is the point in spacetime where all the timelike geodesics start.
- $i^-$ is called the *past timelike infinity*.
    - It is the point in spacetime where all the timelike geodesics start.
- $\mathscr{J}^+$ is called the *future null infinity*.
    - It is the point in spacetime where all the null geodesics end.
- $\mathscr{J}^-$ is called the *past null infinity*.
    - It is the point in spacetime where all the null geodesics start.

## B.3. Bibliographical notes

The author used [2] as a reference while making this appendix. A lot of the calculations were missing in [2]. The algorithm was in a way constructed by the author with the help of a few lectures on General relativity by Friedrich Schuller which are available online on youtube. They can be found on this link

https://www.youtube.com/channel/UCUHKG3S9N_QeIE2jQXd2-VQ .

—***—

# Appendix C

## Singularity Theorems : History

In this section we will summarize some sections of the papers *1965 Singularity theorem* [11] by *David Garfinkle* and *Jose M.M Senovilla* which review the *history, impact and legacy of the Penrose singularity theorem* and *Singularity theorems in General relativity-Achievements and open questions* by *Jose M.M Senovilla* [15]. A lot of examples here will refer contextually to cosmological solutions of Einstein's field equations. It would be recommended if the reader has some idea about the *Cosmological solutions to Einstein's field equations*.

We will be mainly summarizing [11] with comments from [15] as needed.

As the name of the paper suggests, 1960's were important years for Einstein's theory of Gravity and were called the revival year of General relativity. The year 1965 in particular was given an emphasis in the title of the paper as the first papers interpreting the Cosmic microwave background and modern singularity theorems were published in this year. The former being a relic of the Big bang epoch and the later being an essential and effective tool in dealing with singularities. Both the discoveries were made in 1964, exactly 50 years after the postulation of Einstein's field equations ((2) are without cosmological constant),

$$
\begin{aligned}
R_{mn} - \frac{1}{2} R g_{mn} + \Lambda\, g_{mn} \;&=\; \frac{8\pi G}{c^4} T_{mn} \\
&=\; 8\pi T_{mn}
\end{aligned}
\tag{C.0.1}
$$

Penrose theorem is an extremely important milestone in General relativity and Mathematical Physics. There were many different techniques to deal with singularities before Penrose's theorem which we will discuss in the upcoming sections. The reason why Penrose's theorem is so highly regarded is because it introduced the idea of geodesic incompleteness to characterize singularities. It also used the ideas of Cauchy surfaces from Lorentzian geometry which eventually led to the elegant idea of closed trapped surfaces. Trapped surface is an extremely important idea that led to numerous developments in Black hole mechanics and all different fields of relativity.

Singularity theorems was apparently one of the first ideas in classical general relativity which was not foreseen by Einstein from any perspective.

## Pre - Penrose Singularity Theorems

Since the birth of general relativity singularities had been always in the picture. It didn't took long time to chalk out that some of these singularities were coordinate dependent and could be resolved (like the coordinate singularity at the event horizon of a Schwarzschild black hole). Let's look historically how these singularities came in the picture.

## C.1. Friedman closed model and it's de Sitter solution

Long before the idea of singularity theorems in 1922, Friedman looked for solutions of the Einstein's field equations with the form

$$ds^2 = -F dt^2 + a^2(t)(d\chi^2 + \sin^2\chi \, d\Omega^2) \qquad \text{(C.1.1)}$$

where

$$
\begin{aligned}
a(t) &\;:\; \text{Scale factor (Cosmology reference)} \\
F &\;:\; \text{Some arbitrary function} \\
d\Omega^2 &\;:\; \text{Standard metric for a unit round sphere}
\end{aligned}
$$

The energy momentum tensor for this model was described by pressure-less matter i.e. dust. In this case we have $T_{tt}$ as the unique non-identically vanishing component.

Friedman proved that for such a model there were only two possible solutions for such a model,

i. Einstein static universe by

$$F = \text{Constant}$$

ii. De sitter universe ($T_{tt} = 0$ and $\Gamma > 0$)

$$F = c^2 \cos^2\chi$$

He also discovered that there are also dynamical solutions that require $F = c^2$ and a scale factor $a(t)$ satisfying the following set of ODEs

$$\frac{8\pi G}{c^2} T_{tt} + \Lambda = \frac{3}{a^2}(\dot{a} + 1) \qquad \text{(C.1.2)}$$

$$\Lambda = 2\frac{\ddot{a}}{a} + \frac{1}{a^2}(\dot{a}^2 + 1) \qquad \text{(C.1.3)}$$

(Dot represents derivative w.r.t $t$)

After a quick computation one can find the explicit form of $T_{tt}$ to be

$$T_{tt} = \frac{3}{8\pi}\frac{A}{a^3}$$

where $A = $ constant.

Analyzing the different $a(t)$ solutions gives us the condition that when

$$\Lambda < 4\pi T_{tt}$$

we get $a \to 0$ for a finite value of $t$. This is a failure of spacetime itself as spatial part in (C.1.1) *really* vanishes and energy density diverges. At this point Friedman talked about a concept called as *Creation time*. We will bring this concept up which basically tries to explain the situation before in the upcoming section.

While trying a way around this issue, Friedman realized that $A = 0$ solves this giving us another solution with vanishing energy density. This solution is described by

$$a = \lambda \cosh\frac{t}{\lambda}$$

$$\Lambda = 3\lambda^2$$

This solution has a spacetime of positive constant curvature just like the one in the original de Sitter solution. Hence, they must be isometric.

There is a small problem with (C.1.1) with $F = c^2 \cos^2\chi$ at $\chi = \frac{\pi}{2}$. This time, the temporal part of the metric vanishes and that means only $\left(0, \frac{\pi}{2}\right)$ or $\left(\frac{\pi}{2}, \pi\right)$ should be allowed for $\chi$. This makes us omit a point in the manifold with some artificial tampering.

A few workarounds were found but none of them were satisfying enough. Penrose in his 1965 paper *Gravitational collapse and spacetime singularities* took care of this problem with a more deeper interpretation of this kind of behavior.

## C.2. Lemaitre : Extension of Schwarzschild' solution and Big bang models.

In 1927 Lemaitre constructed a model that combined the Einstein static and de Sitter universe in a way that at large past times the model approached the former and in future distant times it approached the latter. He created a general solution of the Field equations for dust and spherical symmetry and found many interesting anomalies, particularly the instability of Einstein static universe, mainly a singularity of the Friedman type models we discussed in the previous section (the one with the idea of *creation time*).

Astonishingly, he gave up on Spherical symmetry as he believed that the singularity at the event horizon was a result of *excess symmetry*. He instead studied the spatially spatial homogeneous but anisotropic models (today known as the Binachi I models).

In the very same paper he proved and understood in detail the non-singular nature of the coordinate dependent singularity of the Schwarzschild event horizon.

He managed to write a general solution of the spherically symmetric vacuum field equations as

$$\mathrm{d}s^2 = -c^2\mathrm{d}\tilde{t}^2 + \left(\frac{\alpha}{r} + \frac{\Lambda}{3}r^2\right)c^2\mathrm{d}\chi^2 + r^2\,\mathrm{d}\Omega^2 \tag{C.2.1}$$

where

$$r^3 = \frac{\alpha}{\lambda}\sinh^2\left(3\lambda c\frac{(\tilde{t} - \chi)}{2}\right) \tag{C.2.2}$$

with $\alpha$ as a constant.

One can easily inspect this and see that this solution is only singular at $r = 0$, therefore successfully removing the so called *Schwarzschild singularity at the event horizon*.

He then found a change of coordinates which brought the line element (C.2.1) into

$$\mathrm{d}s^2 = -\left(1 - \frac{\alpha}{r} - \frac{\Lambda}{3}r^2\right)c^2\,\mathrm{d}\hat{t}^2 + \left(1 - \frac{\alpha}{r} - \frac{\Lambda}{3}r^2\right)^{-1}\mathrm{d}r^2 + r^2\,\mathrm{d}\Omega^2 \tag{C.2.3}$$

We can easily recall from the equation (1.1.26) from chapter (1) which was exactly the same expression with $\Lambda = 0$ as we were not taking cosmological parameters in the Einstein's Field equations while deriving them.

In this solution there appears a hypersurface at $r = \alpha$ just like the one with $\chi = \frac{\pi}{2}$ in the de Sitter spacetime from previous section. Hence the metric is an explicit solution for the Schwarzschild solution for regions $r > \alpha$. He made a distinctive remark saying that the *problem* was due to the assumption of the spherically symmetric spacetime being static. The remarks being

*We show that $(r = \alpha)$ singularity of the Schwarzschild exterior is an apparent singularity, analogous to that which appears at the horizon of the center in the original form of the de Sitter universe.*

   and

*The same singularity as above of the Schwarzschild is thus a fictitious singularity, analogous to that which appears at the horizon of the center in the original form of the de Sitter universe.*

The question *why* does this happen stayed for decades. Actually, in 1924 Eddington had found another extension (today known as the Kerr-Schild form). This was the basis of the Eddington-Finkelstein extension which we discussed in section [1.4]. The same idea was later extrapolated to understand the Kruskal extension which was discussed in section [1.5]. All the extensions have to give up staticity if the regions $r < \alpha$ are to be included, where $4\pi r^2$ is the area of the preferred round spheres. All of these extensions show that there is something unusual going on with round spheres in the $r < \alpha$ regions because their area function $4\pi r^2$ can be seen as a *timelike coordinate* there. We didn't mention this in the previous section be something extremely similar also happens in the de Sitter case where we can explicitly calculate the causal structure of such a function by taking its gradient and observing that it behaves *timelike* in that case.

As a conclusion we can say that, Lemaitre wanted so solve a contradiction between the spherically symmetric Friedman's solutions, where the area of the round spheres can become as small as required. We would require the existence of a minimum value for such an area if one accepts the $r > \alpha$ restriction of the static Schwarzschild solution. This argument can lead to the statement, *Something as massive as the Universe cannot have a radius less than one billion light years* (from the $r > \alpha$ argument). In the next section we will talk about the Oppenheimer-Snyder models where we discuss the arguments in some detail.

## C.3. Oppenheimer-Snyder model

In the previous two models we saw that singularities in the past of our world appeared in the simplest models of the classical General Relativity if the Universe is expanding everywhere. One can consider a time-reversal of this situation : what happens in contracting worlds? This turned out be be of enormous relevance for the study of compact stars as in 1931 Chandrasekhar unexpectedly found out an upper limit for white dwarf stars in equilibrium (even after taking quantum effects into account i.e. the neutron degeneracy pressure). This is known as *Chandrasekhar mass*. Any star with a larger mass than this will inevitably collapse. Later on, the question of massive neutron cores was analyzed by using a cold Fermi gas equation of state and GR. Another mass limit for equilibrium was found and it was concluded that, even allowing for deviations from the Fermi equation of state, a massive enough neutron star will contract indefinitely never reaching equilibrium again.

This made Oppenheimer and Snyder reconsider the solutions that described such physical processes. They proved using general arguments that, in spherical symmetry, values of $r = \alpha$ would eventually be reached, and the light emitted from the star would be more and more red-shifted for external observers. These external observers would only see the star approach $r \to \alpha$ asymptotically and that the entire process will last a finite amount of time for observers co-moving with stellar matter.

Then they constructed an analytical model, in which, in modern language- consists of a portion of the Friedman closed model (cosmological) for dust (i.e. $\Lambda < 0$ and $\dot{a} < 0$) matched with the Schwarzschild solution at the timelike hypersurface defined by $\chi = \chi_0 < \frac{\pi}{2}$ on the interior side- and correspondingly by a hypersurface ruled by timelike geodesics and $r = a(t)\sin\chi_0$ on the vacuum side – proving that the junction requires

$$\alpha = A \sin^3 \chi_0 \tag{C.3.1}$$

Hence,

i. The *Schwarzschild surface* $r = \alpha$ was indeed cross-able by realistic models as dust.

ii. A careful analysis of the model shows that the star will end up in catastrophic singularity where $a(t) \to 0$ and therefore *space* vanishes again.

This was the end of the *pre-singularity theorems.* Shortly after this, as we talked in the introduction, the first *singularity theorem* was published by *Raychaudhuri* and *Komar* independently.

Interested readers can read more about this in from `Page 8`, [11].

## C.4.  Bibliographical notes

This appendix was entirely taken from [11]. This is an excellent paper by *Garfinkle* and *Senovilla*, they have done an outstanding job summarizing the history of this topic.

———————————————————***———————————————————

# Bibliography

[1]    Robert M Wald. *General Relativity, University of Chicago press*. 1984.

[2]    Sean M. Carroll. *Spacetime and geometry: An introduction to general relativity*. 2004.

[3]    Fay Dowker. Lecture Notes : Black Holes. 2014/15. URL = https://www.imperial.ac.uk/media/imperial-college/research-centres-and-groups/theoretical-physics/msc/current/black-holes/bh-notes-2014_15.pdf.

[4]    Eric Poisson. *A Relativist's Toolkit: The Mathematics of Black-Hole Mechanics*. Cambridge University Press, 2004.

[5]    M. P. Hobson, G. P. Efstathiou, and A. N. Lasenby. *General Relativity: An Introduction for Physicists*. Cambridge University Press, 2006.

[6]    James Hartle. *Gravity: An introduction to Einstein's General Relativity*. Pearson Indian edition, 2014.

[7]    Ray.D Inverno. *Introducing Einstein's relativity*. Oxford University press, 1992.

[8]    S. W. Hawking and G. F. R. Ellis. *The Large Scale Structure of Space-Time*. Cambridge Monographs on Mathematical Physics. Cambridge University Press, 2011.

[9]    General relativity in Python. URL = https://www.linuxjournal.com/content/general-relativity-python.

[10]   P.L. Herman. Penrose Diagrams. March 2015. URL = http://people.uncw.edu/hermanr/GRcosmo/penrose.pdf.

[11]   José M. M. Senovilla and David Garfinkle. The 1965 Penrose singularity theorem. *Class. Quant. Grav.*, 32(12):124008, 2015.

[12]   Dana Levanony and Amos Ori. Extended time-travelling objects in Misner space. *Phys. Rev.*, D83:44043, 2011.

[13]   Miguel Alcubierre. The warp drive: hyper fast travel within general relativity. September 2000. URL = https://arxiv.org/pdf/gr-qc/0009013.pdf.

[14]   Francisco S. N. Lobo. Closed timelike curves and causality violation. *Submitted to: Class. Quant. Grav.*, 2008.

[15]   Jose M. M. Senovilla. Singularity theorems in general relativity: Achievements and open questions. *Einstein Stud.*, 12:305–316, 2012.

[16]   Gerard t'Hooft. Lecture Notes : Introduction to the theory of black holes. June 2009. URL = http://www.staff.science.uu.nl/~hooft101/lectures/blackholes/BH_lecturenotes.pdf.

[17]   Valeri P Frolov and Andrei Zelnikov. *Introduction to black hole physics*. Oxford Univ. Press, Oxford, 2011.

[18]   Hauptseminar on Quantum Field theory and Gravity, university of Leipzig.Course ID : 12-PHY-MWPHS4. Notes from the presentations taken during the course.